\documentclass{aa}  

\usepackage{graphicx}
\usepackage{txfonts}
\usepackage{subcaption}
\usepackage{xcolor}
\usepackage{amssymb, amsmath}
\usepackage{placeins}
\usepackage{parskip}
\usepackage{multirow}
\usepackage[colorlinks=true,allcolors=blue]{hyperref}
\begin{document}

    \title{Effect of redshift bin mismatch on the cross correlation between the DESI Legacy Imaging Survey and the \textit{Planck} CMB lensing potential}
    \titlerunning{DESI-LIS $\times$ \textit{Planck} - leakage correction}
%   \subtitle{I. Overviewing the $\kappa$-mechanism}

   \author{Chandra Shekhar Saraf
          \inst{1,2},
          Pawe\l{} Bielewicz\inst{3}
          \and
          Micha\l{} Chodorowski\inst{1}
          }
    \authorrunning{C. S. Saraf et al.}
   \institute{Nicolaus Copernicus Astronomical Centre, Polish Academy of Sciences, ul.~Bartycka 18, Warsaw 00-716, Poland\\
              \email{cssaraf@camk.edu.pl}
         \and     
              Korea Astronomy and Space Science Institute, 776 Daedeok-daero, Yuseong-gu, Daejeon 34055, South Korea\\
         \and
             National Centre for Nuclear Research, ul.~L.~Pasteura 7, Warsaw 02-093, Poland\\
             \email{pawel.bielewicz@ncbj.gov.pl}
             }

   \date{}

% \abstract{}{}{}{}{} 
% 5 {} token are mandatory
 
  \abstract
   {}
   {We study the importance of precise modelling of the photometric redshift error distributions {when estimating parameters from cross-correlation measurements}. {We present a working example of the scattering matrix formalism to correct for the effects of galaxies ending in wrong redshift bins due to their photometric redshift errors.}}
   {We measured the angular galaxy auto-power spectrum and cross-power spectrum in four tomographic bins with {the} redshift intervals $z = [0.0,0.3,0.45,0.6,0.8]$ from the cross-correlation of {the} \textit{Planck} cosmic microwave background lensing potential and {the} photometric galaxy catalogue from the Dark Energy Spectroscopic Instrument Legacy Imaging Survey Data Release 8. We estimated {the} galaxy linear bias and {the} amplitude of cross correlation using maximum likelihood estimation to put constraints on the $\sigma_{8}$ parameter.}
   {We show that the modified Lorentzian function used to fit the photometric redshift error distribution performs well only near the peaks of the distribution. We adopt a sum of Gaussians model to capture the broad tails of the error distribution. Our sum of Gaussians model yields {values of the} cross-correlation amplitude {that are $\sim 2-5\,\sigma$ smaller than those expected based on the $\Lambda$ cold dark matter ($\Lambda$CDM) model}. We compute the $\sigma_{8}$ parameter after correcting for the redshift bin mismatch of objects following the scattering matrix approach. The $\sigma_{8}$ parameter becomes consistent with $\Lambda$CDM model in the last tomographic bin but shows {a tension of $\sim 1-3\,\sigma$} in other redshift bins.}
   {}
    % {We elaborate the importance of precise modelling of the photometric redshift error distributions and present a working example of the scattering matrix formalism to correct for the redshift bin mismatch of galaxies in tomographic analysis. We note that the scattering matrix approach has the potential to address the $\sigma_{8}-\Omega_{m}$ tension in cosmology with tomographic cross correlation measurements.}


   \maketitle
%
%-------------------------------------------------------------------

\section{Introduction}

The standard model of cosmology or the {$\Lambda$ cold dark matter ($\Lambda$CDM) model}, well established based on a vast array of observations, provides us with a comprehensive framework to understand the structure, origin, and evolution of the Universe. The parameters governing the $\Lambda$CDM model have been constrained with unparalleled accuracy by precise measurements of the cosmic microwave background (CMB) (\citealt{BICEP2021}; \citealt{SPT-3G2021}; \citealt{Planck2020VIII}; \citealt{ACTDR42020}; \citealt{Polarbear2020}). However, despite the remarkable success of the $\Lambda$CDM model in providing a satisfactory description of the astrophysical and cosmological probes, it cannot explain some of the key ingredients in our understanding of the Universe, namely dark energy (\citealt{Perlmutter1999}; \citealt{Riess1998}), dark matter \citep{Trimble1987} and inflation \citep{Guth1980}.

The gravitational lensing of the CMB photons carries information about the growth of structure at redshifts of $1-3$. Cross correlations between tracers of the large-scale structure and CMB lensing {are excellent probes of} the evolution of the large-scale gravitational potential and {can be used to} test the validity of {the} $\Lambda$CDM model. Several works over the past decade have established the importance of such cross-correlation measurements in testing the cosmological model (\citealt{Saraf2022}; \citealt{Mitayake2022}; \citealt{Robertson2021}; \citealt{Krolewski2021}; \citealt{Darwish2021} \citealt{Abbott2019}; \citealt{Bianchini2018}; \citealt{Singh2017}; \citealt{Bianchini2016}; \citealt{Bianchini2015}). 
On the other hand, observations of the probes of large-scale structure, such as weak gravitational lensing and galaxy clustering (\citealt{DESY32022}; \citealt{Philcox2022}; \citealt{Secco2022}; \citealt{Amon2022}; \citealt{KiDS2021}; \citealt{Skara2020}; \citealt{Joudaki2017}; \citealt{Macaulay2013}) have consistently pointed out a $2-3\,\sigma$ difference in the strength of matter clustering --- quantified by the $S_{8}\equiv \sigma_{8}\sqrt{\frac{\Omega_{m}}{0.3}}$ parameter --- compared to the conclusions drawn from the CMB-only analysis \citep{Planck2020VI}.

{A large number} of studies that implemented the tomographic cross-correlation measurements --- by dividing the galaxy samples into narrow redshift bins --- reported differences in the values of the $\sigma_{8}, \Omega_{m}$, or $S_{8}$ parameters compared to the values derived from CMB measurements (see \citealt{Alonso2023}; \citealt{Zhengyi2023}; \citealt{Yu2022}; \citealt{White2022}; \citealt{Pandey2022}; \citealt{Chang2023}; \citealt{Sun2022}; \citealt{Krolewski2021}; \citealt{Hang2021}; \citealt{Marques2020}; \citealt{Peacock&Bilicki2018}; \citealt{Giannantonio2016}). Other works like \cite{Bianchini2018}, \cite{Amon2018}, \cite{Blake2016}, \cite{Giannantonio2016}, and \cite{Pullen2016} find consistent deviations in the values of $D_{g}$ and $E_{g}$ statistics when testing the $\Lambda$CDM model with different galaxy surveys. Tomographic measurements with upcoming large-scale structure surveys, such as \textit{Vera C. Rubin} Observatory Legacy Survey of Space and Time (LSST; \citealt{Ivezic2019}; \citealt{LSST2009}), \textit{Euclid} \citep{Euclid2011}, \textit{Nancy Grace Roman} Space Telescope \citep{WFIRST2013}, Dark Energy Spectroscopic Instrument (DESI; \citealt{Dey2019}), and Spectro-Photometer for the History of the Universe, Epoch of Reionization, and Ices Explorer (SPHEREx; \citealt{SPHEREX2014}) will help us {explore} the $\sigma_{8}-\Omega_{m}$ or $S_{8}$ tension in unprecedented detail.

The tomographic cross-correlation measurements with photometric galaxy surveys allow us to study the evolution of $\sigma_{8}$ or $S_{8}$ to high redshifts but are plagued by photometric redshift errors. The errors on {redshift lead to the placement of a fractions of galaxies in wrong redshift bins, referred to here as `the redshift bin mismatch' of galaxies.} The effects of redshift bin mismatch on the estimation of parameters have been explored in a few ways (\citealt{Hang2021} (hereafter H21); \citealt{Stolzner2021}; \citealt{Balaguera2018}). Another attempt to mitigate the bin mismatch of galaxy redshifts and self-calibrate the redshift distributions was proposed by \cite{Zhang2010}, {who used} the scattering matrix formalism. \cite{Zhang2017} proposed an algorithm based on {a} non-negative matrix {factorisation} method to compute the scattering matrix and correct the angular power spectra for the redshift scatter. However, the non-negative matrix {factorisation} method becomes computationally expensive with {increasing redshift bins} and data points in the angular power spectra. We proposed an alternative method in \cite{Saraf2024} for fast and efficient computations of the scattering matrix validated with Monte Carlo simulations of the LSST survey.

In the present paper, we apply the proof of concept presented in \cite{Saraf2024} to the cross correlation between {the} \textit{Planck} CMB lensing potential map and {the} photometric galaxy catalogues from Data Release 8 of the Dark Energy Spectroscopic Instrument Legacy Imaging Survey prepared by H21. {We highlight the importance of precise modelling of the photometric redshift error distributions and quantify the impact of redshift bin mismatch correction on estimation of {the} $\sigma_{8}$ parameter. We show that mitigation of redshift bin mismatch is necessary as a second-stage correction {in order to obtain} unbiased estimates of parameters. We provide a fast and accurate method for redshift bin mismatch correction.} We describe the data used in our analysis in section \ref{sec:data_paper3} and the modelling of the redshift error distributions in section \ref{sec:error_distribution_paper3}. In section \ref{sec:true_redshift_distribution_paper3}, we describe the procedure for estimating {the} redshift distribution of galaxies taking into account the photometric redshift errors. We expand {on the parameter estimation procedure} using maximum likelihood estimation in section \ref{sec:mle_paper3}. The mock catalogues used to validate our analysis pipeline {are} presented in section \ref{sec:simulations_paper3}, and various methods for computing the covariance matrix are {summarised} in section \ref{sec:covariance_matrix}. We present the key results from our analysis and comparisons with H21 in section \ref{sec:results_paper3}. Finally, in section \ref{sec:summary_paper3}, we {summarise} the findings from our work presented in this paper.

%--------------------------------------------------------------------
\section{Data}\label{sec:data_paper3}

\subsection{CMB lensing data}
We used the lensing potential maps from the third data release (PDR3)\footnote{\url{https://pla.esac.esa.int/\#cosmology}} of the \textit{Planck} collaboration \citep{Planck2020VIII}. PDR3 uses the SMICA DX12 CMB maps to reconstruct the lensing potential from the CMB temperature and polarization data covering $\sim 67\%$ of the sky. SMICA (Spectral Matching Independent Component Analysis; \citealt{Delabrouille2003}) is a foreground component separation method that {showed} the best performance during the \textit{Planck} foreground cleaning mock challenge \citep{Planck2014XII}. For our baseline analysis, we used the minimum-variance CMB lensing potential map (hereafter MV map) estimated from the inverse-variance weighting of all lensing estimators based on different correlations of CMB temperature and polarization maps. In addition, \textit{Planck} PDR3 also provides lensing potential maps estimated only from the CMB temperature measurements (hereafter {the} TT map) and with deprojection of the Sunyaev-Zeldovich sources (hereafter, {the} SZ-deproj map). We use the TT and SZ-deproj maps in section \ref{sec:diff_cmb_lensing_maps} to study the dependence of our results on the choice of the CMB lensing potential map.

The lensing potential $\phi$ was converted to lensing convergence $\kappa$, where $\kappa$ is proportional to the {two-dimensional} Laplacian of $\phi$. In spherical harmonic space, this relation can be expressed as \citep{Hu2000}
\begin{equation}
	\kappa_{\ell m} = -\frac{\ell(\ell+1)}{2}\phi_{\ell m}
	\label{eq:lensing_potential_convergence_relation}
\end{equation}
The \textit{Planck} PDR3 package provides spherical harmonics coefficients for the lensing convergence maps at {the} \texttt{HEALPix}\footnote{\url{https://healpix.jpl.nasa.gov/}} \citep{Gorski2005} resolution of ${N}_{\text{side}}=4096$, which we downgraded to ${N}_{\text{side}}=1024$ for our analysis. The data package also provides an estimate of the noise power spectrum $N_{\ell}^{\kappa\kappa}$ for the lensing convergence maps along with binary maps masking the regions of {the} sky not suitable for analysis.

\subsection{Legacy survey data}\label{sec:lis_data}

The Dark Energy Spectroscopic Instrument Legacy Imaging Survey (DESI-LIS; \citealt{Dey2019}) is a combination of three legacy imaging surveys: the Dark Energy Camera Legacy Survey (DECaLS; \citealt{DECaLS2016}), the Beijing-Arizona Sky Survey (BASS; \citealt{BASS2017}) and the Mayall z-band Legacy Survey (MzLS) observed by the Mosaic3 camera \citep{Mosaic3}, with the addition of the Dark Energy Survey (DES; \citealt{DES2005}). The legacy imaging surveys were motivated by the need to provide targets for the ongoing Dark Energy Spectroscopic Instrument (DESI; \citealt{DESI_Collab2016}) survey, {and to supplement} the Sloan Digital Sky Survey (SDSS; e.g. \citealt{Abolfathi2018}; \citealt{Abazajian2009}) photometry data. In our study, we used the galaxy catalogue and photometric redshifts prepared by H21 from the DESI-LIS Data Release 8\footnote{\url{http://legacysurvey.org/dr8/}}, covering a total area of $\sim 17\,800$ deg$^{2}$ with sources observed in three optical bands ($g$, $r$, $z$) and three WISE \citep{Wright2010} bands: $W_{1}$, $W_{2}$, and $W_{3}$. Due to {the} shallower effective depth of the $W_{2}$ and $W_{3}$ bands, H21 used only {the} $W_{1}$ ($3.4\,\mu m$) band, resulting in the following selection criteria {being} applied to the data:
\begin{enumerate}
    \item PSF-type objects are excluded which eliminates most stars and quasars.
    \item Objects are detected in four bands, i.e. \texttt{FLUX G|R|Z|W1} > 0.
    \item \texttt{MW TRANSMISSION G|R|Z|W1} are applied to the fluxes for Galactic extinction correction.
    \item Magnitude cuts are applied with $g<24$, $r<22$, and $W_{1}<19.5$, where all magnitudes are computed by $ m = 22.5-2.5\log_{10}\text{(flux)}$.
\end{enumerate}
{The galaxy density maps are often correlated with {systematics uncertainties such as those related to} survey depth, stellar density, and observational conditions. H21 generated {a} completeness map to account for the foreground contaminations at the map pixel level, including masks for bright stars, globular clusters, and incompleteness in optical bands. {As} DESI-LIS combines data from DECaLS, BASS, MzLS, and DES, there will be additional {systematics errors related to} the differences in their photometric passbands and limiting magnitudes. The magnitude cuts applied to data significantly reduced the photometric variations and correlations with various foreground contaminants (see section 3.4 of H21 for more details).}

H21 used a direct approach to estimate {the} photometric redshifts of galaxies from observed spectroscopy by assigning a redshift to a given location in multi-colour space. The spectroscopic surveys used to calibrate photometric redshifts included {the} Galaxy And Mass Assembly (GAMA) survey Data Release 2 \citep{GAMADR2_2015}, {the} SDSS-III Baryon Oscillation Spectroscopic Survey (BOSS) Data Release 12 \citep{BOSSDR12_2015}, {the} Extended Baryon Oscillation Spectroscopic Survey (eBOSS) Data Release 16 \citep{eBOSSDR16_2020}, {the} VIMOS Public Extragalactic Redshift Survey (VIPERS) Data Release 2 \citep{VIPERSDR2_2018}, and {the} DEEP2 \citep{DEEP2_2013}. Two photometric surveys, {the} Cosmic Evolution Survey (COSMOS) \citep{COSMOS_2009} and {the} Dark Energy Survey (DES) Y1 redMaGiC \citep{REDMAGIC_2018}, were also included with the spectroscopic surveys for their highly accurate photometric redshifts.

The calibration samples, except DES Y1 redMaGiC, were binned into three-dimensional grids of $g-r$, $r-z$, and $z-W_{1}$ with a pixel width of about $0.03$. The ranges of colours used in the three-dimensional grid are $-0.5 < g-r < 2.5$, $-2 < r-z < 3$, and $-2 < z-W_{1} < 4$. Pixels containing more than five objects from the calibration samples were assigned the mean redshift of these objects. The DES Y1 redMaGiC samples were used to fill {in the} empty pixels from the initial binning. When calibrating {the} redshift of galaxies, objects falling in pixels that lack a redshift calibration were excluded. This ensured {the} selection of objects that occupy the same colour space as the calibration sample. To marginalise digitisation artefacts in the redshift distribution, the assigned photometric redshifts were added with a random top-hat dither of $\pm 0.005$. The resulting redshift calibration led to $68\%$ percent of samples having photometric redshifts of within $\pm 0.027$ of their spectroscopic redshifts.
\begin{table*}[hbt!]
    \centering
    \renewcommand{\arraystretch}{1.3}
    \captionsetup{justification=centering}
    \caption{Properties of DESI-LIS tomographic bins.}
    \label{tab:DESI_data}
    \begin{tabular}{lcccccccccc} % six columns, alignment for each
    \hline\hline
    Redshift Bin & $N_{\text{obj}}$ & $\overline{n}$ [gal pix$^{-1}$] & $\overline{n}$ [gal str$^{-1}$] & median $z$ & $x_{0}$ & $\sigma$ & $a$\\
    \hline
	$0<z\leq 0.3$ & $14\,363\,105$ & 2.652 & 2.655$\times$10$^{6}$ & 0.21 & -0.0010 & 0.0122 & 1.257\\
	$0.3<z\leq 0.45$ & $11\,554\,242$ & 2.133 & 2.136$\times$10$^{6}$ & 0.38 & 0.0076  & 0.0151 & 1.104\\
	$0.45<z\leq 0.6$ & $13\,468\,310$ & 2.487 & 2.490$\times$10$^{6}$ & 0.51 & -0.0024 & 0.0155 & 1.476\\
	$0.6<z\leq 0.8$ & $7\,232\,579$ & 1.335 & 1.337$\times$10$^{6}$ & 0.66  & -0.0042 & 0.0265 & 2.019\\
    \hline
    \end{tabular}
    \tablefoot{$N_{\text{obj}}$ is the number of objects, $\overline{n}$ is the mean number of objects, and median $z$ is the median redshift of the tomographic bins. $x_{0},\sigma$ and $a$ are the best fit values of parameters defining the modified Lorentzian fit to the photometric redshift error distribution $p(z_{s}-z_{p}|z_{p})$ (taken from H21).}
\end{table*}

\begin{figure*}[htb!]
    \begin{subfigure}[b]{0.5\linewidth}
        \centering
        \includegraphics[width=\linewidth]{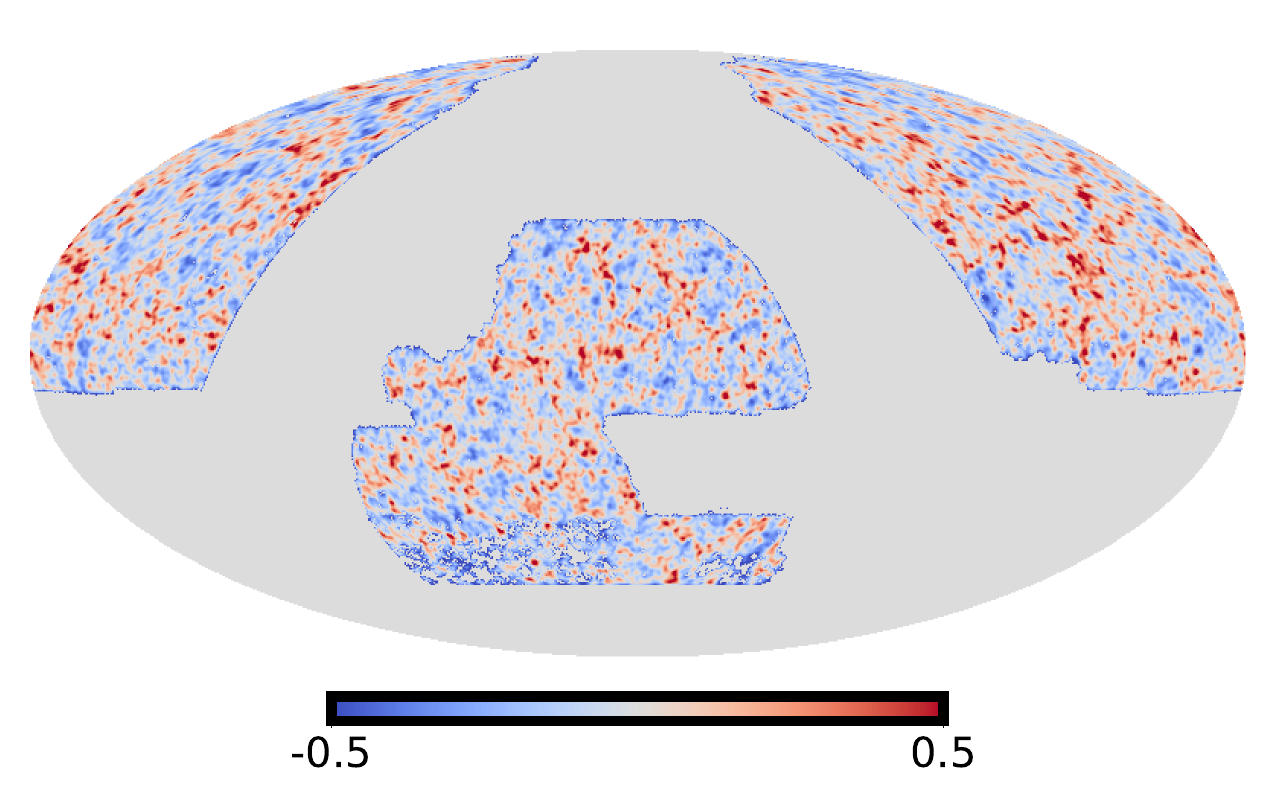}
        \captionsetup{labelformat=empty}
        \caption{\large{$0.0<z\leq0.3$}}
    \end{subfigure}%
    \begin{subfigure}[b]{0.5\linewidth}
        \centering
        \includegraphics[width=\linewidth]{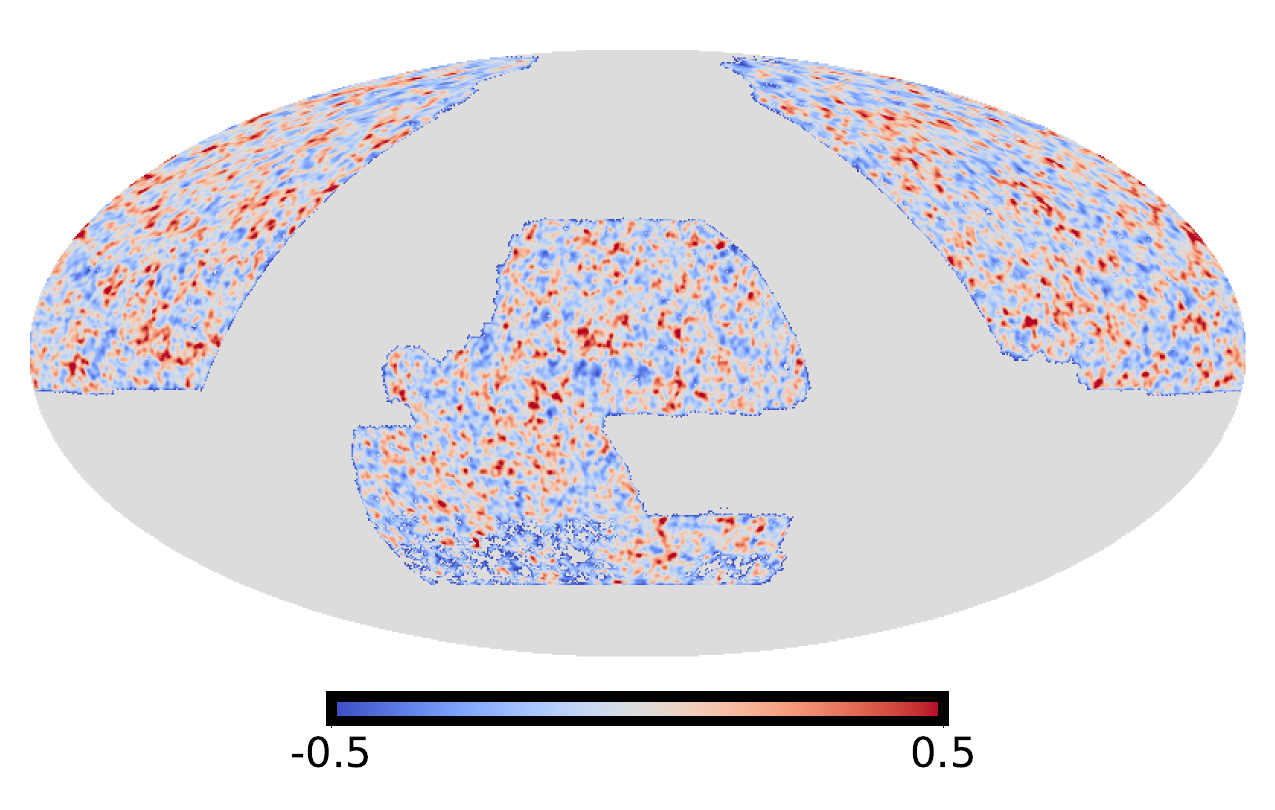}
        \captionsetup{labelformat=empty}
        \caption{\large{$0.3< z\leq0.45$}}
    \end{subfigure}\\
    \begin{subfigure}[b]{0.5\linewidth}
        \centering
        \includegraphics[width=\linewidth]{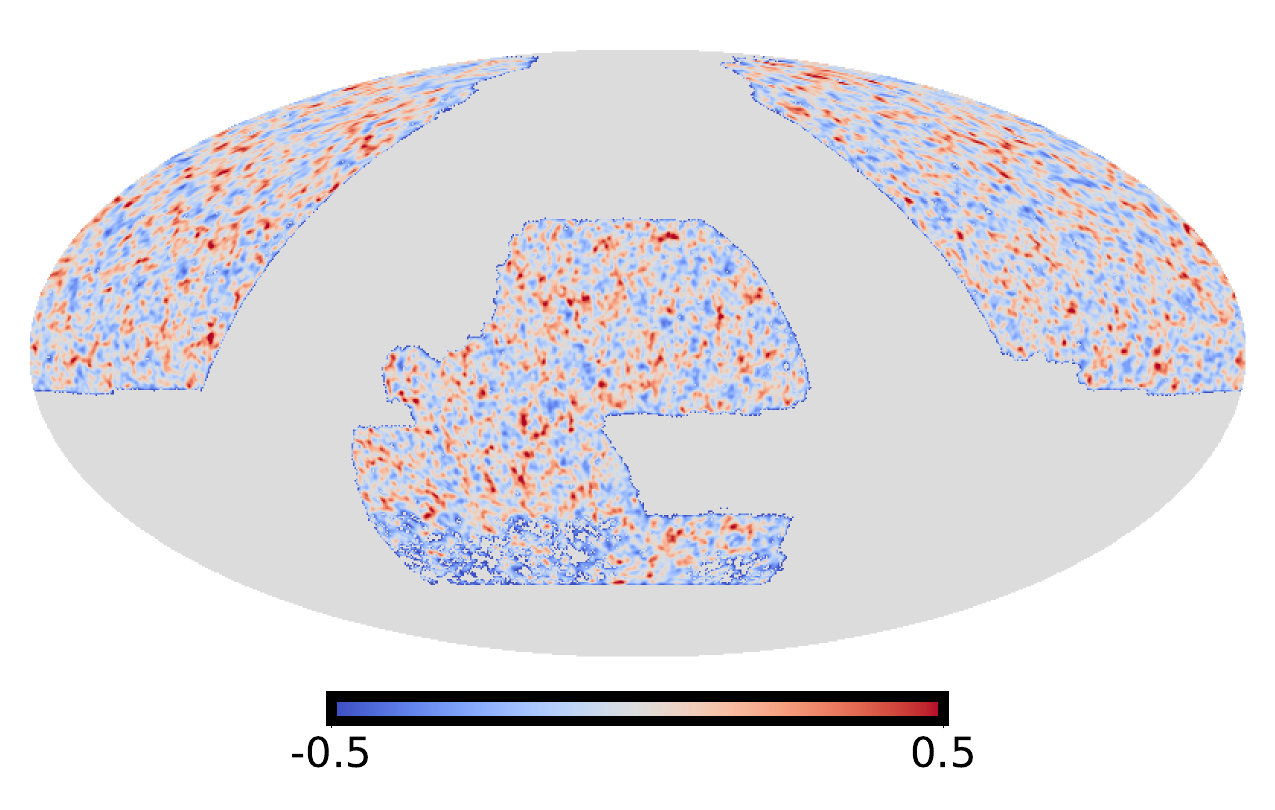}
        \captionsetup{labelformat=empty}
        \caption{\large{$0.45< z\leq0.6$}}
    \end{subfigure}%
    \begin{subfigure}[b]{0.5\linewidth}
        \centering
        \includegraphics[width=\linewidth]{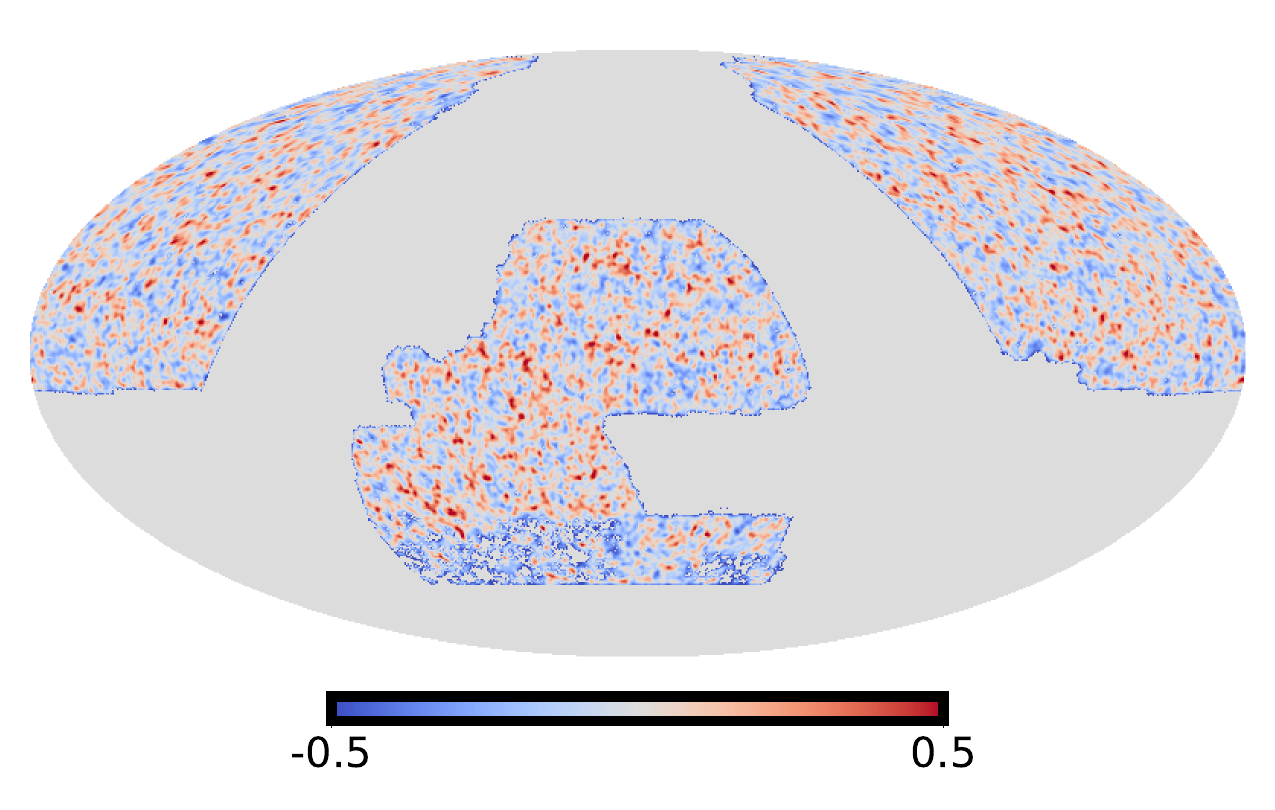}
        \captionsetup{labelformat=empty}
        \caption{\large{$0.6< z\leq0.8$}}
    \end{subfigure}
    \caption{{Galaxy over-density maps from four DESI-LIS tomographic bins. The maps have been smoothed with a Gaussian beam of $60'$ FWHM for illustrative purposes, i.e. to better show the large-scale distribution of galaxies.}}
    \label{fig:filtered_maps_desi_paper3}
\end{figure*}

The galaxies were divided into four tomographic bins based on their photometric redshifts with intervals $(0.0,0.3,0.45,0.6,0.8]$. The galaxy number count maps and the photometric redshift data used in this study are publicly available\footnote{\url{https://gitlab.com/qianjunhang/desi-legacy-survey-cross-correlations}}.
A summary of the four tomographic bins including {the} number of objects and the mean density of objects per pixel and per steradian is given in Table \ref{tab:DESI_data}. The galaxy over-density maps for every tomographic bin were {built} at the \texttt{HEALPix} resolution of ${N}_{\text{side}}=1024$ using the relation
\begin{equation}
	g(\hat{\textbf{n}}) = \frac{n(\hat{\textbf{n}})-\overline{n}}{\overline{n}}
	\label{eq:gal_overdensity_paper3}
\end{equation}
where $n(\hat{\textbf{n}})$ is the completeness-corrected number of galaxies at angular position $\hat{\textbf{n}}$ and $\overline{n}$ is the mean number of objects. The galaxy over-density maps smoothed with a Gaussian beam of $60'$ FWHM (for illustrative purposes only) are shown in Fig. \ref{fig:filtered_maps_desi_paper3}.

%-----------------------------------------------------------------

\section{Distribution of the photometric redshift error}\label{sec:error_distribution_paper3}

Precise modelling of the distribution of photometric redshift error is required to propagate redshift uncertainties in the estimation of parameters. H21 adopted a modified Lorentzian function (Eq. \ref{eq:modified_lorentzian}) to model the redshift error distribution of $\Delta z = z_{s} - z_{p}$ ($s\equiv$ spectroscopic; $p\equiv$ photometric) as a function of $z_{p}$, $p(z_{s} - z_{p}|z_{p})$.
\begin{equation}
    L(x) = \mathcal{N}\bigg[1+\frac{(\Delta z-x_{0})^{2}}{2a\sigma^{2}}\bigg]^{-a}
    \label{eq:modified_lorentzian}
\end{equation}
where $\mathcal{N}$ is the {normalisation} factor and $x_{0},\sigma$ and $a$ are the parameters to be constrained for every tomographic bin. The best fit-values of $x_{0},\sigma$ and $a$ from H21 are quoted in Table \ref{tab:DESI_data}.

\begin{figure*}[hbt!]
    \begin{subfigure}[b]{0.5\linewidth}
        \centering
        \includegraphics[width=8.5cm]{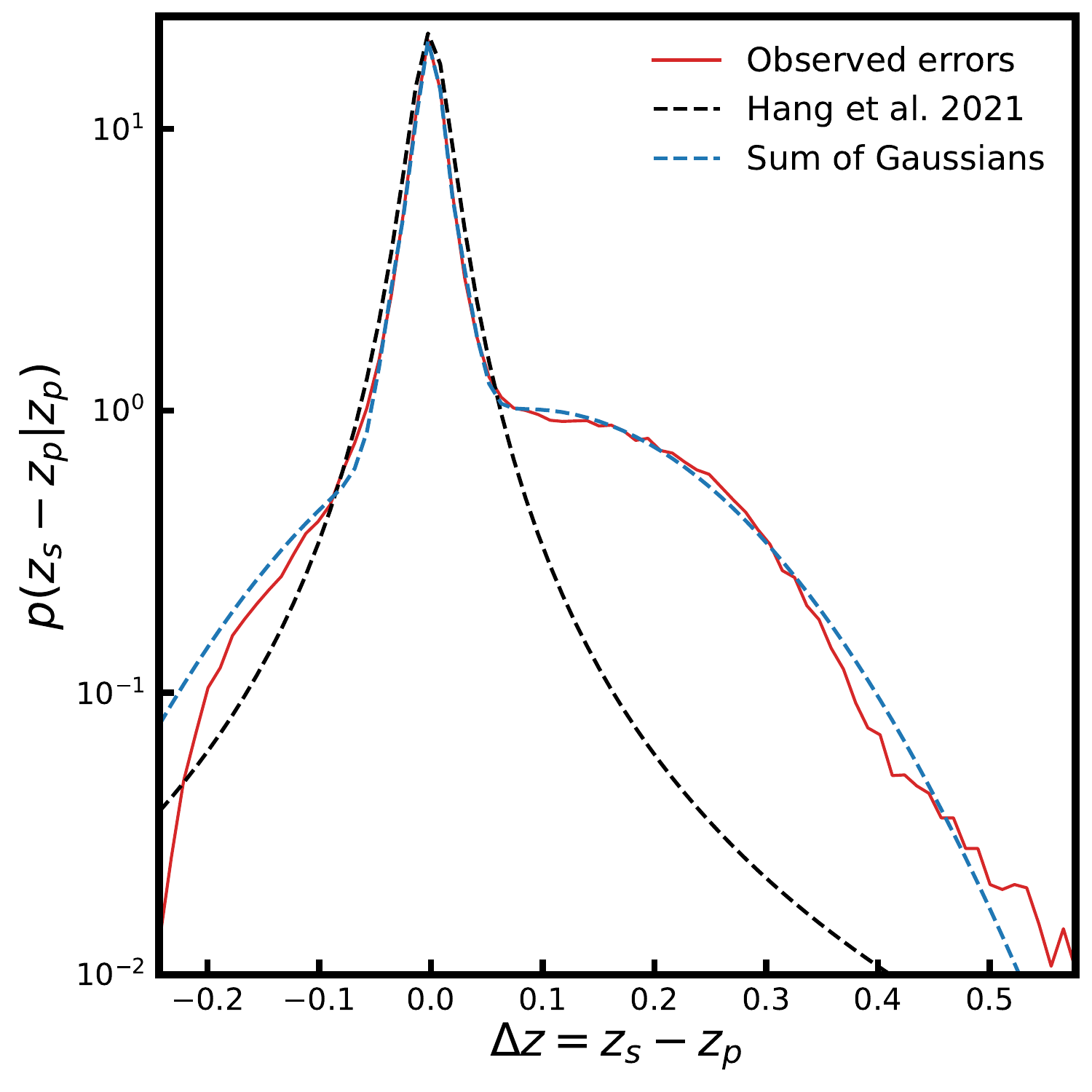}
        \captionsetup{labelformat=empty}
        \caption{\large{$0.0<z\leq0.3$}}
    \end{subfigure}%
    \begin{subfigure}[b]{0.5\linewidth}
        \centering
        \includegraphics[width=8.5cm]{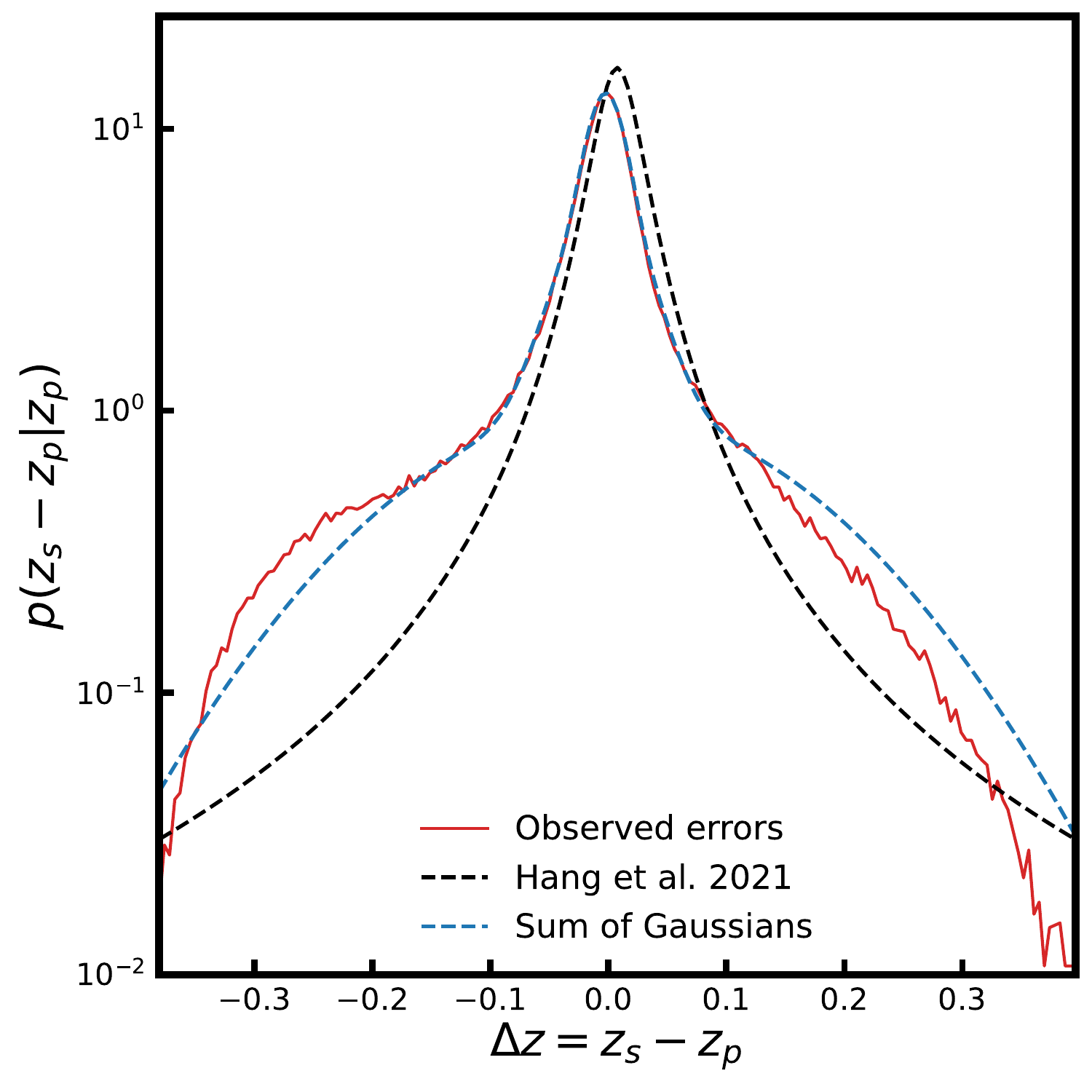}
        \captionsetup{labelformat=empty}
        \caption{\large{$0.3< z\leq0.45$}}
    \end{subfigure}\\
    \begin{subfigure}[b]{0.5\linewidth}
        \centering
        \includegraphics[width=8.5cm]{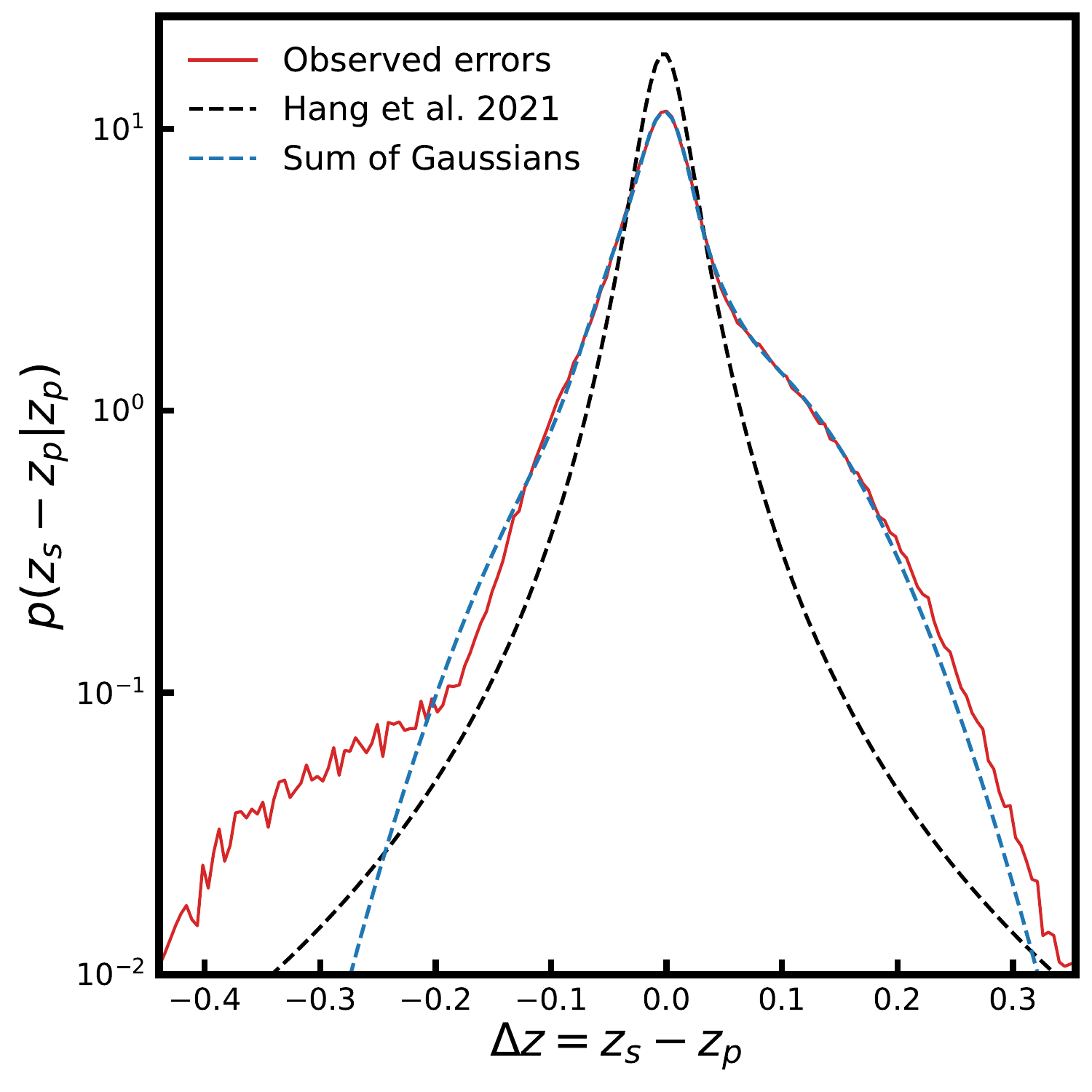}
        \captionsetup{labelformat=empty}
        \caption{\large{$0.45< z\leq0.6$}}
    \end{subfigure}%
    \begin{subfigure}[b]{0.5\linewidth}
        \centering
        \includegraphics[width=8.5cm]{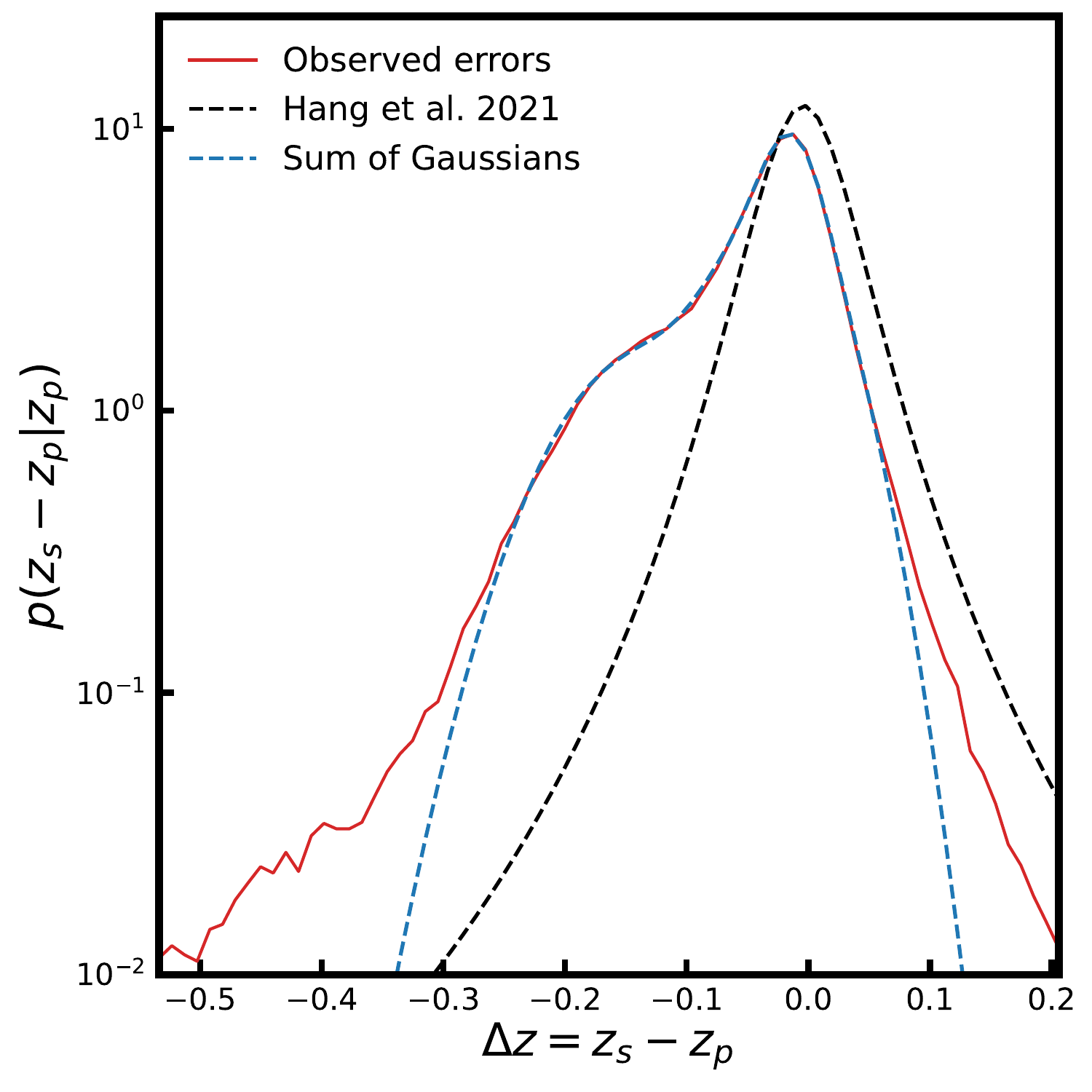}
        \captionsetup{labelformat=empty}
        \caption{\large{$0.6< z\leq0.8$}}
    \end{subfigure}
    \caption{Observed distributions of photometric redshift error (shown by a solid red curve) compared with the best-fit modified Lorentzian function obtained by H21 (dashed black curve) and the sum of three Gaussians (dashed blue curve).}
    \label{fig:photo_err_func_paper3}
\end{figure*}

Fig. \ref{fig:photo_err_func_paper3} shows the photometric error distributions compared to the best-fit modified Lorentzian functions. The modified Lorentzian functions provide a good estimate near the peak of the error distributions but fail to capture the broader tails of the error distribution. We attempted to model the broad wings of the error distributions with a sum of three Gaussians. As shown in Fig. \ref{fig:photo_err_func_paper3}, our sum of Gaussians model better captures the tails of the error distributions, at {the} marginal expense of increasing the number of free parameters in the fitting function. An important point to note is we only seek to fit the broad wings that immediately follow the peak of the error distributions and not the extremities, which is to avoid over-fitting. In section \ref{sec:results_wout_leakage_corr_paper3}, we detail the impact of modelling the wings of the error distributions on the estimation of parameters.

%-----------------------------------------------------------------

\section{Estimation of the true redshift distribution}\label{sec:true_redshift_distribution_paper3}

An estimate of the true redshift distribution can be computed using either the convolution method or the deconvolution method.

\subsection{Convolution method}

For the $i^{\text{th}}$ redshift bin, the true redshift distribution $\frac{\mathrm{d}N^{i}}{\mathrm{d}z_{t}}$ was estimated using the observed photometric redshift distribution $\frac{\mathrm{d}N(z_{p})}{\mathrm{d}z_{p}}$ and the redshift error distribution $p^{i}(z_{s}-z_{p}|z_{p})$ using the relation
\begin{equation}
    \frac{\mathrm{d}N^{i}}{\mathrm{d}z_{t}} = \int\mathrm{d}z_{p}\frac{\mathrm{d}N(z_{p})}{\mathrm{d}z_{p}}W^{i}(z_{p})p^{i}(z_{s}-z_{p}|z_{p})
    \label{eq:true_dist_conv_paper3}
\end{equation}
where $W^{i}(z_{p})$ is the window function defining the boundaries of the $i^{\text{th}}$ redshift bin
\begin{equation}
    W^{i}(z_{p}) = \begin{cases}
        1,& \text{if}\quad z^{i}_{\text{min}}\leq z_{p}<z^{i+1}_{\text{min}}\\
        0,& \text{otherwise}
    \end{cases}
    \label{eq:window_function_paper3}
\end{equation}

\begin{figure*}[hbt!]
    \begin{subfigure}[b]{0.5\linewidth}
        \centering
        \includegraphics[width=8.5cm]{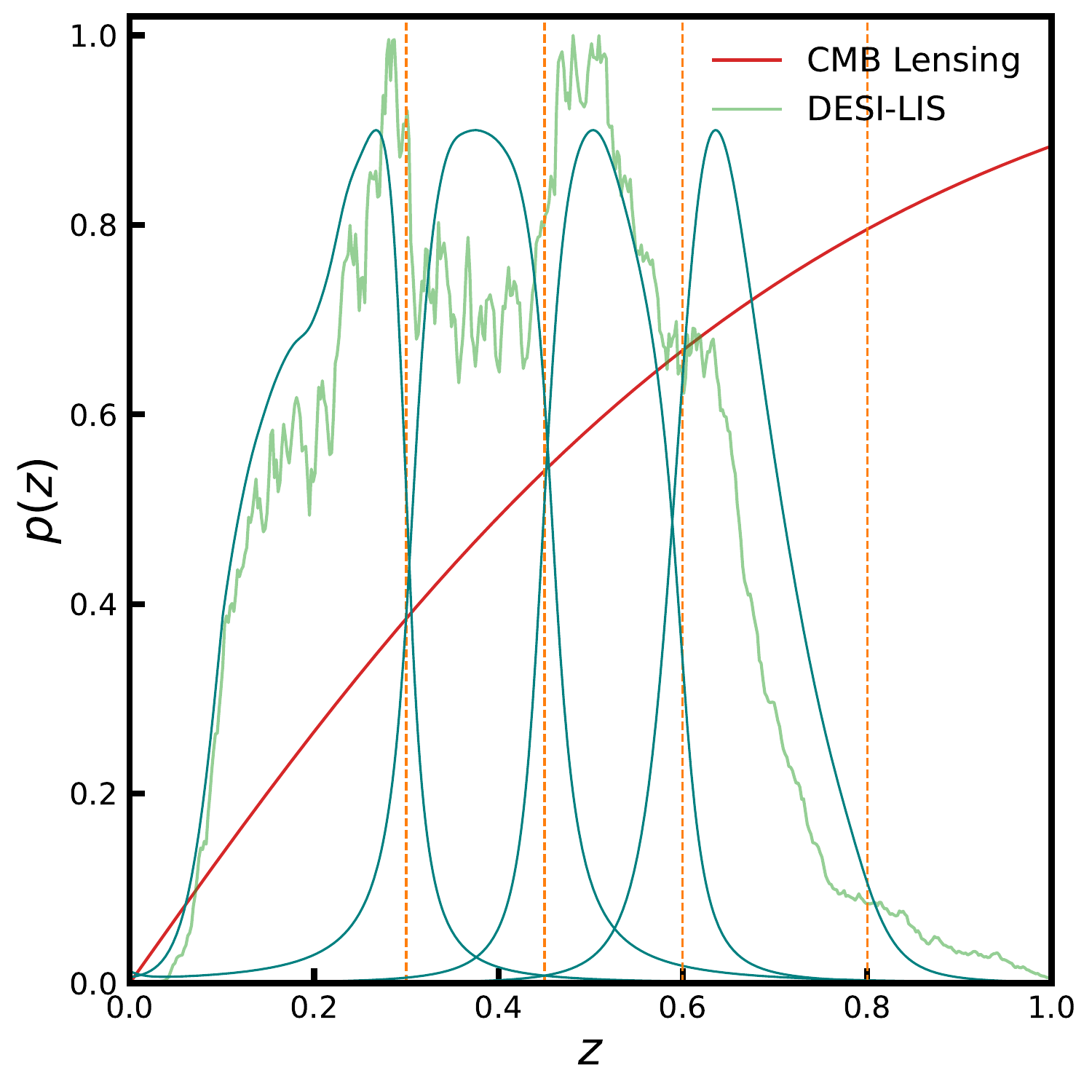}
        \caption{}
    \end{subfigure}%
    \begin{subfigure}[b]{0.5\linewidth}
        \centering
        \includegraphics[width=8.5cm]{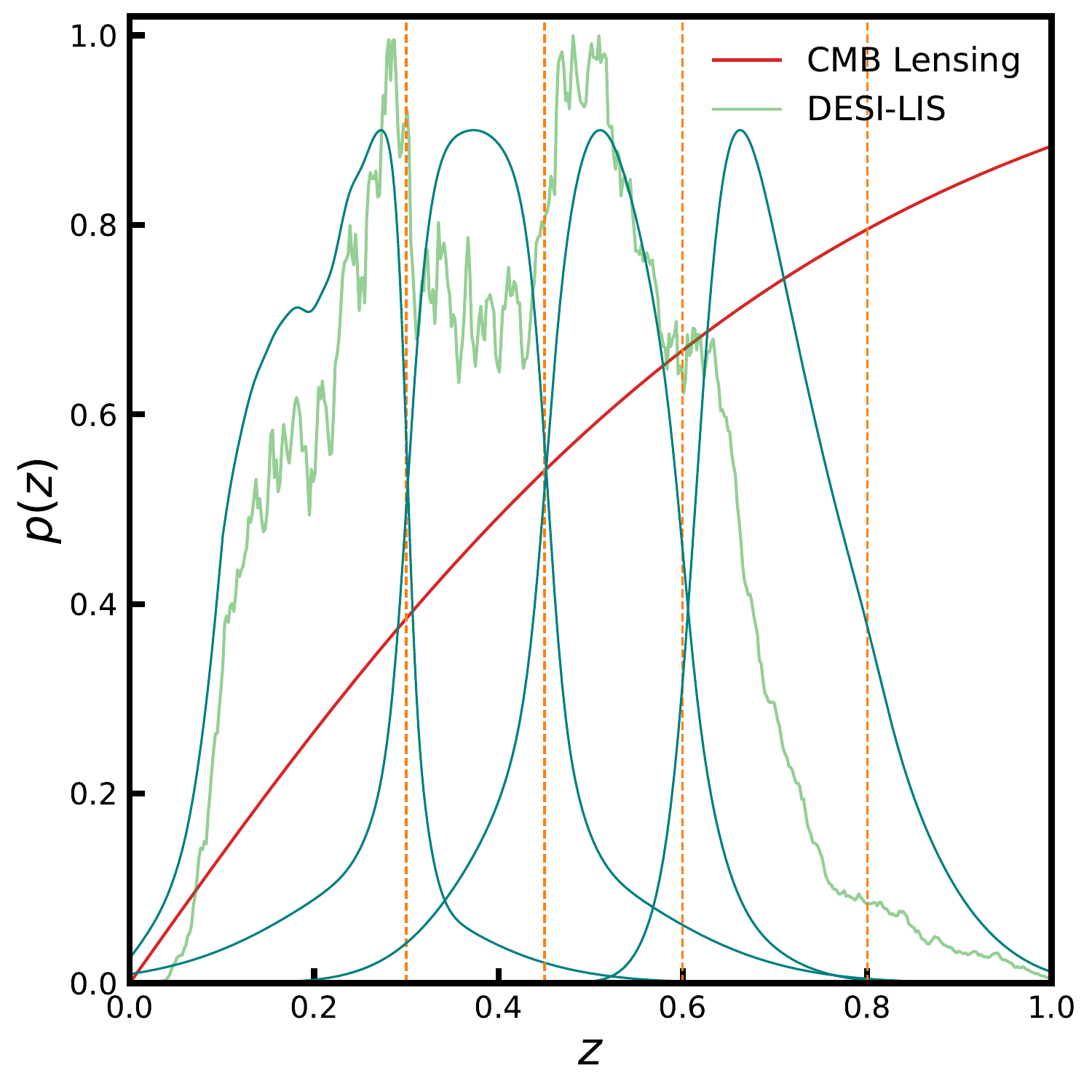}
        \caption{}
    \end{subfigure}
    \caption{Galaxy redshift distributions estimated using Eq.\,(\ref{eq:true_dist_conv_paper3}) for four tomographic bins shown in blue lines estimated using (a) modified Lorentzian and (b) sum of Gaussians fit to the photometric redshift error distribution. The red line marks the CMB lensing kernel and the green line shows the total redshift distribution of DESI-LIS galaxies. The dashed vertical orange lines mark the boundaries for the four tomographic bins. The CMB lensing kernel and redshift distributions are {normalised} to {the} unit maximum for illustration.}
    \label{fig:gal_dist_cmb_lensing_paper3}
\end{figure*}

In Fig. \ref{fig:gal_dist_cmb_lensing_paper3}, we show the galaxy redshift distributions estimated using Eq.\ (\ref{eq:true_dist_conv_paper3}). The left and the right panels show the redshift distributions estimated with {a} modified Lorentzian fit of the error distribution and the sum of Gaussians model, respectively. The sum of Gaussians model produces broader tails in the redshift distributions than the modified Lorentzian function, which suggests stronger leakage of objects across redshift bins.

\subsection{Deconvolution method}\label{sec:true_redshift_distribution_deconv_paper3}

The true redshift distribution $\frac{\mathrm{d}N}{\mathrm{d}z_{t}}$ in the redshift range $0\leq z\leq 0.8$ can be estimated by using the deconvolution method following the Fourier ratio approach \citep{Saraf2024}:
\begin{equation}
    \frac{\mathrm{d}N(z_{t})}{\mathrm{d}z_{t}} = \mathcal{F}^{-1}\bigg[\frac{\mathcal{F}[\frac{\mathrm{d}N(z_{p})}{\mathrm{d}z_{p}}]}{\mathcal{F}[p(z_{p}-z_{s}|z_{s})]}\bigg]
    \label{eq:true_dist_deconv_fourier_method_paper3}
\end{equation}
where $\frac{\mathrm{d}N(z_{p})}{\mathrm{d}z_{p}}$ is the observed photometric redshift distribution and $p(z_{p}-z_{s}|z_{s})$ is the redshift error distribution built from the spectroscopic calibration sample (section \ref{sec:lis_data}). We used the deconvolution method to compute the scattering matrix in section \ref{sec:results_with_leakage_corr_paper3}.

% %-----------------------------------------------------------------

\section{Estimation of parameters}\label{sec:mle_paper3}

We focused on {estimating} two parameters, the linear galaxy bias $b$ and amplitude of cross correlation $A$, from measurements of the galaxy auto-power spectrum and the cross-power spectrum between CMB lensing and galaxy density, using the maximum likelihood estimation method. The galaxy bias accounts for the fact that galaxies are biased tracers of the underlying total matter distribution. The amplitude of cross correlation is a phenomenological parameter rescaling the amplitude of the cross-power spectrum. The cross correlation amplitude can be used to test the validity of the underlying cosmological model, which in our study is the $\Lambda$CDM model (\citealt{Saraf2022}; \citealt{Marques2020}; \citealt{Bianchini2015}). The galaxy auto-power spectrum scales as $b^{2}$, whereas the cross-power spectrum depends on the product of the parameters $b\times A$ and induces a degeneracy in the estimation of parameters. We break this degeneracy using a joint likelihood function of the form
\begin{equation}
\begin{split}
	\mathcal{L}&(\hat{C}_{L}|b,A) = \frac{1}{\sqrt{(2\pi)^{N_{L}}det(\text{Cov}_{LL'})}} \times\\
	& \times \text{exp}\bigg\lbrace -\frac{1}{2}[\hat{C}_{L}-C_{L}(b,A)](\text{Cov}_{LL'})^{-1}[\hat{C}_{L'}-C_{L'}(b,A)]\bigg\rbrace
\end{split}
\label{eq:joint_likeli}
\end{equation}
where $N_{L}$ is the number of data points and $\hat{C}_{L}$ is the measured joint power spectrum, $\hat{C}_{L} = (\hat{C}_{L}^{\kappa g},\hat{C}_{L}^{gg})$. Also, $C_{L}(b,A)$ is the joint theoretical power spectrum template defined as $C_{L}(b,A) = (AC_{L}^{\kappa g}(b),C_{L}^{gg}(b))$, and the covariance matrix is given as
\begin{equation}
	\text{Cov}_{LL'} = 
	\begin{bmatrix}
		\text{Cov}_{LL'}^{\kappa g,\kappa g} & \text{Cov}_{LL'}^{\kappa g,gg} \\
            \vspace{0.1mm}\\
		\text{Cov}_{LL'}^{\kappa g,gg} & \text{Cov}_{LL'}^{gg,gg}	
	\end{bmatrix}
	\label{eq:joint_cov_full}
\end{equation}
In section \ref{sec:covariance_matrix}, we outline different ways to estimate the covariance matrix $\text{Cov}_{LL'}$.

We employed the publicly available \texttt{EMCEE} \citep{EMCEE2013} package to effectively sample the parameter space, and estimated the galaxy bias $b$ and the cross-correlation amplitude $A$. We used flat priors for the parameters with ${b \in [0,10]}$ and ${A \in [-5,5]}$. The other cosmological parameters entering the likelihood estimation through theoretical power spectrum templates were set to their constant values for the fiducial background cosmology described in \cite{Planck2020VI} (i.e.~the flat $\Lambda$CDM cosmology with the best-fit \textit{Planck} + \textit{WP} + highL + lensing parameters, where \textit{WP} stands for the \textit{WMAP} polarisation data at low multipoles, highL is the high-resolution CMB data from the Atacama Cosmology Telescope (ACT) and the South Pole Telescope (SPT), and lensing refers to the inclusion of \textit{Planck} CMB lensing data in the parameter likelihood).

\section{Mock catalogues}\label{sec:simulations_paper3}

{To gain insight into how the observed redshift errors impact the estimation of parameters, we created mock catalogues following the same approach to simulations as in our previous paper, \cite{Saraf2024}. We used the publicly available code \texttt{FLASK} \citep{Flask2016} and created $300$ Monte Carlo simulations of correlated log-normal galaxy density field with observed DESI-LIS physical properties (quoted in Table \ref{tab:DESI_data}) and \textit{Planck} CMB lensing convergence field.} The photometric redshifts were generated from the observed photometric redshift error distributions $p(z_{p}-z_{s}|z_{s})$. We divided the simulated galaxy catalogue into five redshift bins with intervals of $(0.0,0.3,0.45,0.6,0.8,1.0]$. We introduced an additional redshift bin of $0.8\leq z<1.0$ in simulations to account for the scatter of galaxies outside the redshift range considered in {the} analysis, which is $0.0\leq z<0.8$. The theoretical power spectra for galaxy auto-correlation, CMB convergence auto-correlation and their cross correlation were computed from the Limber approximation \citep{Limber1953} using the relation
\begin{equation}
    C_{\ell}^{xy} = \int_{0}^{\chi_{*}}d\chi\frac{W^{x}(\chi)W^{y}(\chi)}{\chi^{2}}P\bigg(k=\frac{\ell+1/2}{\chi},z(\chi)\bigg)
\label{eq:power_spectra_theory}
\end{equation} 
where $\{x,y\}\in\{\kappa,g\}$, $\kappa\equiv$ convergence and $g\equiv$ galaxy over-density. Here, $\chi$ is the comoving distance, and $\chi_{*}$ is the comoving distance to the surface of the last scattering at redshift $z_{*}\simeq 1100$. Also, $P\bigg(k=\frac{\ell+1/2}{\chi},z(\chi)\bigg)$ is the matter power spectrum which is generated using the publicly available cosmology code \texttt{CAMB}\footnote{\url{https://camb.info/}}\citep{Lewis2000} with the \texttt{HALOFIT} prescription to take into account the non-linear distribution of matter. $W^{\kappa}$ and $W^{g}$ are lensing and galaxy kernels given by
\begin{align}
    W^{\kappa}(\chi) &= \frac{3\Omega_{m}}{2c^{2}}H_{0}^{2}(1+z)\chi\frac{\chi_{*}-\chi}{\chi_{*}}\label{eq:lensing_kernel}\\
    W^{g}(\chi) &= b(z(\chi))\frac{H(\chi)}{c}\frac{\mathrm{d}N}{\mathrm{d}z(\chi).}\label{eq:galaxy_kernel}
\end{align}
We used a redshift-dependent galaxy bias (\citealt{Solarz2015}; \citealt{Moscardini1998}; \citealt{Fry1996})
\begin{equation}
    b(z) = 1+\frac{b_{0}-1}{D(z)},
\end{equation}
where $b_{0} = 1.3$ and $D(z)$ is the normalised growth function
\begin{equation}
    D(z) = \exp\bigg\{ -\int\limits_{0}^{z}\frac{[\Omega_{m}(z')]^{\gamma}}{1+z'}\mathrm{d}z' \bigg\}
    \label{eq:growth_function}
\end{equation}
where $\gamma=0.55$ is the growth index for the general relativity \citep{Linder2005}.

We used the CMB convergence noise power spectrum $N_{\ell}^{\kappa\kappa}$ from \textit{Planck} PDR3 to add noise to the simulated CMB lensing convergence maps. To introduce noise in the galaxy density field, we let \texttt{FLASK} Poisson sample the number of galaxies in mock catalogues.

\section{Covariance matrix}\label{sec:covariance_matrix}
In this section, we explore three different ways to estimate the covariance matrix $\text{Cov}_{LL'}$ (Eq. \ref{eq:joint_cov_full}).

\subsection{Analytical covariance matrix}
The analytical covariance matrix between Gaussian fields $A,B,C,$ and $D$ estimated using \cite{Saraf2022} is given by
\begin{equation}
\begin{split}
\text{Cov}_{LL'}^{AB,CD} = &\frac{1}{(2\ell_{L'}+1)\Delta\ell f_{\text{sky}}^{AB}f_{\text{sky}}^{CD}}\bigg[f_{\text{sky}}^{AC,BD}\sqrt{C_{L}^{AC}C_{L'}^{AC}C_{L}^{BD}C_{L'}^{BD}}\\
&+f_{\text{sky}}^{AD,BC}\sqrt{C_{L}^{AD}C_{L'}^{AD}C_{L}^{BC}C_{L'}^{BC}}\bigg]  
\end{split}
\label{eq:error_covariance_different_area}
\end{equation}
where $\{A,B,C,D\}\in\{\kappa,g\}$, $f_{\text{sky}}^{AB}$ is the fraction of sky common to fields $A$ and $B$, and $f^{AC, BD}$ is the composite quantity that acts as a weight when taking into account different observed sky areas for different fields. The analytical covariance confers the advantage that sampling fluctuations, known as {cosmic variance}, are already included in the matrix. It can also be computationally fast to estimate using arbitrary theoretical power spectra.

\subsection{Sample covariance from mock realisations}

From the $300$ simulations of the galaxy density and CMB convergence fields (as described in section \ref{sec:simulations_paper3}), one can compute an unbiased estimator of the covariance matrix as
\begin{equation}
    \text{Cov}_{LL'}^{AB,CD} = \frac{1}{\text{N}_{\text{sim}}-1}\sum_{i=1}^{\text{N}_{\text{sim}}}(\hat{C}_{L}^{AB,i}-\overline{C}_{L}^{AB})(\hat{C}_{L'}^{CD,i}-\overline{C}_{L'}^{CD})
	\label{eq:cov_simul}
\end{equation}
where $\overline{C}_{L}$ is the average power spectrum from $300$ Monte Carlo simulations, $\hat{C}_{L}^{i}$ represents the $i^{th}$ simulated power spectrum estimate, and $\text{N}_{\text{sim}}$ is the number of simulations.

\subsection{Jackknife method}

The advantage of {the} jackknife method is that it is independent of any assumptions regarding the cosmological model. As this method uses data for the estimation of the covariance matrix, it also naturally takes into account the different survey selection effects and any unforeseen systematic errors. However, the jackknife method assumes that the observed data are an accurate representation of the distribution of measurements and is inherently blind to cosmic variance.

To apply the jackknife method, we followed the delete-one approach and {made} $\text{N}_{\text{sub}}=90$ disjoint partitions of the DESI-LIS survey footprint. We then computed the angular power spectra by omitting one subsample at a time. If we use $C_{L,i}$ to denote the angular power spectrum computed by removing the $i^{\text{th}}$ subsample, the covariance matrix can be estimated as \citep{Norberg2009}
\begin{equation}
    \text{Cov}_{LL'}^{AB,CD} = \frac{\text{N}_{\text{sub}}-1}{\text{N}_{\text{sub}}}\sum_{i=1}^{\text{N}_{\text{sub}}}(\hat{C}_{L,i}^{AB}-\overline{C}_{L}^{AB})(\hat{C}_{L',i}^{CD}-\overline{C}_{L'}^{CD})
	\label{eq:cov_jackknife}
\end{equation}
where $\overline{C}_{L}$ is the mean power spectrum over all subsamples.\\

\subsection{From H21}

H21 estimated the covariance matrix from the errors computed directly from the observed data. The observed angular power spectra were binned into groups with multipole bin width $\Delta\ell = 10$
\begin{equation}
    C_{L}\equiv\langle C_{\ell} \rangle = \frac{1}{\Delta\ell}\sum\limits_{\ell'}^{\ell'+\Delta\ell-1}C_{\ell'},\quad \ell' = \Delta\ell,2\Delta\ell,\cdots
\end{equation}
The errors on the binned data points were computed as
\begin{equation}
    \sigma_{\ell} = \frac{1}{f_{\text{sky}}}\sqrt{\frac{\langle C_{\ell}^{2} \rangle - \langle C_{\ell} \rangle^{2}}{\Delta\ell-1}}
    \label{eq:err_data_points_group_hang}
\end{equation}
The diagonal covariance matrix is then given by $\text{Cov}_{LL'} = \text{diag}(\sigma_{\ell}^{2})$.
\begin{figure*}[htb!]
    \begin{subfigure}[b]{\linewidth}
        \centering
        \captionsetup{labelformat=empty}
        \caption{\large{galaxy $\times$ galaxy}}
        \includegraphics[width=\linewidth]{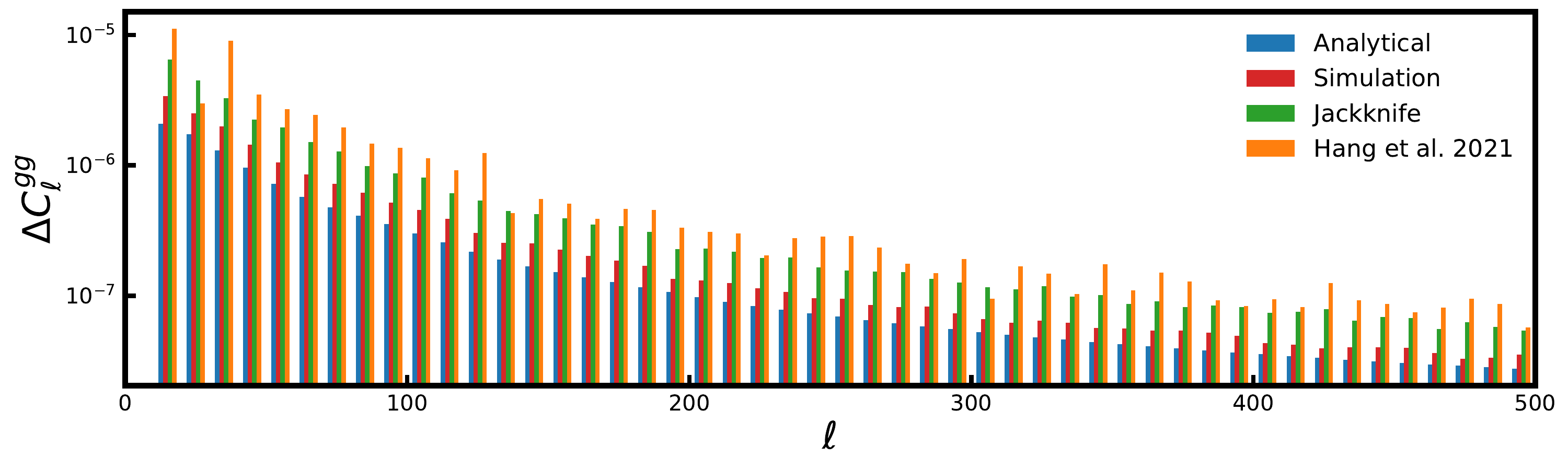}
    \end{subfigure}\\
    \begin{subfigure}[b]{\linewidth}
        \centering
        \captionsetup{labelformat=empty}
        \caption{\large{galaxy $\times$ CMB lensing}}
        \includegraphics[width=\linewidth]{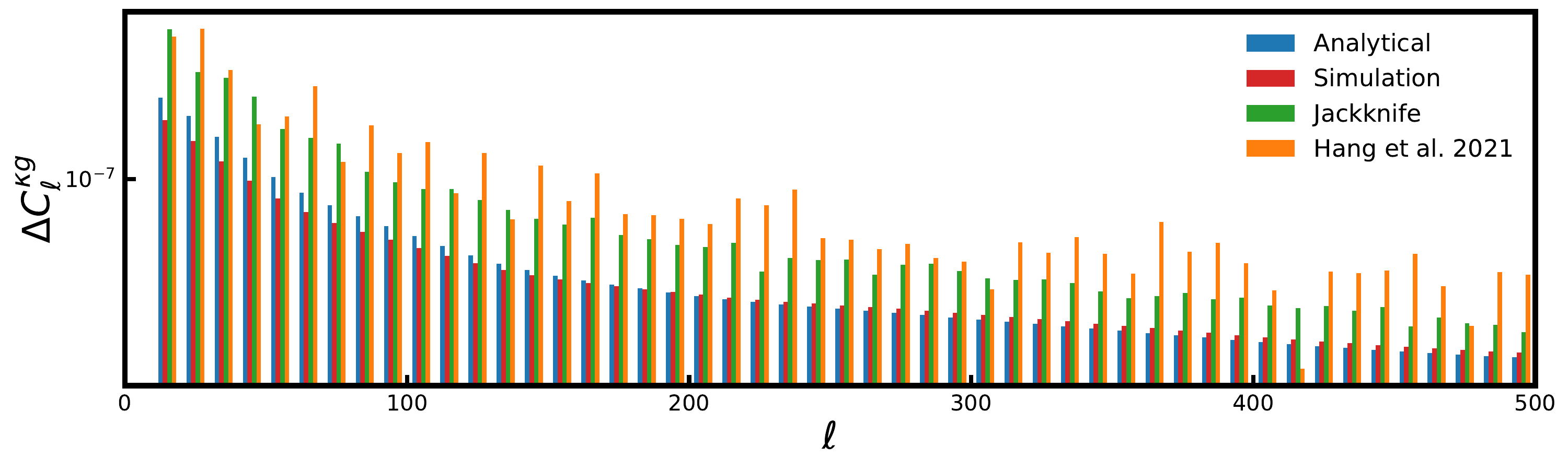}
    \end{subfigure}
    \caption{{Comparison of angular power spectrum errors} computed using the analytical method, mock catalogues, the jackknife method and the procedure from H21. {Top}: Errors on the galaxy auto-power spectrum from redshift bin $0.45\leq z<0.6$. {Bottom}: Errors on the cross-power spectrum with CMB lensing.}
    \label{fig:comp_cov_matrices}
\end{figure*}

In Fig. \ref{fig:comp_cov_matrices} we show the power spectrum errors estimated as {the} square root of the diagonal elements of the covariance matrix from different methods for the redshift slice $0.45\leq z<0.6$. We find that the internal methods, namely the jackknife method and the covariance matrix from H21 showed larger contributions to the diagonal elements of the covariance matrix. The analytical covariance and the sample covariance from mock realisations are underestimated, suggesting that the nuances of the data are not captured by these methods. We find similar results for other redshift bins and chose to use the covariance matrix from H21 for the estimation of parameters.

%-----------------------------------------------------------------

\section{Results: DESI-LIS \texorpdfstring{$\times$}{Lg} \textit{Planck}}\label{sec:results_paper3}

We followed the analysis choice of H21 and {binned} the measured galaxy auto-power spectra and cross-power spectra between the DESI-LIS photometric galaxy catalogue and the \textit{Planck} minimum-variance CMB lensing potential map with a bin width of $\Delta\ell=10$ in the multipole range $10<\ell<500$. In this section, we present our estimates of galaxy bias and cross-correlation amplitude, taking into account the effects of redshift bin mismatch of objects across tomographic bins.

\subsection{Mock catalogues}
\begin{figure*}[hbt!]
    \begin{subfigure}[b]{0.5\linewidth}
        \centering
        \includegraphics[width=8.5cm]{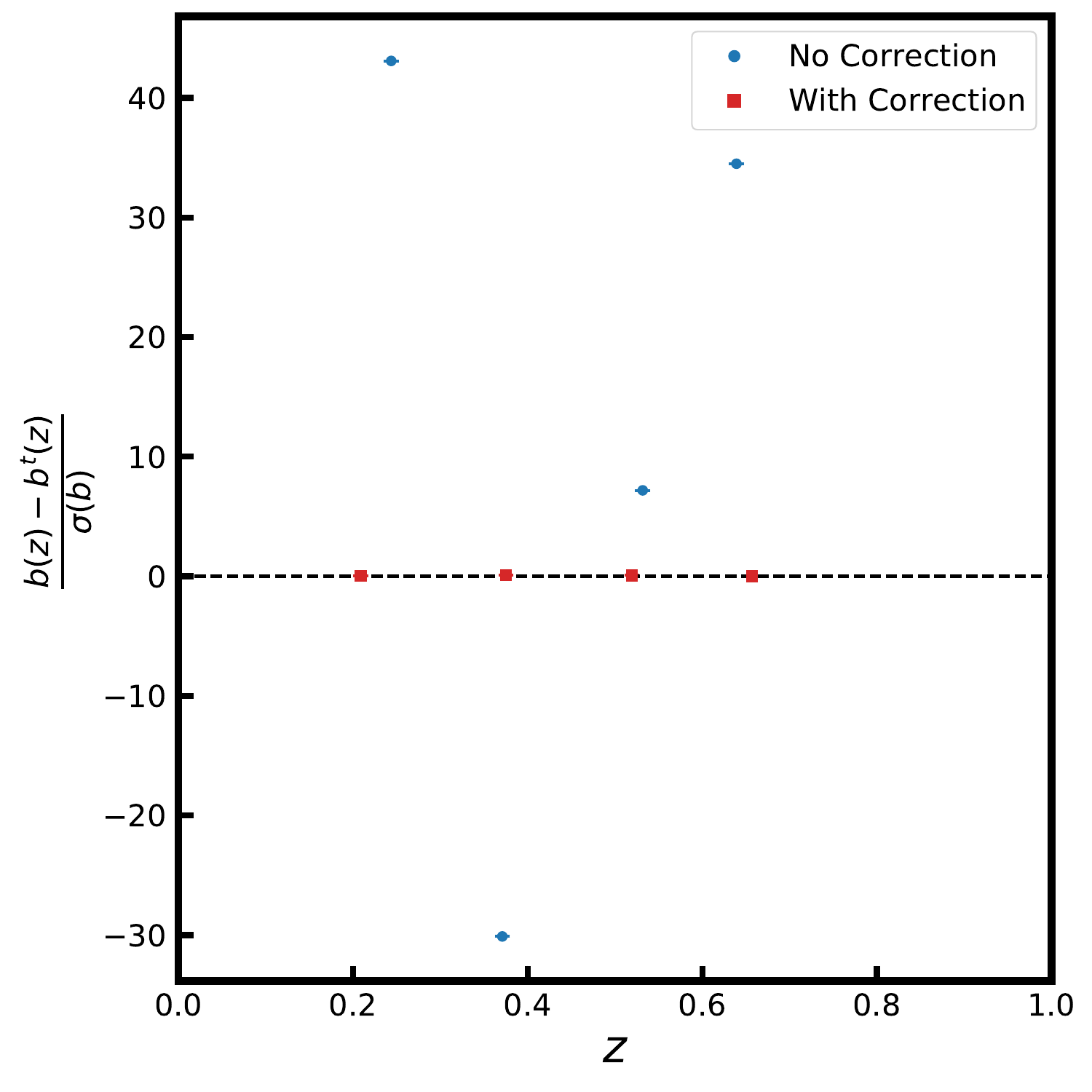}
        \caption{}
    \end{subfigure}%
    \begin{subfigure}[b]{0.5\linewidth}
        \centering
        \includegraphics[width=8.5cm]{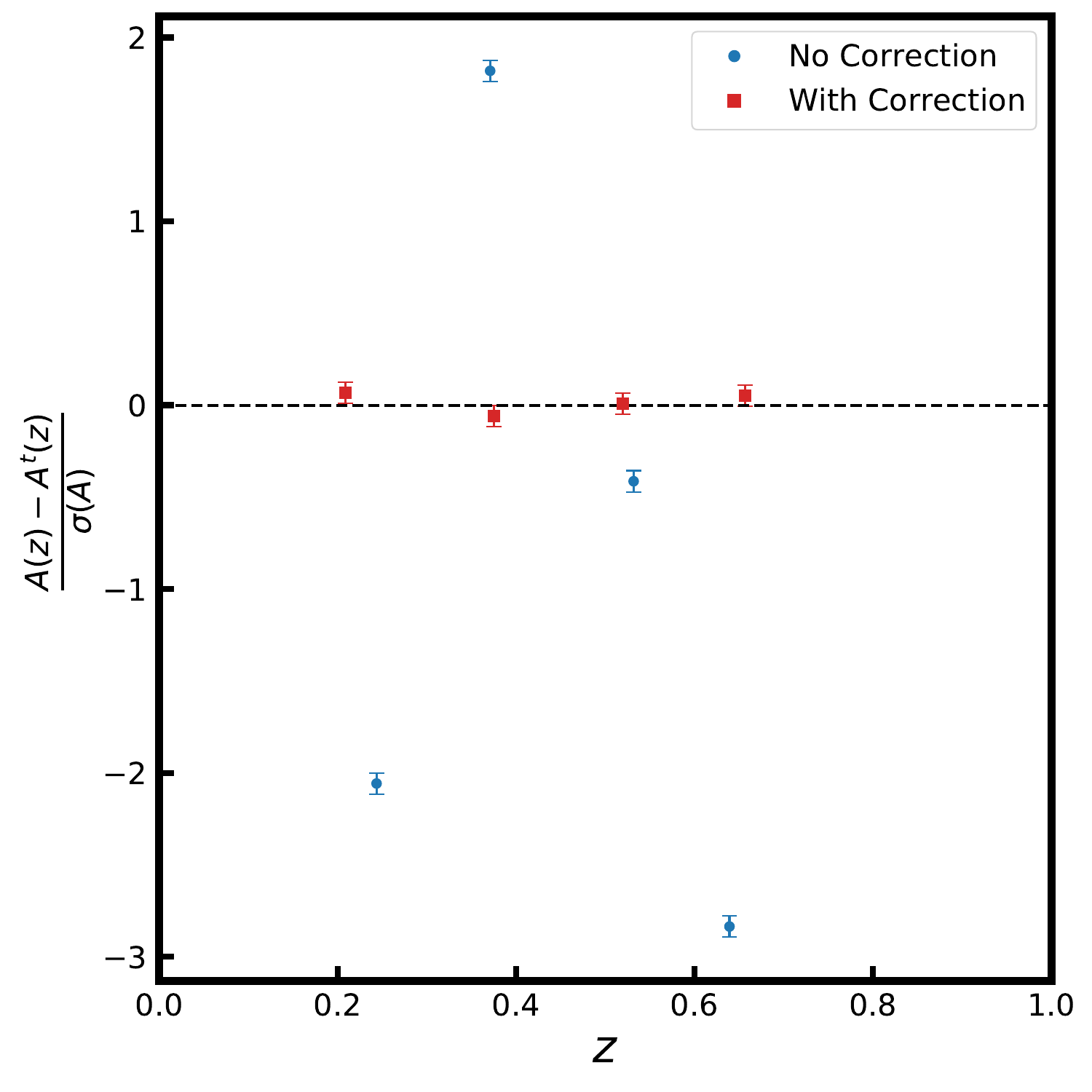}
        \caption{}
    \end{subfigure}
    \caption{{Comparison of the relative difference for (a) galaxy bias and (b) cross-correlation amplitude estimated from the average power spectra of $300$ Monte Carlo simulations. The blue circles and red squares denote the values before and after correction for redshift bin mismatch, respectively. The error bars on the data points correspond to the standard error from the average spectra, and the standard deviations $\sigma(b)$ and $\sigma(A)$ correspond to a single realisation.}}
    \label{fig:comp_params_simulations_paper3}
\end{figure*}

In section \ref{sec:simulations_paper3}, we describe the mock catalogues prepared using the code \texttt{FLASK} with the DESI-LIS observed photometric redshift errors. We validate our pipeline and estimate the impact of {the} redshift bin mismatch of DESI-LIS galaxies in this section. We follow the procedure from \cite{Saraf2024} to estimate galaxy bias and cross-correlation amplitude from simulated photometric datasets. {In Fig. \ref{fig:comp_params_simulations_paper3}, we show the deviations of the estimated values of the galaxy linear bias (left panel) and the amplitude of cross-correlation (right panel) from their fiducial values used in simulations in terms of standard deviations. The values of parameters are computed from the average power spectra of $300$ simulations before (blue circles) and after (red squares) accounting for {the} redshift bin mismatch of objects. The average power spectra and relative differences between the averages and best-fit theoretical spectra are shown in Appendix \ref{sec_appndx:appndx_a} (Figs. \ref{fig:power_spectra_simualtions_desi_avg_gg_and_kg} and \ref{fig:power_spectra_simualtions_desi_avg_gg_and_kg_rel_err}, respectively).} Before leakage correction, the galaxy bias {was} higher, with $7-42\,\sigma$ deviations in three out of four tomographic bins. The second redshift bin gives a $\sim 30\sigma$ lower estimate of the galaxy bias. Such large deviations in terms of errors are a consequence of small uncertainties on the galaxy bias constrained from the galaxy auto-power spectrum. We find the amplitude of cross-correlation to be smaller than unity by $\sim 2\sigma, \sim 0.4\sigma, \text{ and } \sim 3\sigma$ in three redshift bins, and higher by $\sim 2\sigma$ in the second redshift bin. The estimated parameters closely follow their expected values after accounting for the leakage across redshift bins.

\subsection{DESI-LIS $\times$ \textit{Planck}}
In this section, we present the measurements of angular power spectra and parameter estimates from cross correlation of the DESI-LIS photometric galaxy catalogue with \textit{Planck} CMB lensing potential maps.

\subsubsection{Before leakage correction}\label{sec:results_wout_leakage_corr_paper3}

H21 estimated the galaxy bias directly from the measured galaxy-auto power spectrum, $\hat{C}_{\ell}^{gg}$, across all tomographic bins by adopting a two-parameter bias model for the linear and non-linear regimes:
\begin{equation}
    C_{\ell}^{\text{gg,th}} = b_{1}^{2}\,C_{\ell}^{\text{lin,th}} + b_{2}^{2}\,\Delta C_{\ell}^{\text{nl,th}}
\end{equation}
where the linear power spectrum $C_{\ell}^{\text{lin,th}}$ and the non-linear correction $\Delta C_{\ell}^{\text{nl,th}}$ are computed using \texttt{CAMB}. The galaxy bias for individual redshift bins {is} given by
\begin{equation}
    b_{1}^{2} = \sum\limits_{\ell}w(\ell)\frac{\hat{C}_{\ell}^{\text{gg}}}{C_{\ell}^{\text{lin,th}}}; \qquad b_{2}^{2} = \sum\limits_{\ell}w(\ell)\frac{\hat{C}_{\ell}^{\text{gg}}}{C_{\ell}^{\text{nl,th}}}
    \label{eq:galaxy_bias_hang}
\end{equation}
with
\begin{equation}
    w(\ell) = \frac{1/\sigma_{\ell}^{2}}{\sum\limits_{\ell}(1/\sigma_{\ell}^{2})}
\end{equation}
where $\sigma_{\ell}$ are the errors on the galaxy auto-power spectra (Eq. \ref{eq:err_data_points_group_hang}). The choice of a two-parameter bias model is motivated {by} the fact that the galaxy auto-power spectra cannot be fit well by a constant bias beyond $\ell\simeq 250$ (see section 4.1.1 of H21).

To account for the photometric redshift uncertainties, H21 extended the set of free parameters to \{$A,x_{0}^{i}, a^{i}$\}, where $x_{0}^{i}$ and $a^{i}$ are nuisance parameters that define the modified Lorentzian model for the error distribution in the $i^{\text{th}}$ redshift bin. The constraints on the nuisance parameters were derived from the galaxy--galaxy cross power spectra between tomographic bins. The cross-correlation amplitudes $A$ for redshift bins were estimated from the galaxy-CMB lensing cross-power spectrum (with an estimate of galaxy linear bias from Eq. \ref{eq:galaxy_bias_hang}). The H21 best-fit values of $x_{0}^{i}, a^{i}$ are quoted in Table \ref{tab:DESI_data} and the values of galaxy linear bias $b_{1}$ --- to which we compare estimates of the galaxy bias for our model --- and cross correlation amplitude $A$ are mentioned in Table \ref{tab:best_fit_params_bA_nocorr_paper3}. We find our estimates of the cross-correlation amplitude to be in tension with respect to the $\Lambda$CDM expectations at $1.5-2.5\,\sigma$.

In \cite{Saraf2024}, we showed that estimations of galaxy bias from the galaxy auto-power spectrum alone in tomographic analyses produced biased results due to redshift bin mismatch of objects, even with a precise model of the photometric redshift error distribution. We aim to correct the scatter of objects across redshift bins through the scattering matrix formalism. For the DESI-LIS datasets, we estimated the effective maximum wave number $k_{\text{max}}\simeq 0.2$ (or $\ell\simeq 500$), where we expect the non-linear corrections to be mild. Thus, instead of {the} H21 two-bias model, we used a single-parameter redshift-dependent model of galaxy bias, $b(z) = 1+\frac{b_{0}-1}{D(z)}$, and a {non-linear} theoretical power spectrum computed using \texttt{CAMB} with the \texttt{HALOFIT} prescription.

Before correcting the redshift bin mismatch, we compared our estimates of the galaxy linear bias and cross-correlation amplitude to the estimates from H21 in Table \ref{tab:best_fit_params_bA_nocorr_paper3}. For the modified Lorentzian error distribution, we find good agreement between the galaxy bias estimates, whereas the cross-correlation amplitudes differ by $\sim 1\,\sigma$ in all redshift bins. We find the amplitude to be consistent with the $\Lambda$CDM model in the last two tomographic bins and a $\sim 3\,\sigma$ deviation in the other two redshift bins. It is worth noting that, although we estimated the covariance matrix in the same way as H21 in this analysis, we obtain smaller errors {in} galaxy bias.

We also quote in Table \ref{tab:best_fit_params_bA_nocorr_paper3} the best-fit values of parameters $b$ and $A$ adopting the sum of Gaussian model for the photometric redshift error distribution. We consistently estimate higher values of galaxy bias and lower values for {the} amplitude of the cross-correlation across all redshift bins compared to the modified Lorentzian fit, except for the first bin where the amplitude is found to be higher than the modified Lorentzian fit by $\sim 1.5\,\sigma$. The tension in the amplitude with respect to the $\Lambda$CDM model rises to $2.3-5\,\sigma$ with the sum of Gaussians approach, with the strongest deviation observed in the second redshift bin.

\begin{figure*}[hbt!]
    \begin{subfigure}{\linewidth}
        \centering
        \includegraphics[width=17cm]{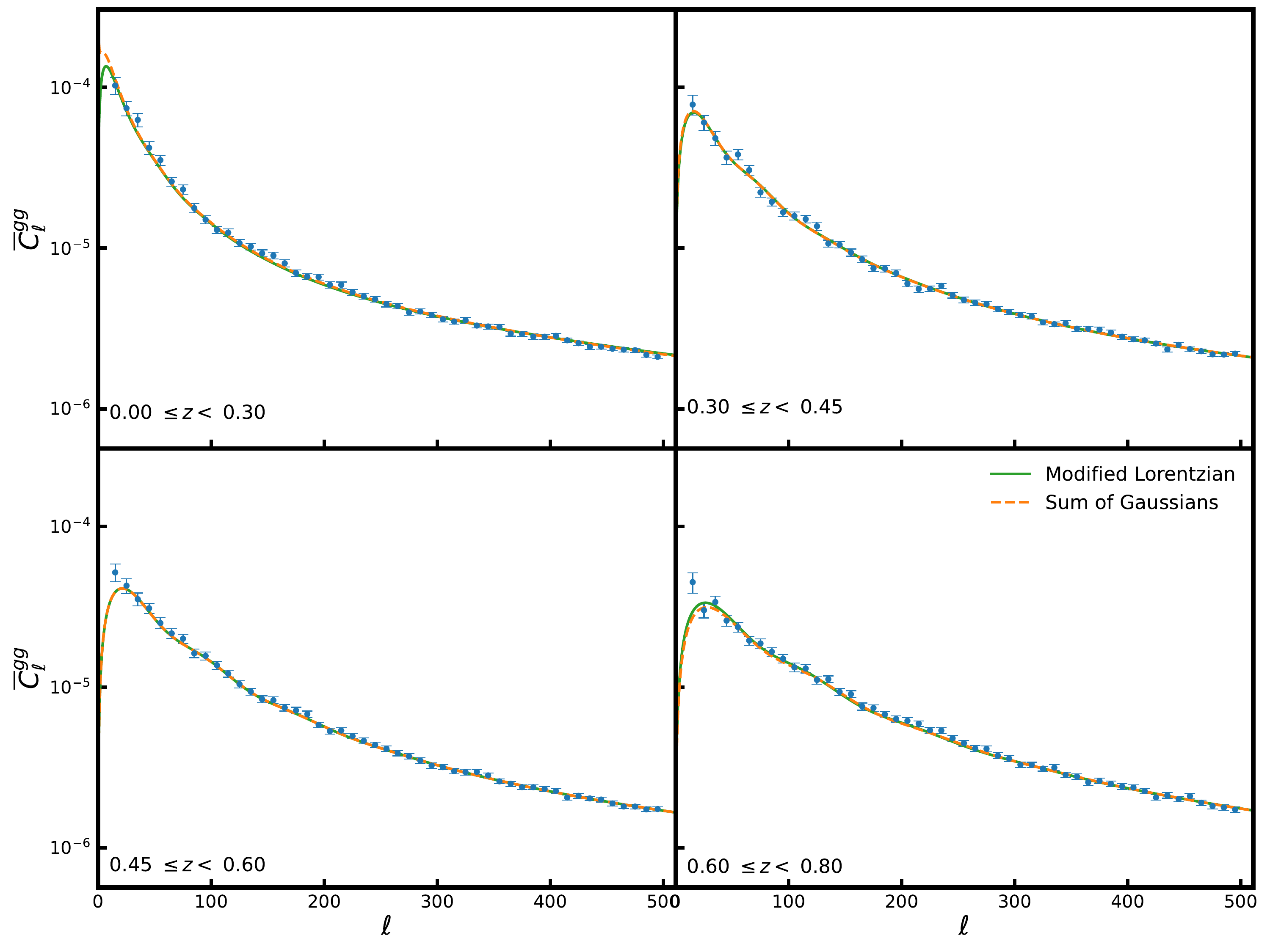}
        \captionsetup{labelformat=empty}
    \end{subfigure}%
    \caption{Galaxy auto-power spectrum measured from four DESI-LIS tomographic bins. The green solid line is the theoretical power spectrum computed following our best-fit estimates of parameters $b$ and $A$ using the modified Lorentzian fit to the error distribution. The orange dashed line is the theoretical power spectrum corresponding to the sum of Gaussians fit to the error distribution. {The standard error bars are computed from the covariance matrix used in the likelihood function, i.e. Eq.\,(\ref{eq:err_data_points_group_hang}).}}
    \label{fig:power_spectra_desi_gg_comp_mod_lor_vs_ngauss}
\end{figure*}

\begin{figure*}[hbt!]
    \begin{subfigure}{\linewidth}
        \centering
        \includegraphics[width=17cm]{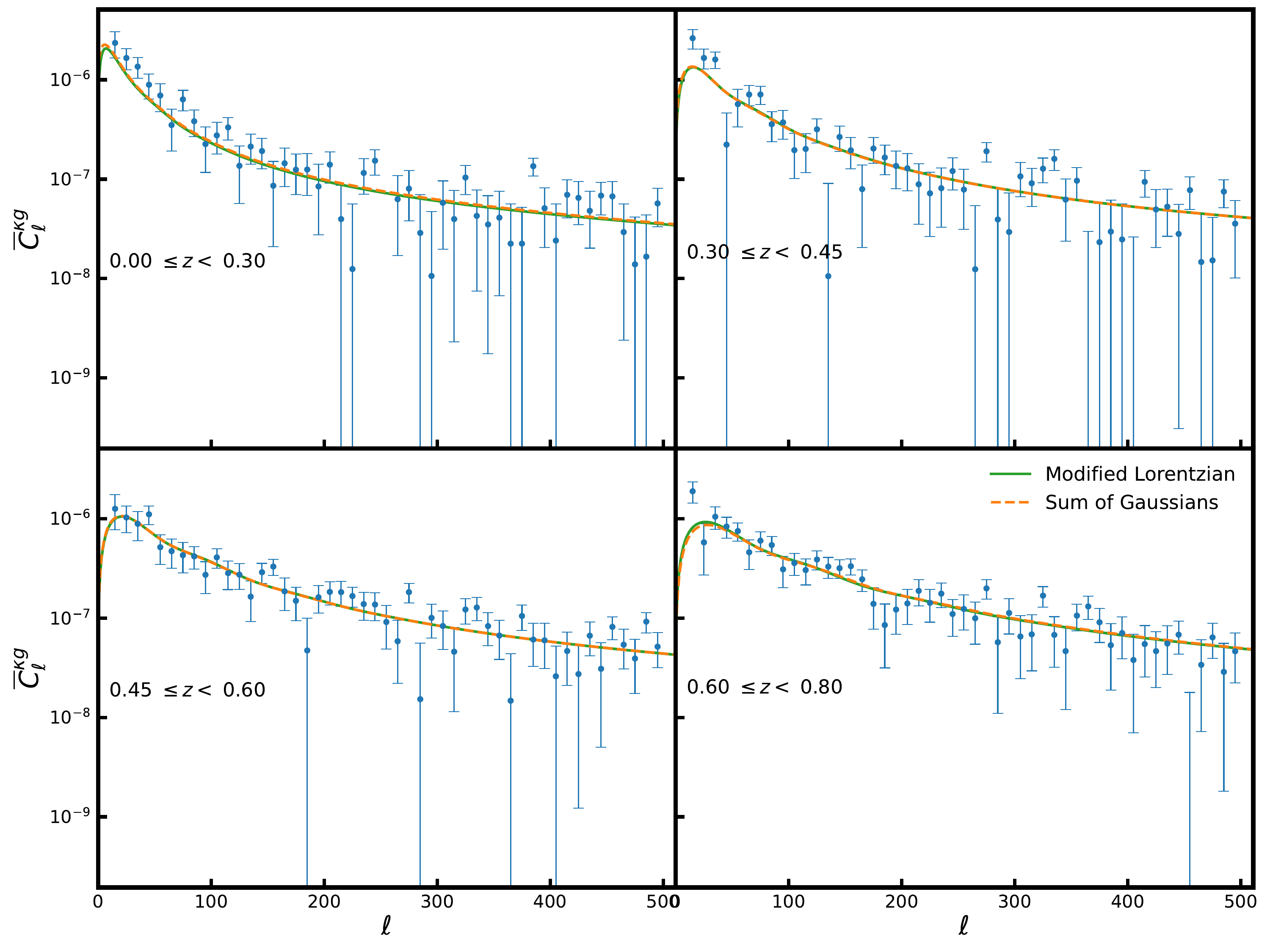}
        \captionsetup{labelformat=empty}
    \end{subfigure}%
    \caption{Cross-power spectrum measured from four DESI-LIS tomographic bins. The green solid line is the theoretical power spectrum computed following our best-fit estimates of parameters $b$ and $A$ using the modified Lorentzian fit to the error distribution. The orange dashed line is the theoretical power spectrum corresponding to the sum of Gaussians fit to the error distribution. {The standard error bars are computed from the covariance matrix used in the likelihood function, i.e. Eq.\,(\ref{eq:err_data_points_group_hang})}.}
    \label{fig:power_spectra_desi_kg_comp_mod_lor_vs_ngauss}
\end{figure*}

We computed the reduced $\chi^{2}-$values for the best-fit theoretical power spectrum to the measured angular cross-power spectrum with $\nu = 49-2=47$ degrees of freedom to establish the goodness of fit with modified Lorentzian and {the} sum of Gaussian models. The $\chi^{2}-$values show that both error distribution models can be used for the analysis of the data. Figures \ref{fig:power_spectra_desi_gg_comp_mod_lor_vs_ngauss} and \ref{fig:power_spectra_desi_kg_comp_mod_lor_vs_ngauss} show the measured galaxy auto-power spectra and cross-power spectra for the four tomographic bins, with the best-fit theoretical power spectra computed using modified Lorentzian (solid green line) and {the} sum of Gaussians (dashed orange line) fit to the error distributions. As the sum of Gaussians model better captures the peculiarities of the error distributions (as shown in Fig. \ref{fig:photo_err_func_paper3}), we treat the parameters estimated under this model as our baseline results.
\begin{table*}[t!]
    \renewcommand{\arraystretch}{1.5}
    \centering
    \caption{Best-fit galaxy linear bias $b$ and amplitude of cross correlation $A$ without redshift bin leakage correction.}
    \label{tab:best_fit_params_bA_nocorr_paper3}
    \begin{tabular}{ccc||ccc||ccc}
    \hline\hline
    Bin & \multicolumn{2}{c}{From Hang et al.} & \multicolumn{6}{c}{This work}\\
    \cline{2-9}
     & & & \multicolumn{3}{c}{Modified Lorentzian} & \multicolumn{3}{c}{Sum of Gaussians}\\
    \cline{4-9} 
     & $b_{1}$ & $A$ & $b$ & $A$ & $\chi_{r}^{2}$ & $b$ & $A$ & $\chi_{r}^{2}$ \\
    \hline\hline
    $(0.0, 0.3]$ & $1.25^{+0.01}_{-0.01}$ & $0.91^{+0.05}_{-0.05}$ & $1.216^{+0.006}_{-0.006}$ & $0.848^{+0.063}_{-0.064}$ & $1.136$ & $1.251^{+0.006}_{-0.006}$ & $0.856^{+0.062}_{-0.062}$ & $1.049$ \\
    $(0.3, 0.45]$ & $1.56^{+0.02}_{-0.02}$ & $0.80^{+0.04}_{-0.04}$ & $1.591^{+0.007}_{-0.007}$ & $0.838^{+0.054}_{-0.054}$ & $1.039$ & $1.756^{+0.007}_{-0.007}$ & $0.756^{+0.049}_{-0.049}$ & $1.033$ \\
    $(0.45, 0.6]$ & $1.53^{+0.01}_{-0.01}$ & $0.94^{+0.04}_{-0.04}$ & $1.526^{+0.006}_{-0.006}$ & $0.978^{+0.055}_{-0.055}$ & $0.824$ & $1.739^{+0.006}_{-0.006}$ & $0.857^{+0.048}_{-0.048}$ & $0.823$ \\
    $(0.6, 0.8]$ & $1.83^{+0.02}_{-0.02}$ & $0.91^{+0.04}_{-0.04}$ & $1.840^{+0.008}_{-0.008}$ & $0.959^{+0.048}_{-0.048}$ & $1.193$ & $2.107^{+0.009}_{-0.009}$ & $0.862^{+0.043}_{-0.043}$ & $1.228$ \\
    \hline
    \end{tabular}
    \tablefoot{The first set of parameters $b_{1}$ and $A$ are taken from H21. The second and third sets of $b$ and $A$ are estimated using modified Lorentzian (with parameters from H21) and sum of Gaussians fit to the error distribution $p(z_{s}-z_{p}|z_{p})$, respectively. $\chi^{2}_{r}$ is the reduced chi-square values for the cross-power spectrum with $\nu = 47$ degrees of freedom.}
\end{table*}
 
\subsubsection{With leakage correction}\label{sec:results_with_leakage_corr_paper3}

\begin{figure*}[hbt!]
    \begin{subfigure}[b]{0.5\linewidth}
        \centering
        \includegraphics[width=8.5cm]{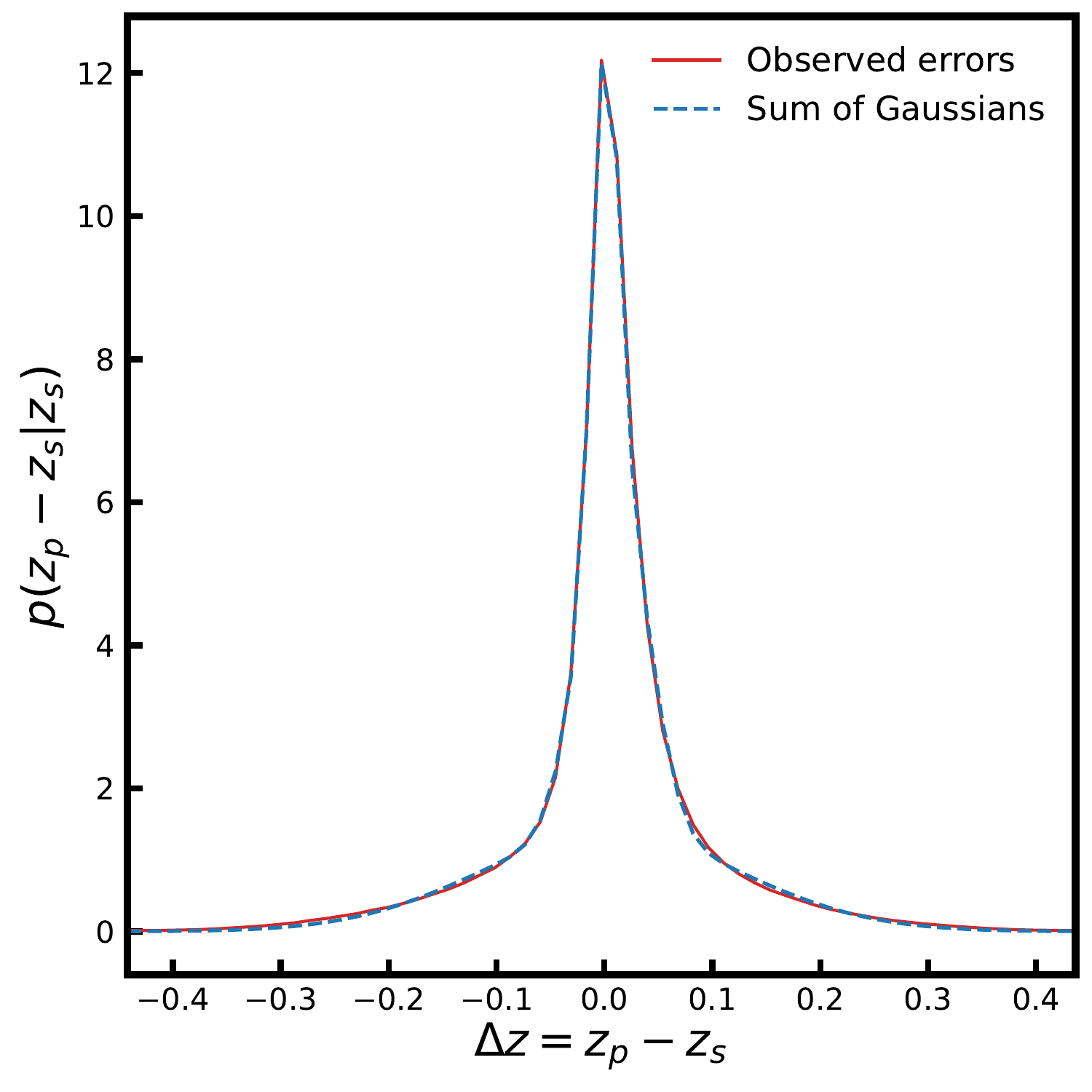}
        \caption{}
    \end{subfigure}%
    \begin{subfigure}[b]{0.5\linewidth}
        \centering
        \includegraphics[width=8.5cm]{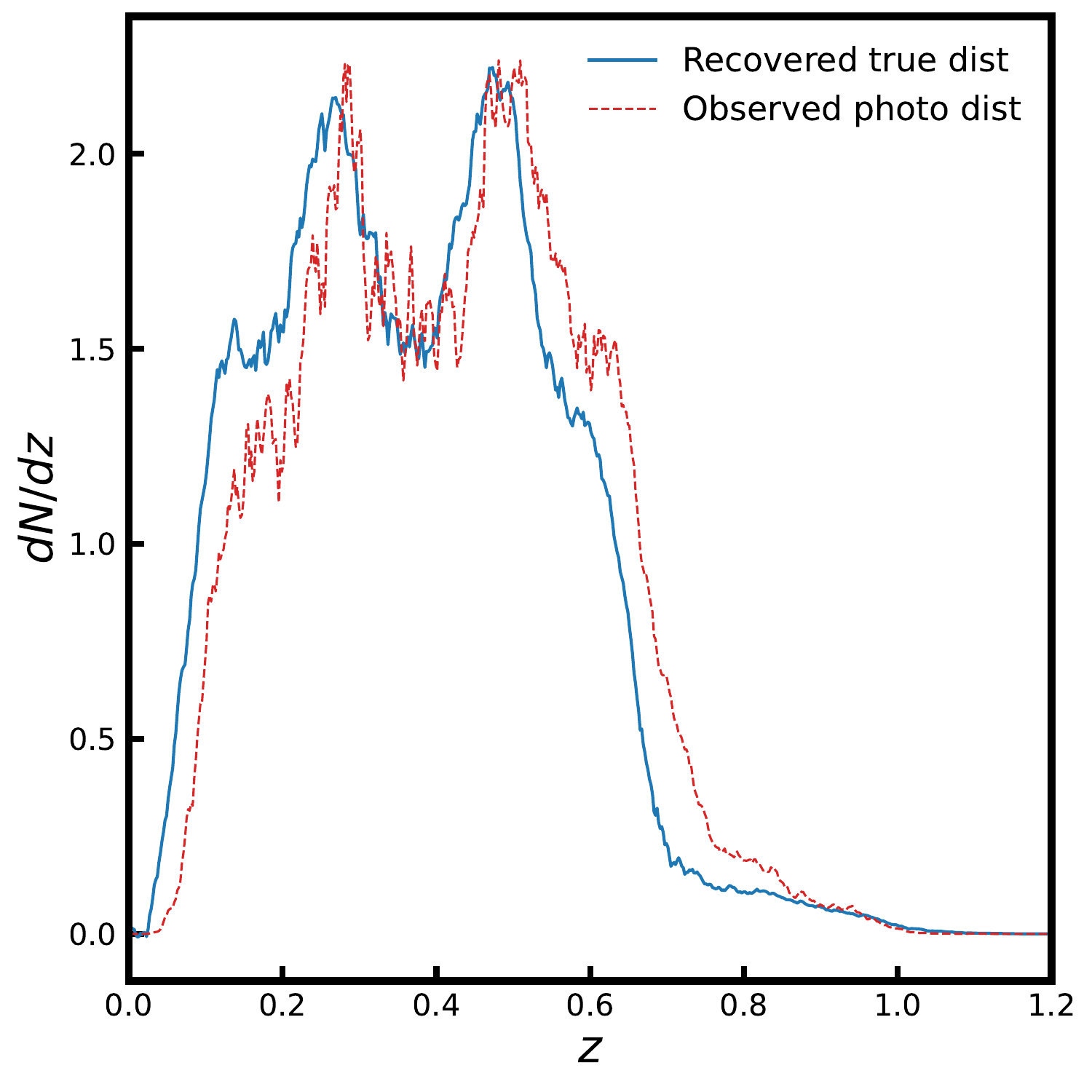}
        \caption{}
    \end{subfigure}
    \caption{(a) Observed redshift error distribution fit with the sum of Gaussians model. (b) Comparison of the photometric redshift distribution and the true redshift distribution recovered using the deconvolution method.}
    \label{fig:deconvolution_paper3}
\end{figure*}

As mentioned above, the parameters estimated in a tomographic cross-correlation study will be biased unless the power spectra are corrected for the redshift bin mismatch of objects. In \cite{Saraf2024}, we proposed a fast and efficient scattering matrix method to correct the mismatch of objects. Computation of {the} scattering matrix requires estimation of the true redshift distribution. We use the deconvolution method described in section \ref{sec:true_redshift_distribution_deconv_paper3} to estimate the true redshift distribution from the observed photometric redshift distribution and a model of the error distribution $p(z_{p}-z_{s}|z_{s})$. In Fig. \ref{fig:deconvolution_paper3} we show the sum of Gaussians fit to $p(z_{p}-z_{s}|z_{s})$ and the true redshift distribution recovered using the deconvolution method.

\begin{figure*}[hbt!]
    \begin{subfigure}[b]{0.5\linewidth}
        \centering
        \includegraphics[width=8.5cm]{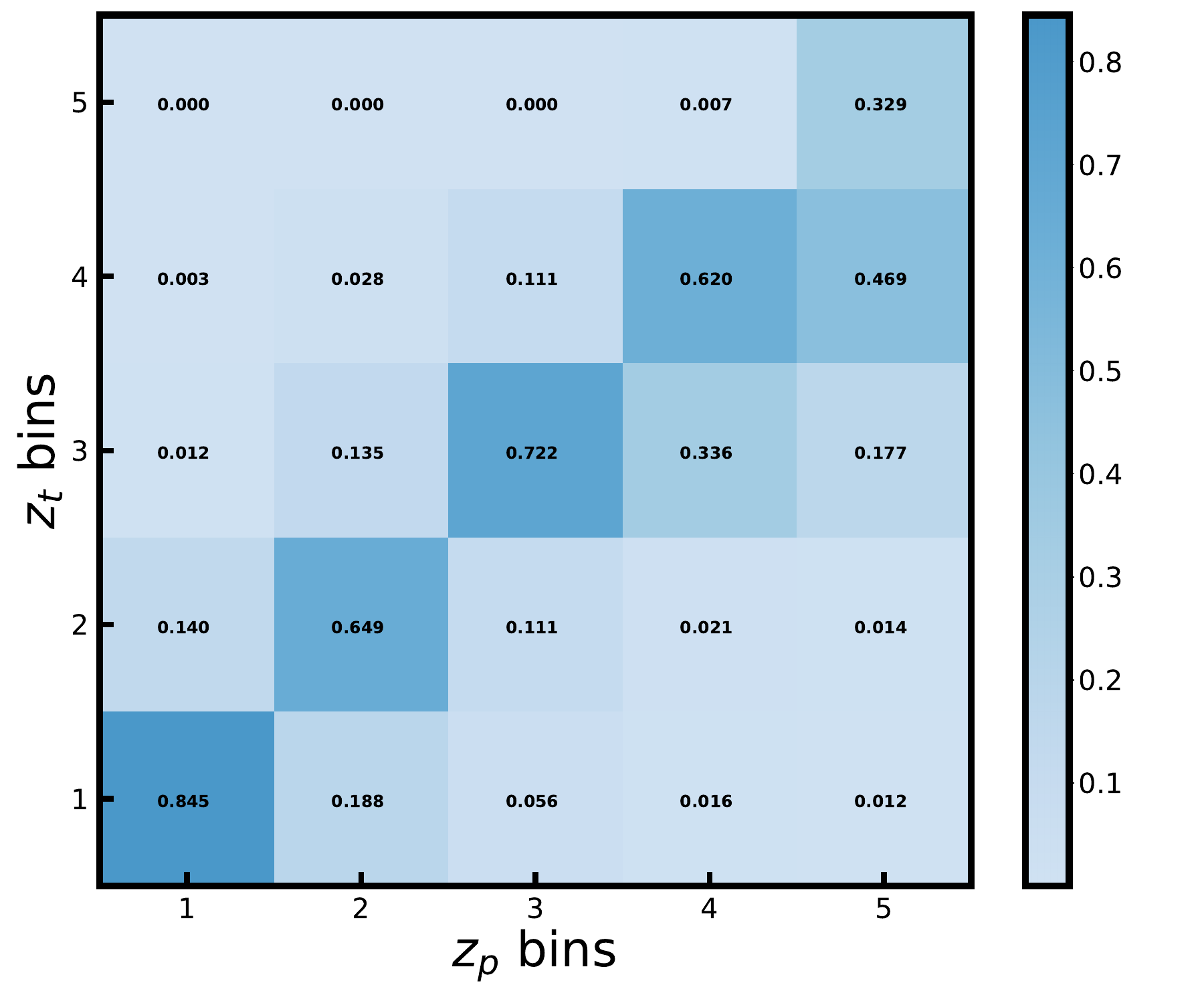}
        \caption{}
    \end{subfigure}%
    \begin{subfigure}[b]{0.5\linewidth}
        \centering
        \includegraphics[width=8.5cm]{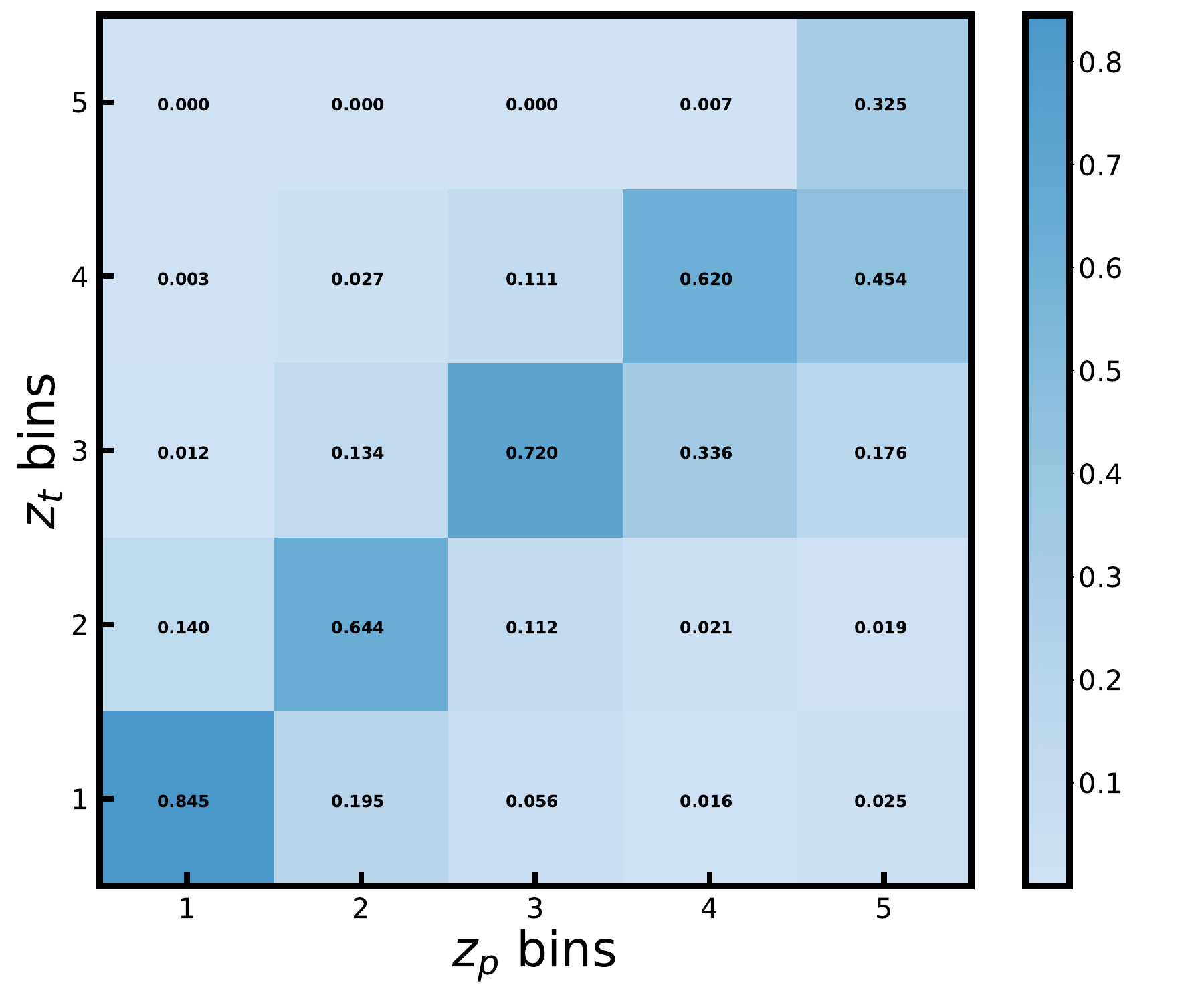}
        \caption{}
    \end{subfigure}
    \caption{Comparison of scattering matrices. (a) The scattering matrix estimated from DESI-LIS data. (b) The mean true scattering matrix estimated from $300$ Monte Carlo simulations of DESI-LIS data (described in section \ref{sec:simulations_paper3}). $z_{p}$ and $z_{t}$ bins represent the photometric and true redshift bins, respectively.}
    \label{fig:scattering_matrix_paper3}
\end{figure*}

Due to photometric redshift errors, some galaxies from $z<0.8$ will be wrongly placed at higher redshifts. To account for this loss of objects, we extended {the} estimation of the auto- and cross-power spectra as well as the scattering matrix for the tomographic bin $0.8<z<1.0$, following the procedure outlined in section \ref{sec:data_paper3}. In Fig. \ref{fig:scattering_matrix_paper3}, we compare the scattering matrix computed for the DESI-LIS photometric catalogue with the mean true scattering matrix estimated from $300$ Monte Carlo simulations described in section \ref{sec:simulations_paper3}. The similarities in the scattering matrices from data and simulations are expected given that we used the observed photometric error distribution in our simulations. On these grounds, we expect the scattering matrix estimated from {the} data to be robust in correcting the power spectra.

The leakage-corrected galaxy auto- and cross-power spectra $\hat{C}^{gg,\text{tr}}$ and $\hat{C}^{\kappa g,\text{tr}}$ were computed from the matrix relations \citep{Saraf2024}
\begin{align}
    \hat{C}^{gg,\text{tr}} &= P^{\text{T}^{-1}}\hat{C}^{gg,\text{ph}}P^{-1}\label{eq:scattering_relation_gg_matrix_paper3}\\
    \hat{C}^{\kappa g,\text{tr}} &= P^{\text{T}^{-1}}\hat{C}^{\kappa g,\text{ph}}
    \label{eq:scattering_relation_kg_matrix_paper3}
\end{align}
where $P$ and $P^{\text{T}}$ are the scattering matrix and its transpose, and $\hat{C}^{gg,\text{ph}}$ and $\hat{C}^{\kappa g,\text{ph}}$ are the power spectra estimated from the DESI-LIS catalogue and its cross correlation with the \textit{Planck} CMB lensing map (presented in section \ref{sec:results_wout_leakage_corr_paper3}). To compute the parameters from the estimates of the true power spectra using {the} maximum likelihood estimation method, we used the power spectrum template for each tomographic bin given by
\begin{align}
    C^{gg,\text{th}}_{i}(\ell) &= \int_{0}^{\chi_{*}}\frac{d\chi}{{\chi^{2}}}b^{2}(z)\bigg(\frac{\mathrm{d}N^{i}(z_{t})}{\mathrm{d}z_{t}}\bigg)^{2}P\bigg(k=\frac{\ell+1/2}{\chi},z(\chi)\bigg)\label{eq:gg_th_wout_corr_paper2}\\
    C^{\kappa g,\text{th}}_{i}(\ell) &= A\int_{0}^{\chi_{*}}\frac{d\chi}{{\chi^{2}}}W^{\kappa}(\chi)\,b(z)\frac{\mathrm{d}N^{i}(z_{t})}{\mathrm{d}z_{t}}P\bigg(k=\frac{\ell+1/2}{\chi},z(\chi)\bigg)\label{eq:kg_th_wout_corr_paper2}
\end{align}
where $b(z)$ and $A$ are free parameters and $\frac{\mathrm{d}N^{i}(z_{t})}{\mathrm{d}z_{t}}$ is given by
\begin{equation}
    \frac{\mathrm{d}N^{i}}{\mathrm{d}z_{t}} = \frac{\mathrm{d}N}{\mathrm{d}z_{t}}W^{i}(z_{t})
    \label{eq:true_dist_binned}
\end{equation}
where $\frac{\mathrm{d}N}{\mathrm{d}z_{t}}$ is the true redshift distribution computed using the deconvolution method and $W^{i}(z_{t})$ is window function given by Eq. (\ref{eq:window_function_paper3}).

\begin{table*}[hbt!]
    \renewcommand{\arraystretch}{1.5}
    \centering
    \caption{Best-fit galaxy linear bias $b$ and amplitude of cross correlation $A$.}
    \label{tab:best_fit_params_bA_paper3}
    \begin{tabular}{ccc||ccc||ccc}
    \hline\hline
    Bin & \multicolumn{2}{c}{From Hang et al.} & \multicolumn{6}{c}{This work}\\
    \cline{2-9}
     & & & \multicolumn{3}{c}{Before correction} & \multicolumn{3}{c}{After correction}\\
    \cline{4-9} 
     & $b_{1}$ & $A$ & $b$ & $A$ & $\chi_{r}^{2}$ & $b$ & $A$ & $\chi_{r}^{2}$ \\
    \hline\hline
    $(0.0, 0.3]$ & $1.25^{+0.01}_{-0.01}$ & $0.91^{+0.05}_{-0.05}$ & $1.251^{+0.006}_{-0.006}$ & $0.856^{+0.062}_{-0.062}$ & $1.049$ & $1.190^{+0.006}_{-0.006}$ & $0.915^{+0.060}_{-0.059}$ & $1.010$ \\
    $(0.3, 0.45]$ & $1.56^{+0.02}_{-0.02}$ & $0.80^{+0.04}_{-0.04}$ & $1.756^{+0.007}_{-0.007}$ & $0.756^{+0.049}_{-0.049}$ & $1.033$ & $1.628^{+0.007}_{-0.007}$ & $0.789^{+0.052}_{-0.052}$ & $1.251$ \\
    $(0.45, 0.6]$ & $1.53^{+0.01}_{-0.01}$ & $0.94^{+0.04}_{-0.04}$ & $1.739^{+0.006}_{-0.006}$ & $0.857^{+0.048}_{-0.048}$ & $0.823$ & $1.571^{+0.006}_{-0.006}$ & $0.941^{+0.047}_{-0.047}$ & $0.953$ \\
    $(0.6, 0.8]$ & $1.83^{+0.02}_{-0.02}$ & $0.91^{+0.04}_{-0.04}$ & $2.107^{+0.009}_{-0.009}$ & $0.862^{+0.043}_{-0.043}$ & $1.228$ & $1.693^{+0.008}_{-0.008}$ & $1.009^{+0.050}_{-0.050}$ & $1.013$ \\
    \hline
    \end{tabular}
    \tablefoot{The first set of parameters $b_{1}$ and $A$ are taken from H21. The second and third sets of $b$ and $A$ are estimated using sum of Gaussians fit to the redshift error distribution, before and after correcting for redshift bin mismatch of objects, respectively. $\chi^{2}_{r}$ is the reduced chi-square values for the cross-power spectrum with $\nu = 47$ degrees of freedom.}
\end{table*}
\begin{figure*}[hbt!]
    \begin{subfigure}[b]{\linewidth}
        \centering
        \includegraphics[width=17cm]{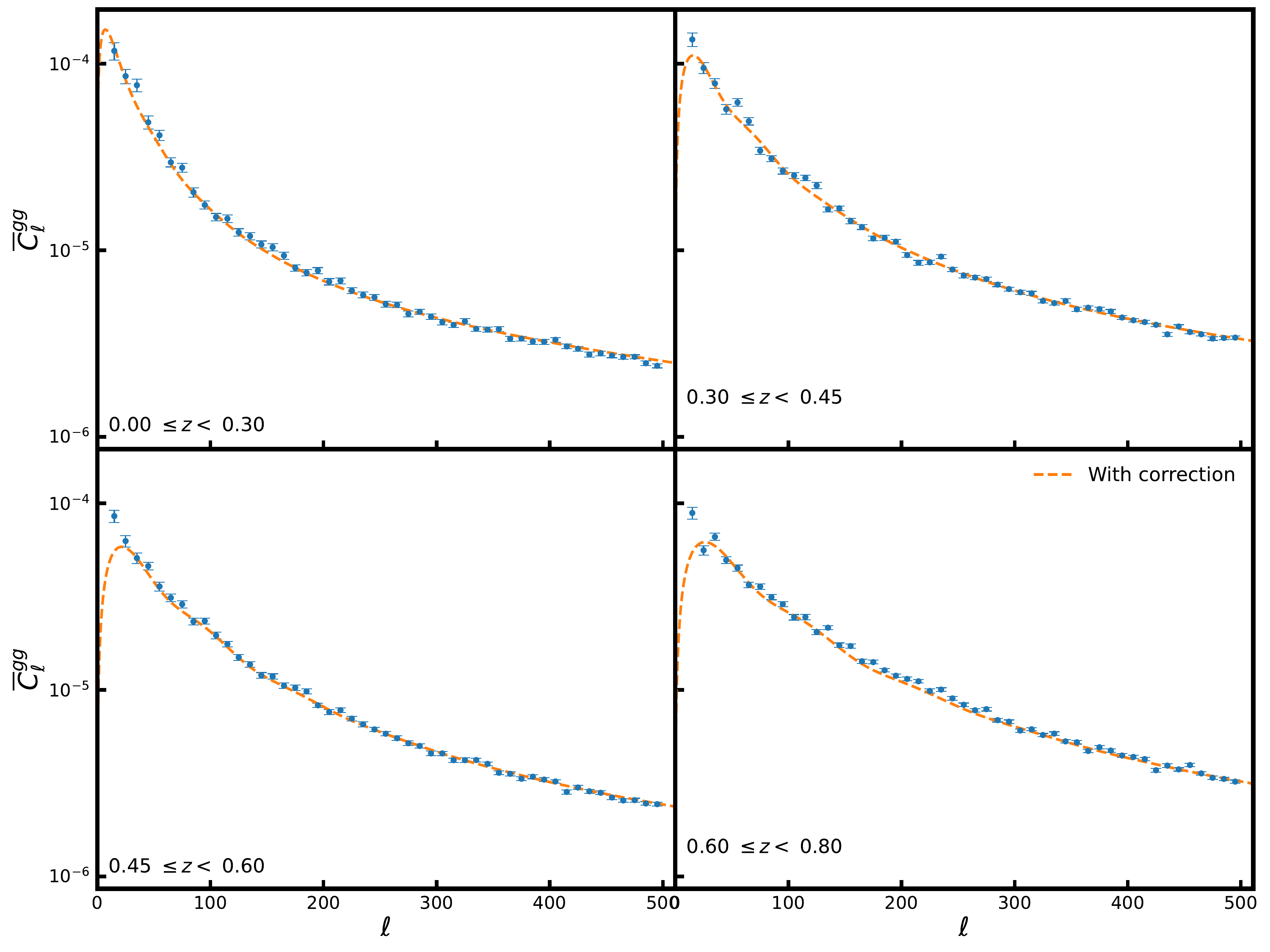}
        \captionsetup{labelformat=empty}
    \end{subfigure}%
    \caption{Galaxy auto-power spectrum measured from four DESI-LIS tomographic bins with the MV map after leakage correction. The orange dashed line is the theoretical power spectrum computed with the best-fit values of parameters quoted in Table \ref{tab:best_fit_params_bA_paper3}. {The standard error bars are computed from the covariance matrix used in the likelihood function; i.e. Eq.\,(\ref{eq:err_data_points_group_hang}).}}
    \label{fig:power_spectra_desi_gg_comp_with_wout_corr}
\end{figure*}
\begin{figure*}[hbt!]
    \begin{subfigure}[b]{\linewidth}
        \centering
        \includegraphics[width=17cm]{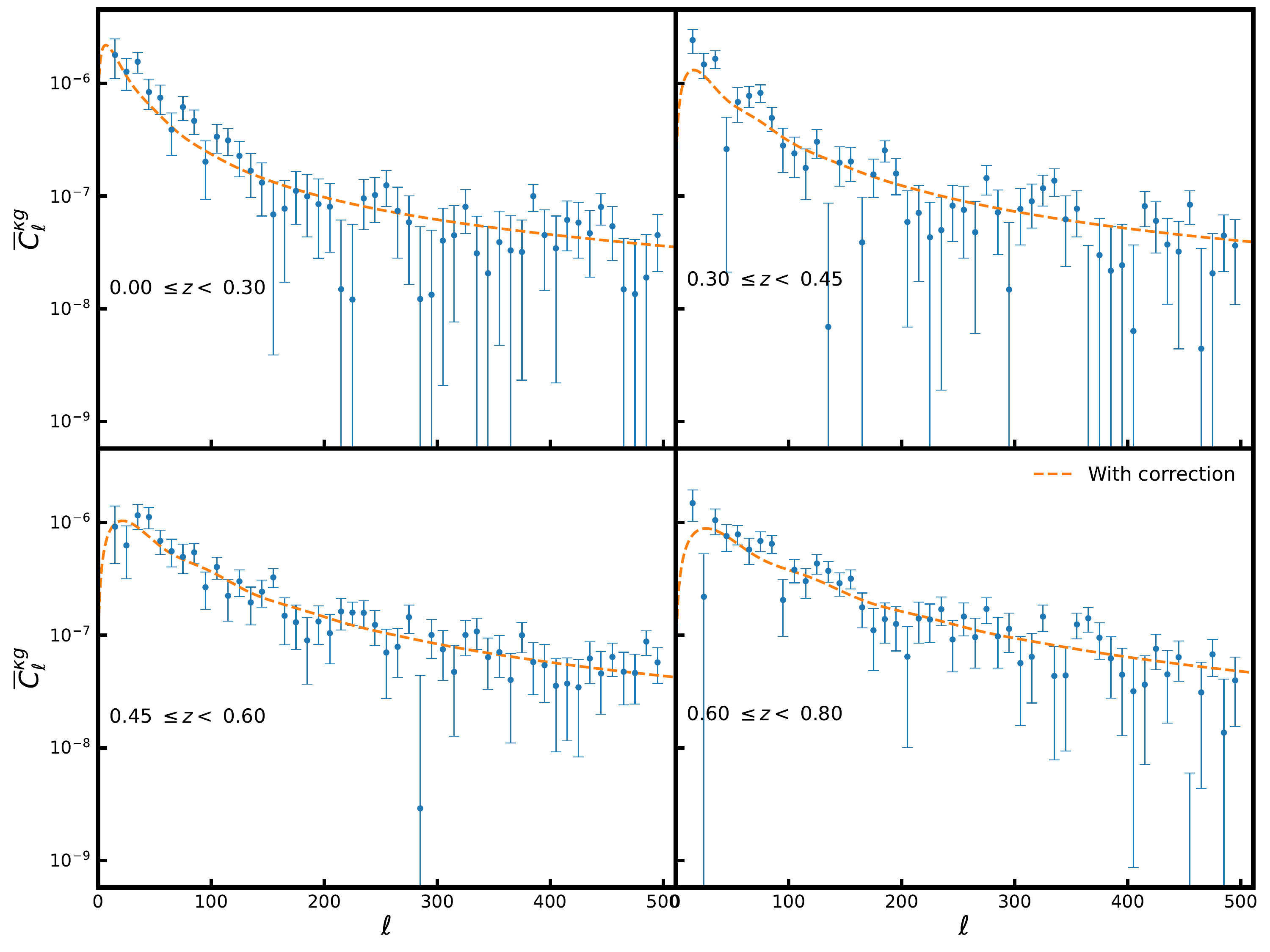}
        \captionsetup{labelformat=empty}
    \end{subfigure}%
    \caption{Cross-power spectrum measured from four DESI-LIS tomographic bins with the MV map after leakage correction. The orange dashed line is the theoretical power spectrum computed with the best-fit values of parameters quoted in Table \ref{tab:best_fit_params_bA_paper3}. {The standard error bars are computed from the covariance matrix used in the likelihood function; i.e. Eq.\,(\ref{eq:err_data_points_group_hang}).}}
    \label{fig:power_spectra_desi_kg_comp_with_wout_corr}
\end{figure*}

\begin{figure*}[hbt!]
    \begin{subfigure}[b]{0.5\linewidth}
        \centering
        \includegraphics[width=8.5cm]{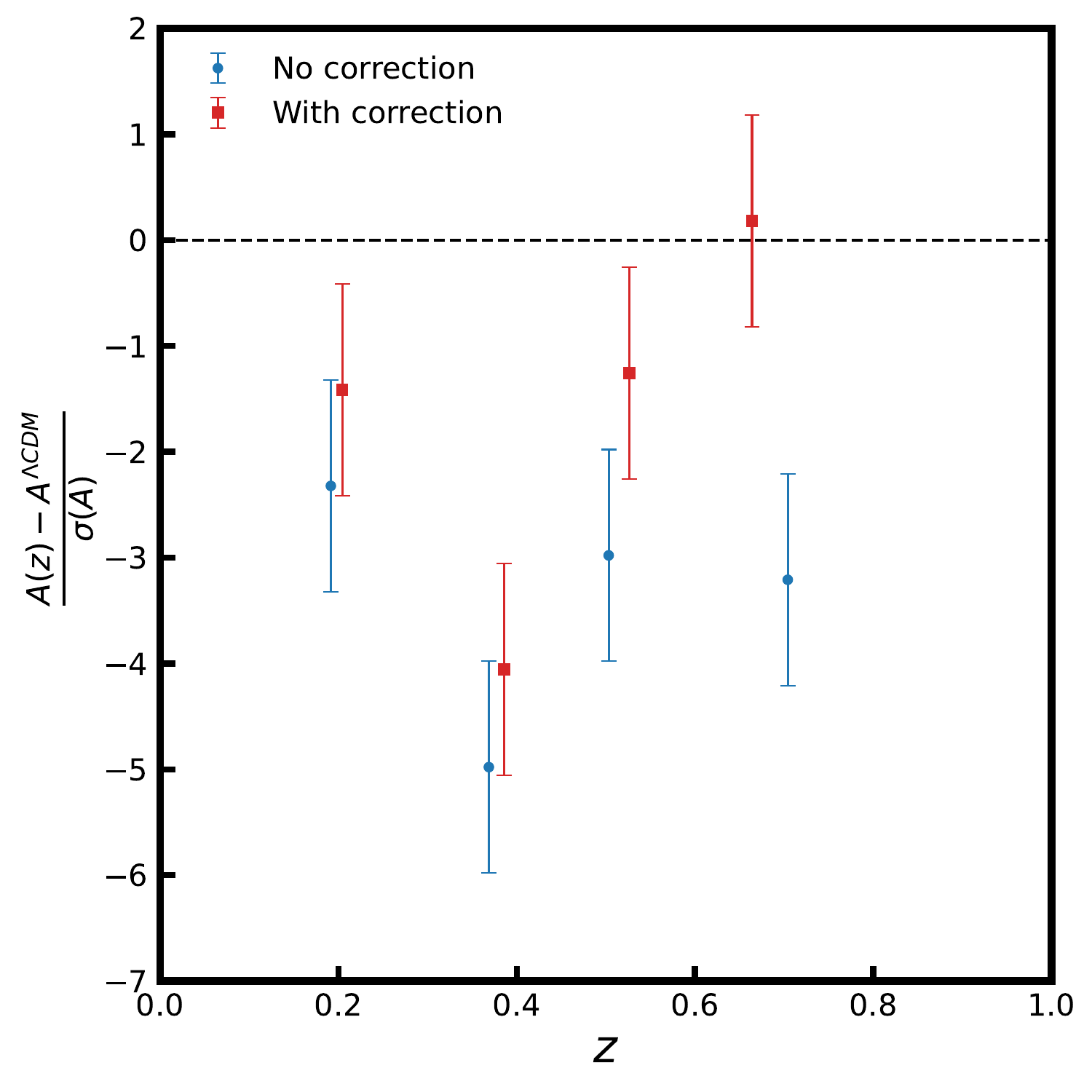}
        \caption{}
    \end{subfigure}%
    \begin{subfigure}[b]{0.5\linewidth}
        \centering
        \includegraphics[width=8.5cm]{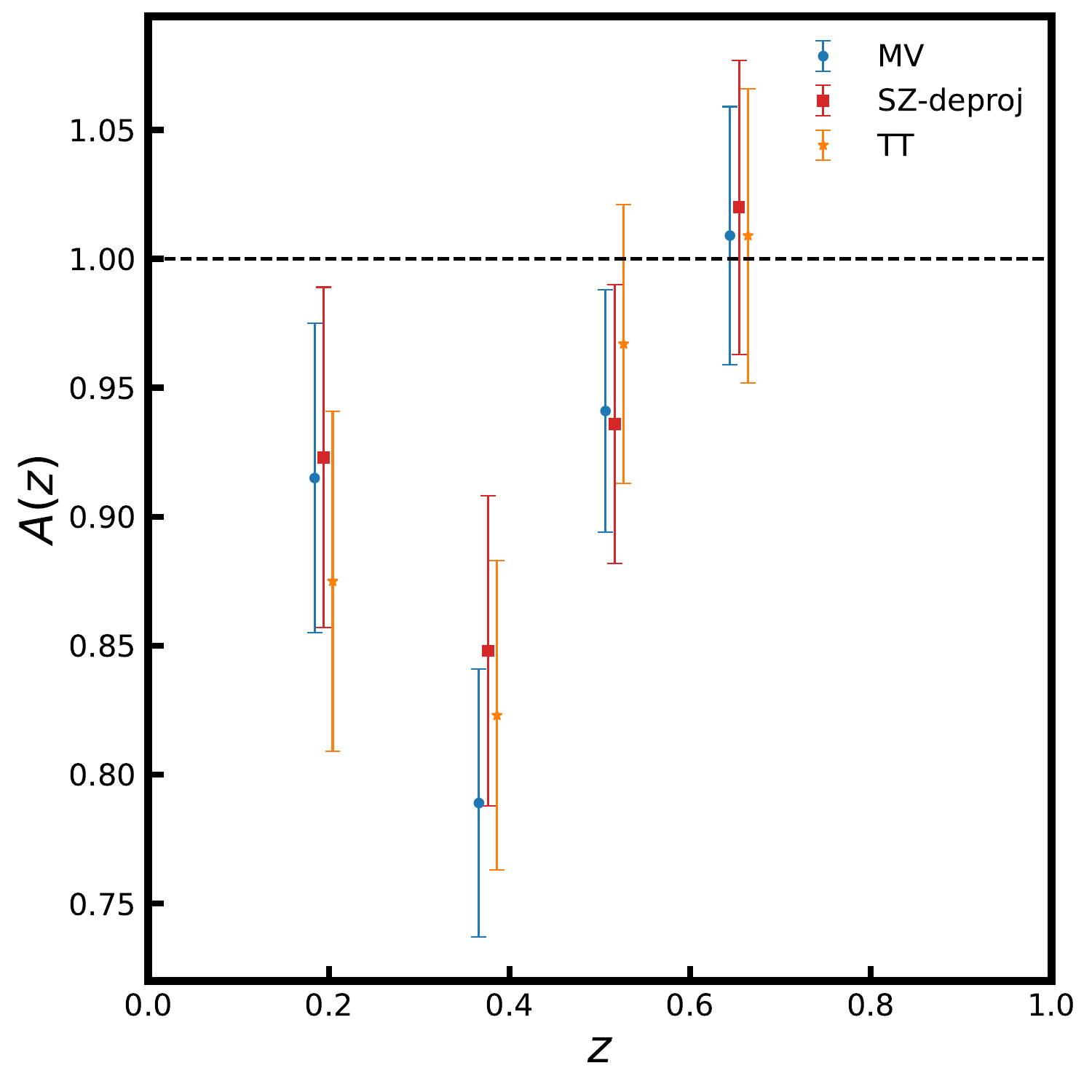}
        \caption{}
    \end{subfigure}
    \caption{Effect of the redshift bin mismatch on the amplitude of cross-correlation. (a) Deviation in the amplitude of cross correlation estimated from the DESI-LIS galaxy catalogue with and without leakage correction from its true value ($A=1$) in terms of the standard deviation of the amplitude. (b) Comparison of redshift-bin-mismatch-corrected amplitudes from cross correlation with MV, SZ-deproj, and TT CMB lensing convergence maps.}
    \label{fig:comp_amplitude_paper3}
\end{figure*}

In Figures \ref{fig:power_spectra_desi_gg_comp_with_wout_corr} and \ref{fig:power_spectra_desi_kg_comp_with_wout_corr}, we show the galaxy auto-power spectra and cross-power spectra for the four tomographic bins corrected for the redshift bin mismatch of galaxies and the corresponding best-fit theoretical power spectra. In Table \ref{tab:best_fit_params_bA_paper3}, we compare the parameters $b$ and $A$ computed before and after leakage correction for our baseline results using the sum of Gaussians fit to the redshift error distributions. The cross-correlation amplitude is notably higher in all tomographic bins after correction for redshift bin mismatch and becomes consistent with unity in the last tomographic bin. The tension in the amplitude remains at $\sim 4\,\sigma$ in the second redshift bin, but reduces to $1-1.5\,\sigma$ in other redshift bins. We note that the amplitude of cross correlation {is} consistent with the estimates of H21, except for the last redshift bin where we quote {a} higher value of $A$ by $\sim 2.5\,\sigma$. The behaviour in the last bin can be understood as an effect of including the objects scattered outside the redshift range $0\leq z< 0.8$, which were not considered by H21. The galaxy bias, however, differs significantly between our estimates. {We find lower values of the galaxy bias after correcting our baseline results with the scattering matrix formalism. We report values of galaxy bias that are slightly different from those of H21 which could result from our different assumptions in the model for galaxy bias evolution.} As the amplitude of cross correlation {is} consistent, we believe that the need for a {two-parameter} bias model (Eq. \ref{eq:galaxy_bias_hang}) may also have been created as a result of {the} redshift bin mismatch of objects. In Fig. \ref{fig:comp_amplitude_paper3}, we present the deviations of the estimated amplitude of cross correlation from its expected value ($A=1$) in terms of {the} standard deviation of the amplitude. By countering the impact of redshift bin mismatch, we reduce the tension in amplitude from $4-6\,\sigma$ to $\sim 2\,\sigma$, with complete agreement within errors for the last tomographic bin.

We note the peculiar behaviour of the second redshift bin --- correction with scattering matrix does not affect the amplitude of cross correlation whereas our simulations predict an increase in the galaxy bias (or a corresponding reduction in the amplitude) after correction of the redshift bin mismatch (see Fig. \ref{fig:comp_params_simulations_paper3}). A possible explanation for the different behaviour can be found in section 3.2 of H21 where the authors mention that, for their redshift calibration, a small proportion of objects with spectroscopic redshifts of $0.2<z<0.4$ were assigned photometric redshifts of $0.4<z<0.6$. This systematic in the {H21} redshift calibration will impact the photometric redshift error distribution and we might underestimate the scatter of objects from the second redshift bin.

Having estimates of the galaxy bias from four redshift bins we can compute the galaxy bias at redshift zero, $b_0$, from our model of galaxy bias $b(z) = 1+\frac{b_{0}-1}{D(z)}$. We find $b_{0}= 1.529\pm 0.012 \text{ and } 1.401\pm 0.006$ without and with correction for redshift bin mismatch, respectively. Accounting for the scatter of objects between redshift bins leads to {a} significantly lower value of $b_{0}$ and will result in different inferences about the relation between the dark matter and luminous matter.

\subsubsection{Using different CMB lensing potential maps}\label{sec:diff_cmb_lensing_maps}

\begin{figure*}[hbt!]
    \begin{subfigure}[b]{\linewidth}
        \centering
        \includegraphics[width=17cm]{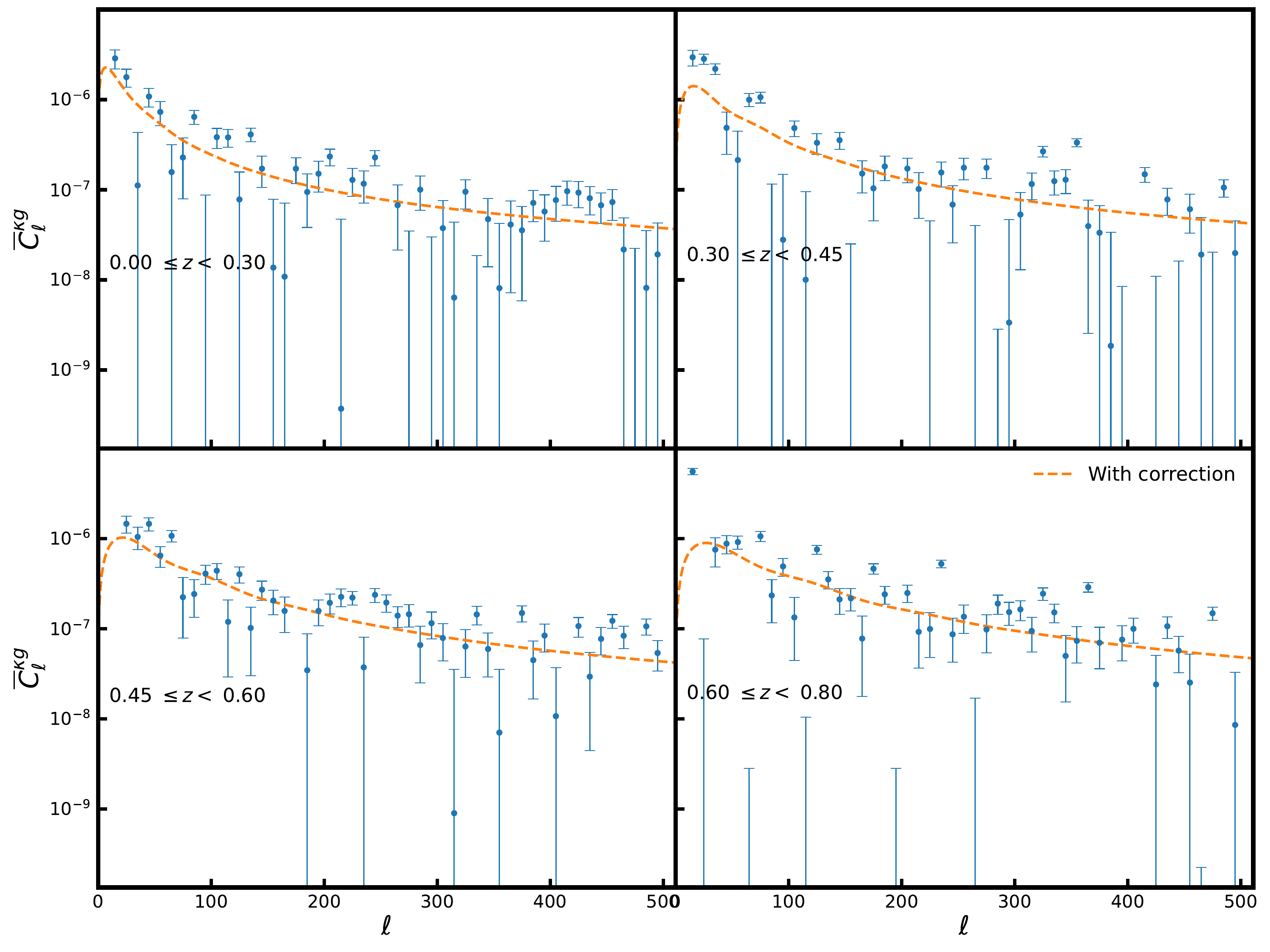}
        \captionsetup{labelformat=empty}
    \end{subfigure}%
    \caption{Cross-power spectrum measured from four DESI-LIS tomographic bins with the SZ-deproj map after leakage correction. The orange dashed line is the theoretical power spectrum computed with the best-fit values of parameters quoted in Table \ref{tab:best_fit_params_bA_paper3}. {The standard error bars are computed from the covariance matrix used in the likelihood function; i.e. Eq.\,(\ref{eq:err_data_points_group_hang}).}}
    \label{fig:power_spectra_desi_kg_comp_with_wout_corr_sz}
\end{figure*}

\begin{figure*}[hbt!]
    \begin{subfigure}[b]{\linewidth}
        \centering
        \includegraphics[width=17cm]{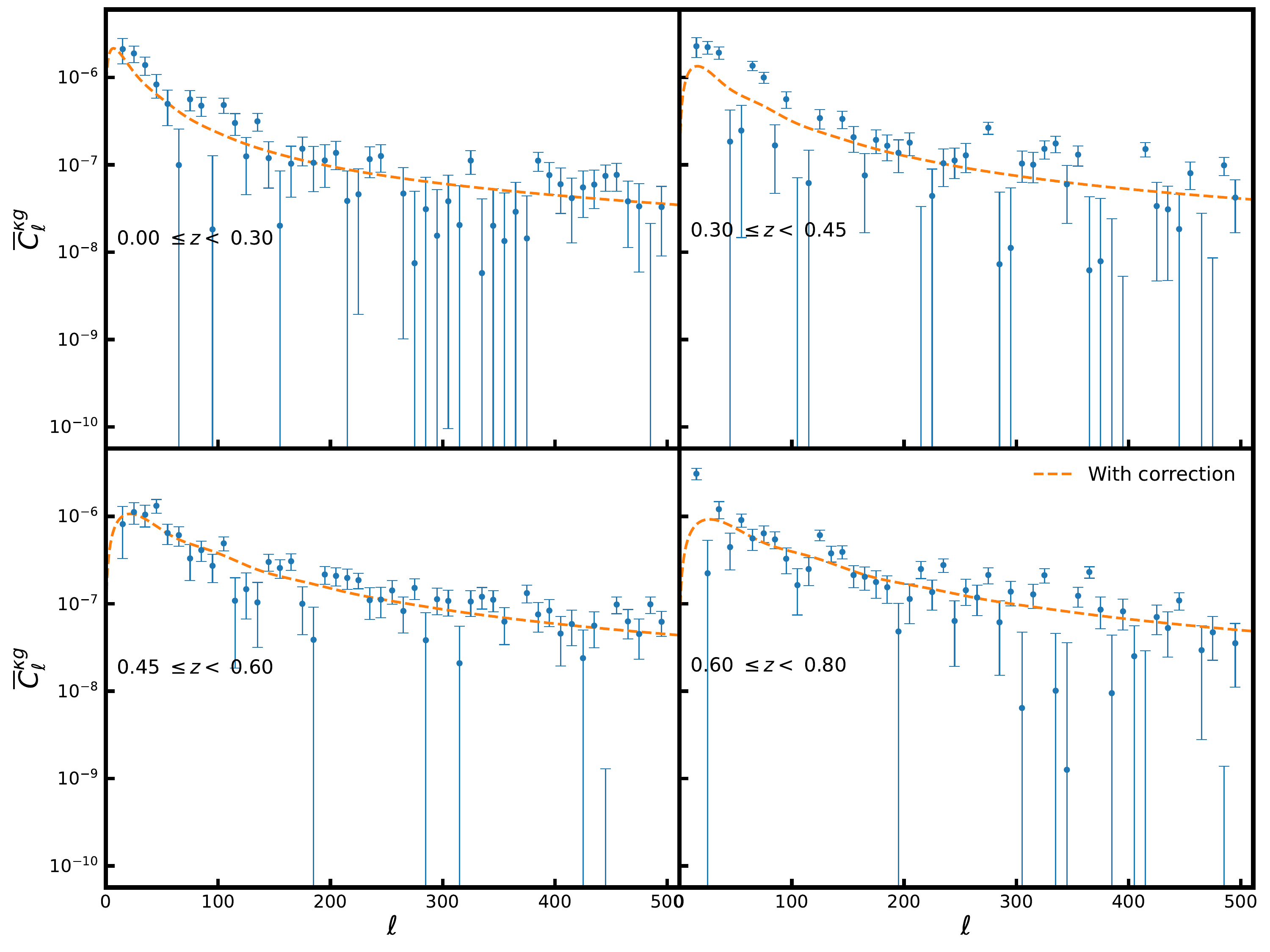}
        \captionsetup{labelformat=empty}
    \end{subfigure}%
    \caption{Cross-power spectrum measured from four DESI-LIS tomographic bins with the TT map after leakage correction. The orange dashed line is the theoretical power spectrum computed with the best-fit values of parameters quoted in Table \ref{tab:best_fit_params_bA_paper3}. {The standard error bars are computed from the covariance matrix used in the likelihood function; i.e. Eq.\,(\ref{eq:err_data_points_group_hang}).}}
    \label{fig:power_spectra_desi_kg_comp_with_wout_corr_tt}
\end{figure*}

Although accounting for leakage reduces the deviations observed on the amplitude of cross correlation for our baseline analysis, there remains a $\sim 2\,\sigma$ tension concerning the prediction of the standard cosmological model. There are several galaxy survey systematic errors, such as catastrophic errors, photometric calibration errors, and magnification, that can affect the estimation of power spectra in a cross correlation analysis. We have not considered these systematic errors with DESI-LIS datasets, because the goal of this study is to convey the importance of redshift mismatch correction in the estimation of parameters. However, one of the important conclusions from \cite{Saraf2022} was that different CMB lensing convergence maps produce significantly different values of the cross-correlation amplitude parameter. Hence, in this section, we compared estimates of galaxy bias and cross-correlation amplitude obtained from the SZ-deproj and TT CMB lensing convergence maps.
\begin{table*}[hbt!]
    \renewcommand{\arraystretch}{1.5}
    \centering
    \caption{Galaxy linear bias and cross-correlation amplitude with MV, SZ-deproj and TT-only convergence maps.}
    \label{tab:likeli_result_comparing_mv_sz_tt_desi_with_corr}
    \begin{tabular}{l||ccc||ccc||ccc}
    \hline\hline
    Bin & \multicolumn{3}{c}{MV} & \multicolumn{3}{c}{SZ-deproj} & \multicolumn{3}{c}{TT}\\
    \cline{2-10}
    & $b$ & $A$ & $\chi_{r}^{2}$ & $b$ & $A$ & $\chi_{r}^{2}$ & $b$ & $A$ & $\chi_{r}^{2}$ \\
    \hline
    $(0.0, 0.3]$ & $1.190^{+0.006}_{-0.006}$ & $0.915^{+0.060}_{-0.059}$ & $1.010$ & $1.190^{+0.003}_{-0.003}$ & $0.923^{+0.066}_{-0.067}$ & $2.531$ & $1.189^{+0.003}_{-0.003}$ & $0.875^{+0.066}_{-0.067}$ & $1.781$ \\
    $(0.3, 0.45]$ & $1.628^{+0.007}_{-0.007}$ & $0.789^{+0.052}_{-0.052}$ & $1.251$ & $1.628^{+0.004}_{-0.004}$ & $0.848^{+0.060}_{-0.060}$ & $4.173$ & $1.628^{+0.004}_{-0.004}$ & $0.823^{+0.060}_{-0.060}$ & $3.706$ \\
    $(0.45, 0.6]$ & $1.571^{+0.006}_{-0.006}$ & $0.941^{+0.047}_{-0.047}$ & $0.953$ & $1.571^{+0.004}_{-0.004}$ & $0.936^{+0.054}_{-0.055}$ & $2.718$ & $1.571^{+0.004}_{-0.004}$ & $0.967^{+0.054}_{-0.054}$ & $1.748$ \\
    $(0.6, 0.8]$ & $1.693^{+0.008}_{-0.008}$ & $1.009^{+0.050}_{-0.050}$ & $1.013$ & $1.693^{+0.004}_{-0.004}$ & $1.020^{+0.057}_{-0.057}$ & $4.014$ & $1.693^{+0.004}_{-0.004}$ & $1.009^{+0.056}_{-0.057}$ & $3.218$ \\
    \hline
    \end{tabular}
    \tablefoot{The parameters are computed for DESI-LIS tomographic bins taking into account the effects of leakage correction.}
\end{table*}

In Table \ref{tab:likeli_result_comparing_mv_sz_tt_desi_with_corr}, we compare the best-fit values of parameters $b$ and $A$ estimated for cross correlation of DESI-LIS tomographic bins with MV, SZ-deproj, and TT CMB lensing convergence maps after leakage correction. As we can see from the right panel of Fig. \ref{fig:comp_amplitude_paper3}, the amplitude estimated with the SZ-deproj map {is} found to be in agreement with {the} MV map, except in the second redshift bin, where {the} SZ-deproj map prefers a $\sim 1\,\sigma$ higher value than the MV map. The TT map, on the other hand, is consistent with the MV map overall, while alleviating the tension on the cross-correlation amplitude for the third and fourth tomographic bins. Figures \ref{fig:power_spectra_desi_kg_comp_with_wout_corr_sz} and \ref{fig:power_spectra_desi_kg_comp_with_wout_corr_tt} show the redshift-bin-mismatch-corrected cross-power spectra for the four tomographic bins with SZ-deproj and TT maps, respectively.

\subsection{Estimation of the \texorpdfstring{$\sigma_{8}$}{Lg} parameter}
\begin{figure*}[hbt!]
    \begin{subfigure}[b]{0.5\linewidth}
        \centering
        \includegraphics[width=8.5cm]{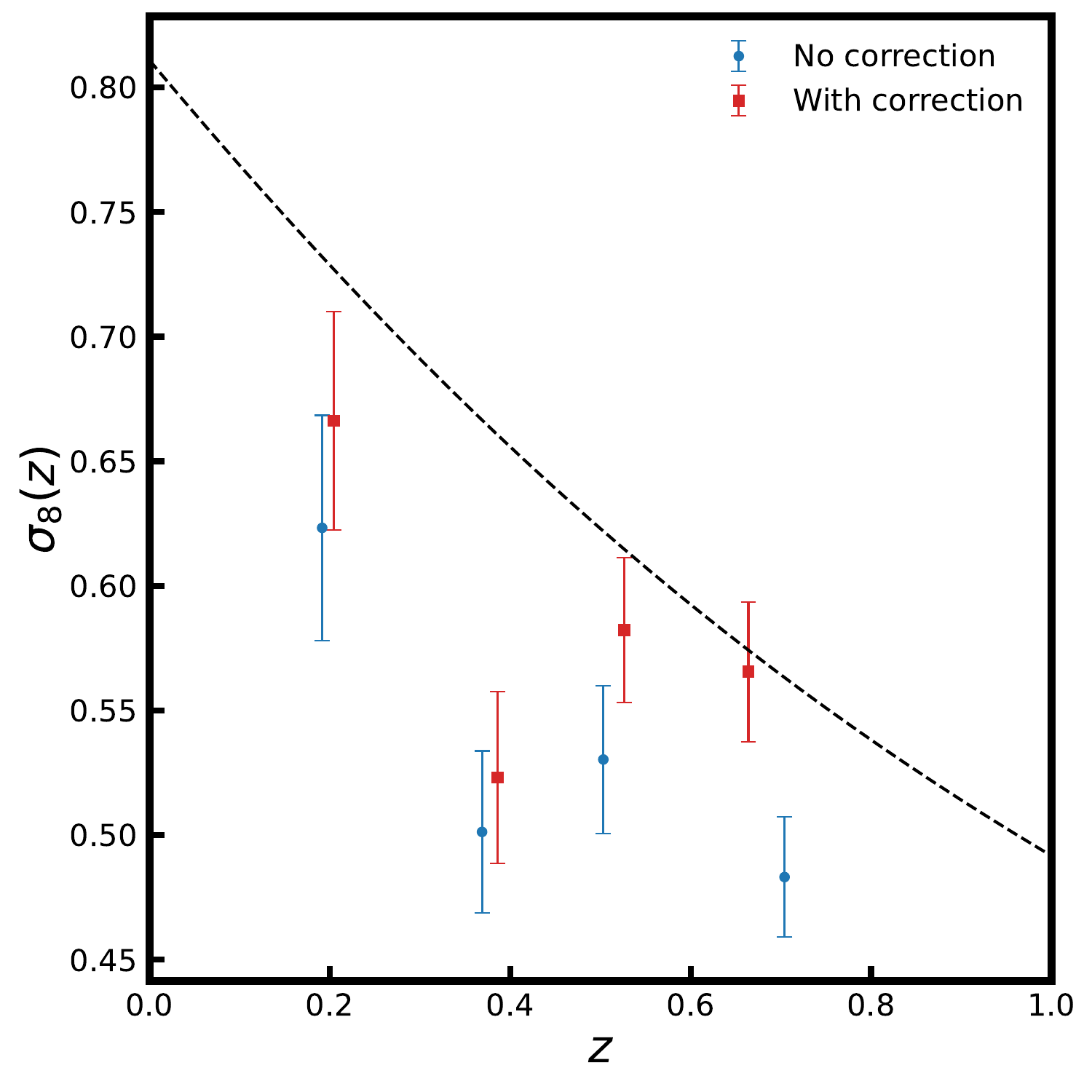}
        \caption{}
    \end{subfigure}%
    \begin{subfigure}[b]{0.5\linewidth}
        \centering
        \includegraphics[width=8.5cm]{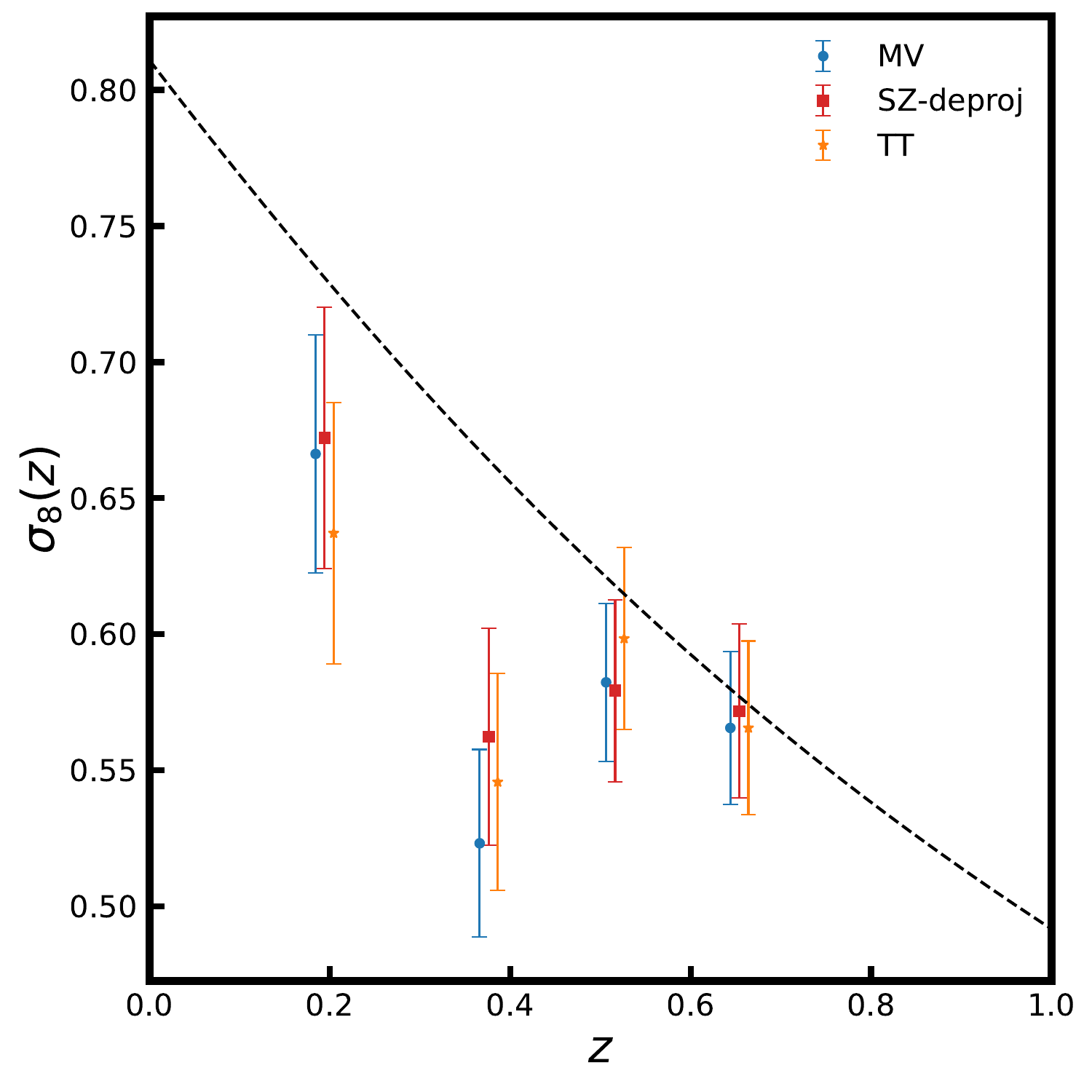}
        \caption{}
    \end{subfigure}
    \caption{Estimates of the $\sigma_{8}$ parameter. (a) Comparison of the $\sigma_{8}$ parameter computed using cross-correlation measurement with the MV map before and after leakage correction. (b) Comparison of the $\sigma_{8}$ parameter after leakage correction computed from cross correlations with MV, SZ-deproj, and TT CMB lensing maps. The dashed line represents the redshift evolution of the $\sigma_{8}$ parameter for the $\Lambda$CDM cosmology.}
    \label{fig:sigma8_evolution_paper3}
\end{figure*}

The amplitude of cross-correlation can be translated to the $\sigma_{8}$ parameter using the relation \citep{Peacock&Bilicki2018}
\begin{equation}
    \sigma_{8}(z) = A(z)\,\sigma_{8,0}\,D(z)
    \label{eq:sigma8_evolution}
\end{equation}
where $\sigma_{8,0}$ is the value of the $\sigma_{8}$ parameter at redshift $z=0$. In the left panel of Fig. \ref{fig:sigma8_evolution_paper3}, we show the impact of redshift bin mismatch on the $\sigma_{8}$ parameter estimated from our baseline analysis for the cross-correlation measurements between the DESI-LIS photometric galaxy catalogue and the \textit{Planck} MV CMB lensing potential. The dashed black line is the redshift evolution of $\sigma_{8}$ for our fiducial cosmology. The $\sigma_{8}$ parameter, after correcting for the redshift bin mismatch of objects, agrees better with the predictions of the $\Lambda$CDM model overall, becoming fully consistent in the last tomographic bin. {However, there remains a $\sim 1.4\,\sigma$ discrepancy in the first redshift bin, a $\sim 3\,\sigma$ difference for the second redshift bin, and a $\sim 1.2\,\sigma$ discrepancy for the third redshift bin in the $\sigma_{8}$ parameter compared to the $\Lambda$CDM model. The observed tension in the $\sigma_{8}$ parameter with respect to the standard cosmological model is consistent with other measurements (\citealt{Nakoneczny2024}; \citealt{Chang2023}; \citealt{White2022}; \citealt{Sun2022}; \citealt{Krolewski2021}). }

{A possible reason for this discrepancy with the DESI-LIS dataset could be systematic uncertainty in the redshift calibration performed by H21, as already {mentioned} in section \ref{sec:results_with_leakage_corr_paper3}. In the estimation of photometric redshifts by H21, a proportion of objects with spectroscopic redshifts between $0.2<z<0.4$ were assigned photometric redshifts between $0.4<z<0.6$. This systematic error will significantly impact the bin mismatch correction based on the scattering matrix and estimation of the $\sigma_{8}$ parameter. In addition, there can be residual photometric calibration errors across the three legacy imaging surveys (DECaLS, BASS, MzLS), which can alter the measured angular power spectrum. However, the goal of this study is to present the scattering matrix approach for the correction of redshift bin mismatch in observational data. A detailed analysis of other systematic uncertainties and their impact on the $\sigma_{8}$ tension is outside the scope of this work and we reserve such analyses for future studies.}

{If these systematic errors were not able to explain the observed offsets in the $\sigma_{8}$ parameter, then we could conclude that there is a mild tension with the $\Lambda$CDM model. A wide range of scenarios are being considered in the literature to explain the $\sigma_{8}-\Omega_{m}$ tension in cosmology, such as the impact of redshift space distortions \citep{Macaulay2013}, a redshift dependence of the $S_{8}$ parameter \citep{Adil2024}, or modifications to the standard cosmological model (see \citealt{Abdalla2022} for a comprehensive review).}

In the right panel of Fig. \ref{fig:sigma8_evolution_paper3}, we compare the $\sigma_{8}$ parameter computed from the cross-correlation measurements with MV, SZ-deproj, and TT CMB lensing maps. The estimates of $\sigma_{8}$ parameters from different CMB lensing maps are consistent with each other. {In Table \ref{tab:likeli_result_comparing_bias_sigma8_mv_sz_tt_desi_with_corr}, we compare values of galaxy bias and $\sigma_{8}$ for the four DESI-LIS tomographic bins obtained before and after correction for redshift bin mismatch of objects.}

\begin{table*}[hbt!]
    \renewcommand{\arraystretch}{1.5}
    \centering
    \caption{{Galaxy linear bias $b$ and $\sigma_{8}$ from cross-correlation of the DESI-LIS galaxies with the \textit{Planck} MV CMB lensing potential map.}}
    \label{tab:likeli_result_comparing_bias_sigma8_mv_sz_tt_desi_with_corr}
    \begin{tabular}{l||c||cc||cc||cc||cc}
    \hline\hline
     & & \multicolumn{2}{c}{$0.0<z\leq 0.3$} & \multicolumn{2}{c}{$0.3<z\leq 0.45$} & \multicolumn{2}{c}{$0.45<z\leq 0.6$} & \multicolumn{2}{c}{$0.6<z\leq 0.8$}\\
    \cline{3-10}
    & leakage & $b$ & $\sigma_{8}$  & $b$ & $\sigma_{8}$  & $b$ & $\sigma_{8}$  & $b$ & $\sigma_{8}$\\
    & correction & & ($0.731$) & & ($0.664$) & & ($0.618$) & & ($0.577$) \\[1ex]
    \hline
    H21 & no & $1.25^{+0.01}_{-0.01}$ & $0.663^{+0.036}_{-0.036}$ & $1.56^{+0.02}_{-0.02}$ & $0.530^{+0.026}_{-0.026}$ & $1.53^{+0.01}_{-0.01}$ & $0.582^{+0.025}_{-0.025}$ & $1.83^{+0.02}_{-0.02}$ & $0.510^{+0.022}_{-0.022}$ \\[1ex]
    \hline
    \multirow{2}{3em}{This work} & no & $1.251^{+0.006}_{-0.006}$ & $0.623^{+0.045}_{-0.045}$ & $1.756^{+0.007}_{-0.007}$ & $0.501^{+0.032}_{-0.032}$ & $1.739^{+0.006}_{-0.006}$ & $0.530^{+0.030}_{-0.030}$ & $2.107^{+0.009}_{-0.009}$ & $0.483^{+0.024}_{-0.024}$ \\[1ex]
    
    & yes & $1.190^{+0.006}_{-0.006}$ & $0.669^{+0.044}_{-0.044}$ & $1.628^{+0.007}_{-0.007}$ & $0.524^{+0.034}_{-0.034}$ & $1.571^{+0.006}_{-0.006}$ & $0.581^{+0.029}_{-0.029}$ & $1.693^{+0.008}_{-0.008}$ & $0.582^{+0.029}_{-0.029}$ \\[1ex]
%    & SZ-deproj & $1.190^{+0.003}_{-0.003}$ & $0.923^{+0.066}_{-0.067}$ & $1.628^{+0.004}_{-0.004}$ & $0.848^{+0.060}_{-0.060}$ & $1.571^{+0.004}_{-0.004}$ & $0.936^{+0.054}_{-0.055}$ & $1.693^{+0.004}_{-0.004}$ & $1.020^{+0.057}_{-0.057}$ \\[1ex]
%    & TT & $1.189^{+0.003}_{-0.003}$ & $0.875^{+0.066}_{-0.067}$ & $1.628^{+0.004}_{-0.004}$ & $0.823^{+0.060}_{-0.060}$ & $1.571^{+0.004}_{-0.004}$ & $0.967^{+0.054}_{-0.054}$ & $1.693^{+0.004}_{-0.004}$ & $1.009^{+0.056}_{-0.057}$ \\[1ex]
    \hline\hline
    \end{tabular}
    \tablefoot{{The parameters are computed for DESI-LIS tomographic bins taking into account the effects of leakage correction. The $\sigma_{8}$ values in the brackets are values expected within the $\Lambda$CDM cosmological model.}}
\end{table*}

%-----------------------------------------------------------------

\section{Summary}\label{sec:summary_paper3}

We performed a tomographic cross-correlation analysis between the minimum-variance CMB lensing convergence map from \cite{Planck2020VIII} and photometric galaxy catalogues from the Data Release 8 of Dark Energy Spectroscopic Instrument Legacy Imaging Survey prepared by H21. We divided the galaxy density field into four redshift slices covering $0<z<0.8$, with a photometric redshift precision of $\frac{\sigma_{z}}{1+z}$ in the range $0.012-0.015$.

H21 used {a} modified Lorentzian function to approximate the photometric redshift error distribution. Here, we show that the modified Lorentzian function performs well only near the peaks of the error distribution but fails to capture the tails of the error distribution. Although the number of objects in the tails of the error distribution will be smaller than in the peak, we show that broad tails significantly affect the estimation of parameters by adopting a better fitted sum of Gaussians model for the redshift error distributions. We find that our `sum of Gaussians' model gives $\sim 1-1.5\,\sigma$ smaller values for the cross-correlation amplitude and significantly higher estimates of the galaxy linear bias than H21. As the sum of Gaussians model better captures the peculiarities of the error distributions, we treat the parameters estimated under this model as our baseline results.

We implemented the scattering matrix formalism presented in \cite{Saraf2024} to correct for the scatter of objects across redshift bins causing biased estimation of parameters. For our baseline analysis with {the} sum of Gaussians fit to the redshift error distributions, we find a $\sim 1-2\,\sigma$ higher values of the amplitude of cross correlation from three out of four tomographic bins after correction for leakage, whereas no significant change was observed in the second redshift bin. This may be caused by a small proportion of objects with redshifts between $0.2<z<0.4$, that have assigned photometric redshifts between $0.4<z<0.6$, as mentioned by H21 in section 3.2 {of their paper}. Due to this systematic error in the redshift calibration, we may underestimate the scattering matrix elements for the second redshift bin. Our estimates of cross-correlation amplitude are consistent with the $\Lambda$CDM expectations, except for the last redshift bin where we get a value that is higher by $\sim 2.5\,\sigma$, making it completely consistent with $\Lambda$CDM expectations. The difference in the last bin could be the result of taking into account objects with redshift of higher than 0.8 in our analysis, which were not considered by H21. However, the galaxy bias differs significantly from H21, primarily due to the different models of galaxy bias.

As shown by \cite{Saraf2022}, different CMB lensing maps produce significantly different cross-correlation amplitudes. To test this dependence, we estimated galaxy linear bias and cross-correlation amplitude using Sunyaev-Zeldovich deprojected (SZ-deproj) and temperature-only (TT) reconstructions of the CMB lensing convergence maps. We found the results from SZ-deproj and TT maps to be consistent with the MV map overall. The SZ-deproj map yields an $\sim 1\,\sigma$ higher value for cross-correlation amplitude in the second tomographic bin and {the} TT map completely resolves the deviation on the amplitude with respect to the $\Lambda$CDM model for {the} last two tomographic bins, but with poorer $\chi^{2}$ values. Finally, we estimated the impact of the redshift mismatch correction on the $\sigma_{8}$ parameter. {We} find that the corrected power spectra yields estimates that are consistent with the standard cosmological model for the fourth bin, smaller by around $\sim 1\,\sigma$ for the first and third bins, and smaller by $\sim 3\,\sigma$ for the second bin.

In this study, we demonstrate the importance of precise modelling of the photometric redshift error distributions when estimating parameters such as $\sigma_{8}$ from cross-correlation measurements and present a working example of scattering matrix formalism to correct for the redshift bin mismatch of objects in tomographic cross-correlation analysis. The $\sigma_{8}-\Omega_{m}$ tension is one of the pressing challenges of modern cosmology. {Upcoming large-scale structure surveys such as \textit{Vera C. Rubin} Observatory Legacy Survey of Space and Time (\citealt{LSST2019}; \citealt{LSST2009}), are expected to have smaller redshift errors. This will enable us to put tighter constraints on the $\sigma_{8}-\Omega_{m}$ tension with future tomographic cross-correlation measurements. However, the statistical errors of these surveys will also be small and the net outcome of the bin mismatch systematic error will depend on the interplay between redshift errors and survey statistical errors. From the results for simulations of {the} LSST survey presented in our previous publication \cite{Saraf2024}, we can expect the redshift bin mismatch to remain a significant systematic error even for future photometric datasets. Here we show that the method developed in \cite{Saraf2024} based on the use of the scattering matrix offers a robust way to mitigate bin mismatch, {which} should enable unbiased inference of cosmological parameters from future tomographic measurements.}

%-----------------------------------------------------------------

\begin{acknowledgements}
      {We thank the anonymous referee whose comments helped to increase the quality of the manuscript.} We thank Qianjun Hang for providing DESI-Legacy Imaging Survey datasets. CSS thanks Maciej Bilicki for providing feedback on the data analysis in the paper. The authors thank Agnieszka Pollo for various discussion on the content of this paper. The authors acknowledge the use of \texttt{CAMB}, \texttt{HEALPix}, \texttt{EMCEE} and \texttt{FLASK} software packages. The work has been supported by the Polish Ministry of Science and Higher Education grant DIR/WK/2018/12 and is based on observations obtained with Planck (http://www.esa.int/Planck), an ESA science mission with instruments and contributions directly funded by ESA Member States, NASA, and Canada.
\end{acknowledgements}

   \bibliographystyle{aa} % style aa.bst
   \bibliography{references} % your references Yourfile.bib

\begin{thebibliography}{85}
\expandafter\ifx\csname natexlab\endcsname\relax\def\natexlab#1{#1}\fi

\bibitem[{{Abazajian} {et~al.}(2009){Abazajian}, {Adelman-McCarthy}, {Ag{\"u}eros}, {Allam}, {Allende Prieto}, {An}, {Anderson}, {Anderson}, {Annis}, {Bahcall}, {Bailer-Jones}, {Barentine}, {Bassett}, {Becker}, {Beers}, {Bell}, {Belokurov}, {Berlind}, {Berman}, {Bernardi}, {Bickerton}, {Bizyaev}, {Blakeslee}, {Blanton}, {Bochanski}, {Boroski}, {Brewington}, {Brinchmann}, {Brinkmann}, {Brunner}, {Budav{\'a}ri}, {Carey}, {Carliles}, {Carr}, {Castander}, {Cinabro}, {Connolly}, {Csabai}, {Cunha}, {Czarapata}, {Davenport}, {de Haas}, {Dilday}, {Doi}, {Eisenstein}, {Evans}, {Evans}, {Fan}, {Friedman}, {Frieman}, {Fukugita}, {G{\"a}nsicke}, {Gates}, {Gillespie}, {Gilmore}, {Gonzalez}, {Gonzalez}, {Grebel}, {Gunn}, {Gy{\"o}ry}, {Hall}, {Harding}, {Harris}, {Harvanek}, {Hawley}, {Hayes}, {Heckman}, {Hendry}, {Hennessy}, {Hindsley}, {Hoblitt}, {Hogan}, {Hogg}, {Holtzman}, {Hyde}, {Ichikawa}, {Ichikawa}, {Im}, {Ivezi{\'c}}, {Jester}, {Jiang}, {Johnson}, {Jorgensen}, {Juri{\'c}}, {Kent}, {Kessler}, {Kleinman}, {Knapp},
  {Konishi}, {Kron}, {Krzesinski}, {Kuropatkin}, {Lampeitl}, {Lebedeva}, {Lee}, {Lee}, {French Leger}, {L{\'e}pine}, {Li}, {Lima}, {Lin}, {Long}, {Loomis}, {Loveday}, {Lupton}, {Magnier}, {Malanushenko}, {Malanushenko}, {Mandelbaum}, {Margon}, {Marriner}, {Mart{\'\i}nez-Delgado}, {Matsubara}, {McGehee}, {McKay}, {Meiksin}, {Morrison}, {Mullally}, {Munn}, {Murphy}, {Nash}, {Nebot}, {Neilsen}, {Newberg}, {Newman}, {Nichol}, {Nicinski}, {Nieto-Santisteban}, {Nitta}, {Okamura}, {Oravetz}, {Ostriker}, {Owen}, {Padmanabhan}, {Pan}, {Park}, {Pauls}, {Peoples}, {Percival}, {Pier}, {Pope}, {Pourbaix}, {Price}, {Purger}, {Quinn}, {Raddick}, {Re Fiorentin}, {Richards}, {Richmond}, {Riess}, {Rix}, {Rockosi}, {Sako}, {Schlegel}, {Schneider}, {Scholz}, {Schreiber}, {Schwope}, {Seljak}, {Sesar}, {Sheldon}, {Shimasaku}, {Sibley}, {Simmons}, {Sivarani}, {Allyn Smith}, {Smith}, {Smol{\v{c}}i{\'c}}, {Snedden}, {Stebbins}, {Steinmetz}, {Stoughton}, {Strauss}, {SubbaRao}, {Suto}, {Szalay}, {Szapudi}, {Szkody}, {Tanaka},
  {Tegmark}, {Teodoro}, {Thakar}, {Tremonti}, {Tucker}, {Uomoto}, {Vanden Berk}, {Vandenberg}, {Vidrih}, {Vogeley}, {Voges}, {Vogt}, {Wadadekar}, {Watters}, {Weinberg}, {West}, {White}, {Wilhite}, {Wonders}, {Yanny}, {Yocum}, {York}, {Zehavi}, {Zibetti}, \& {Zucker}}]{Abazajian2009}
{Abazajian}, K.~N., {Adelman-McCarthy}, J.~K., {Ag{\"u}eros}, M.~A., {et~al.} 2009, \apjs, 182, 543

\bibitem[{{Abbott} {et~al.}(2019){Abbott}, {Abdalla}, {Alarcon}, {Allam}, {Annis}, {Avila}, {Aylor}, {Banerji}, {Banik}, {Baxter}, {Bechtol}, {Becker}, {Benson}, {Bernstein}, {Bertin}, {Bianchini}, {Blazek}, {Bleem}, {Bridle}, {Brooks}, {Buckley-Geer}, {Burke}, {Carlstrom}, {Carnero Rosell}, {Carrasco Kind}, {Carretero}, {Castander}, {Cawthon}, {Chang}, {Chang}, {Cho}, {Choi}, {Chown}, {Crawford}, {Crites}, {Crocce}, {Cunha}, {D'Andrea}, {da Costa}, {Davis}, {de Haan}, {DeRose}, {Desai}, {De Vicente}, {Diehl}, {Dietrich}, {Dobbs}, {Dodelson}, {Doel}, {Drlica-Wagner}, {Eifler}, {Elvin-Poole}, {Everett}, {Flaugher}, {Fosalba}, {Friedrich}, {Frieman}, {Garc{\'\i}a-Bellido}, {Gatti}, {Gaztanaga}, {George}, {Gerdes}, {Giannantonio}, {Gruen}, {Gruendl}, {Gschwend}, {Gutierrez}, {Halverson}, {Harrington}, {Hartley}, {Holder}, {Hollowood}, {Holzapfel}, {Honscheid}, {Hou}, {Hoyle}, {Hrubes}, {Huterer}, {Jain}, {James}, {Jarvis}, {Jeltema}, {Johnson}, {Johnson}, {Kent}, {Kirk}, {Knox}, {Kokron}, {Krause}, {Kuehn},
  {Lahav}, {Lee}, {Leitch}, {Li}, {Lima}, {Lin}, {Luong-Van}, {MacCrann}, {Maia}, {Manzotti}, {Marrone}, {Marshall}, {Martini}, {McMahon}, {Menanteau}, {Meyer}, {Miquel}, {Mocanu}, {Mohr}, {Muir}, {Natoli}, {Nicola}, {Nord}, {Omori}, {Padin}, {Pandey}, {Plazas}, {Porredon}, {Prat}, {Pryke}, {Rau}, {Reichardt}, {Rollins}, {Romer}, {Roodman}, {Ross}, {Rozo}, {Ruhl}, {Rykoff}, {Samuroff}, {S{\'a}nchez}, {Sanchez}, {Sayre}, {Scarpine}, {Schaffer}, {Secco}, {Serrano}, {Sevilla-Noarbe}, {Sheldon}, {Shirokoff}, {Simard}, {Smith}, {Soares-Santos}, {Sobreira}, {Staniszewski}, {Stark}, {Story}, {Suchyta}, {Swanson}, {Tarle}, {Thomas}, {Troxel}, {Tucker}, {Vanderlinde}, {Vieira}, {Vielzeuf}, {Vikram}, {Walker}, {Wechsler}, {Weller}, {Williamson}, {Wu}, {Yanny}, {Zahn}, {Zhang}, {Zuntz}, {DES}, \& {SPT Collaborations}}]{Abbott2019}
{Abbott}, T.~M.~C., {Abdalla}, F.~B., {Alarcon}, A., {et~al.} 2019, \prd, 100, 023541

\bibitem[{{Abbott} {et~al.}(2022){Abbott}, {Aguena}, {Alarcon}, {Allam}, {Alves}, {Amon}, {Andrade-Oliveira}, {Annis}, {Avila}, {Bacon}, {Baxter}, {Bechtol}, {Becker}, {Bernstein}, {Bhargava}, {Birrer}, {Blazek}, {Brandao-Souza}, {Bridle}, {Brooks}, {Buckley-Geer}, {Burke}, {Camacho}, {Campos}, {Carnero Rosell}, {Carrasco Kind}, {Carretero}, {Castander}, {Cawthon}, {Chang}, {Chen}, {Chen}, {Choi}, {Conselice}, {Cordero}, {Costanzi}, {Crocce}, {da Costa}, {da Silva Pereira}, {Davis}, {Davis}, {De Vicente}, {DeRose}, {Desai}, {Di Valentino}, {Diehl}, {Dietrich}, {Dodelson}, {Doel}, {Doux}, {Drlica-Wagner}, {Eckert}, {Eifler}, {Elsner}, {Elvin-Poole}, {Everett}, {Evrard}, {Fang}, {Farahi}, {Fernandez}, {Ferrero}, {Fert{\'e}}, {Fosalba}, {Friedrich}, {Frieman}, {Garc{\'\i}a-Bellido}, {Gatti}, {Gaztanaga}, {Gerdes}, {Giannantonio}, {Giannini}, {Gruen}, {Gruendl}, {Gschwend}, {Gutierrez}, {Harrison}, {Hartley}, {Herner}, {Hinton}, {Hollowood}, {Honscheid}, {Hoyle}, {Huff}, {Huterer}, {Jain}, {James}, {Jarvis},
  {Jeffrey}, {Jeltema}, {Kovacs}, {Krause}, {Kron}, {Kuehn}, {Kuropatkin}, {Lahav}, {Leget}, {Lemos}, {Liddle}, {Lidman}, {Lima}, {Lin}, {MacCrann}, {Maia}, {Marshall}, {Martini}, {McCullough}, {Melchior}, {Mena-Fern{\'a}ndez}, {Menanteau}, {Miquel}, {Mohr}, {Morgan}, {Muir}, {Myles}, {Nadathur}, {Navarro-Alsina}, {Nichol}, {Ogando}, {Omori}, {Palmese}, {Pandey}, {Park}, {Paz-Chinch{\'o}n}, {Petravick}, {Pieres}, {Plazas Malag{\'o}n}, {Porredon}, {Prat}, {Raveri}, {Rodriguez-Monroy}, {Rollins}, {Romer}, {Roodman}, {Rosenfeld}, {Ross}, {Rykoff}, {Samuroff}, {S{\'a}nchez}, {Sanchez}, {Sanchez}, {Sanchez Cid}, {Scarpine}, {Schubnell}, {Scolnic}, {Secco}, {Serrano}, {Sevilla-Noarbe}, {Sheldon}, {Shin}, {Smith}, {Soares-Santos}, {Suchyta}, {Swanson}, {Tabbutt}, {Tarle}, {Thomas}, {To}, {Troja}, {Troxel}, {Tucker}, {Tutusaus}, {Varga}, {Walker}, {Weaverdyck}, {Wechsler}, {Weller}, {Yanny}, {Yin}, {Zhang}, {Zuntz}, \& {DES Collaboration}}]{DESY32022}
{Abbott}, T.~M.~C., {Aguena}, M., {Alarcon}, A., {et~al.} 2022, \prd, 105, 023520

\bibitem[{{Abdalla} {et~al.}(2022){Abdalla}, {Abell{\'a}n}, {Aboubrahim}, {Agnello}, {Akarsu}, {Akrami}, {Alestas}, {Aloni}, {Amendola}, {Anchordoqui}, {Anderson}, {Arendse}, {Asgari}, {Ballardini}, {Barger}, {Basilakos}, {Batista}, {Battistelli}, {Battye}, {Benetti}, {Benisty}, {Berlin}, {de Bernardis}, {Berti}, {Bidenko}, {Birrer}, {Blakeslee}, {Boddy}, {Bom}, {Bonilla}, {Borghi}, {Bouchet}, {Braglia}, {Buchert}, {Buckley-Geer}, {Calabrese}, {Caldwell}, {Camarena}, {Capozziello}, {Casertano}, {Chen}, {Chluba}, {Chen}, {Chen}, {Chudaykin}, {Cicoli}, {Copi}, {Courbin}, {Cyr-Racine}, {Czerny}, {Dainotti}, {D'Amico}, {Davis}, {de Cruz P{\'e}rez}, {de Haro}, {Delabrouille}, {Denton}, {Dhawan}, {Dienes}, {Di Valentino}, {Du}, {Eckert}, {Escamilla-Rivera}, {Fert{\'e}}, {Finelli}, {Fosalba}, {Freedman}, {Frusciante}, {Gazta{\~n}aga}, {Giar{\`e}}, {Giusarma}, {G{\'o}mez-Valent}, {Handley}, {Harrison}, {Hart}, {Hazra}, {Heavens}, {Heinesen}, {Hildebrandt}, {Hill}, {Hogg}, {Holz}, {Hooper}, {Hosseininejad}, {Huterer},
  {Ishak}, {Ivanov}, {Jaffe}, {Jang}, {Jedamzik}, {Jimenez}, {Joseph}, {Joudaki}, {Kamionkowski}, {Karwal}, {Kazantzidis}, {Keeley}, {Klasen}, {Komatsu}, {Koopmans}, {Kumar}, {Lamagna}, {Lazkoz}, {Lee}, {Lesgourgues}, {Levi Said}, {Lewis}, {L'Huillier}, {Lucca}, {Maartens}, {Macri}, {Marfatia}, {Marra}, {Martins}, {Masi}, {Matarrese}, {Mazumdar}, {Melchiorri}, {Mena}, {Mersini-Houghton}, {Mertens}, {Milakovi{\'c}}, {Minami}, {Miranda}, {Moreno-Pulido}, {Moresco}, {Mota}, {Mottola}, {Mozzon}, {Muir}, {Mukherjee}, {Mukherjee}, {Naselsky}, {Nath}, {Nesseris}, {Niedermann}, {Notari}, {Nunes}, {{\'O} Colg{\'a}in}, {Owens}, {{\"O}z{\"u}lker}, {Pace}, {Paliathanasis}, {Palmese}, {Pan}, {Paoletti}, {Perez Bergliaffa}, {Perivolaropoulos}, {Pesce}, {Pettorino}, {Philcox}, {Pogosian}, {Poulin}, {Poulot}, {Raveri}, {Reid}, {Renzi}, {Riess}, {Sabla}, {Salucci}, {Salzano}, {Saridakis}, {Sathyaprakash}, {Schmaltz}, {Sch{\"o}neberg}, {Scolnic}, {Sen}, {Sehgal}, {Shafieloo}, {Sheikh-Jabbari}, {Silk}, {Silvestri}, {Skara},
  {Sloth}, {Soares-Santos}, {Sol{\`a} Peracaula}, {Songsheng}, {Soriano}, {Staicova}, {Starkman}, {Szapudi}, {Teixeira}, {Thomas}, {Treu}, {Trott}, {van de Bruck}, {Vazquez}, {Verde}, {Visinelli}, {Wang}, {Wang}, {Wang}, {Watkins}, {Watson}, {Webb}, {Weiner}, {Weltman}, {Witte}, {Wojtak}, {Yadav}, {Yang}, {Zhao}, \& {Zumalac{\'a}rregui}}]{Abdalla2022}
{Abdalla}, E., {Abell{\'a}n}, G.~F., {Aboubrahim}, A., {et~al.} 2022, Journal of High Energy Astrophysics, 34, 49

\bibitem[{{Abolfathi} {et~al.}(2018){Abolfathi}, {Aguado}, {Aguilar}, {Allende Prieto}, {Almeida}, {Ananna}, {Anders}, {Anderson}, {Andrews}, {Anguiano}, {Arag{\'o}n-Salamanca}, {Argudo-Fern{\'a}ndez}, {Armengaud}, {Ata}, {Aubourg}, {Avila-Reese}, {Badenes}, {Bailey}, {Balland}, {Barger}, {Barrera-Ballesteros}, {Bartosz}, {Bastien}, {Bates}, {Baumgarten}, {Bautista}, {Beaton}, {Beers}, {Belfiore}, {Bender}, {Bernardi}, {Bershady}, {Beutler}, {Bird}, {Bizyaev}, {Blanc}, {Blanton}, {Blomqvist}, {Bolton}, {Boquien}, {Borissova}, {Bovy}, {Bradna Diaz}, {Brandt}, {Brinkmann}, {Brownstein}, {Bundy}, {Burgasser}, {Burtin}, {Busca}, {Ca{\~n}as}, {Cano-D{\'\i}az}, {Cappellari}, {Carrera}, {Casey}, {Cervantes Sodi}, {Chen}, {Cherinka}, {Chiappini}, {Choi}, {Chojnowski}, {Chuang}, {Chung}, {Clerc}, {Cohen}, {Comerford}, {Comparat}, {Correa do Nascimento}, {da Costa}, {Cousinou}, {Covey}, {Crane}, {Cruz-Gonzalez}, {Cunha}, {da Silva Ilha}, {Damke}, {Darling}, {Davidson}, {Dawson}, {de Icaza Lizaola}, {de la Macorra}, {de
  la Torre}, {De Lee}, {de Sainte Agathe}, {Deconto Machado}, {Dell'Agli}, {Delubac}, {Diamond-Stanic}, {Donor}, {Downes}, {Drory}, {du Mas des Bourboux}, {Duckworth}, {Dwelly}, {Dyer}, {Ebelke}, {Davis Eigenbrot}, {Eisenstein}, {Elsworth}, {Emsellem}, {Eracleous}, {Erfanianfar}, {Escoffier}, {Fan}, {Fern{\'a}ndez Alvar}, {Fernandez-Trincado}, {Fernando Cirolini}, {Feuillet}, {Finoguenov}, {Fleming}, {Font-Ribera}, {Freischlad}, {Frinchaboy}, {Fu}, {G{\'o}mez Maqueo Chew}, {Galbany}, {Garc{\'\i}a P{\'e}rez}, {Garcia-Dias}, {Garc{\'\i}a-Hern{\'a}ndez}, {Garma Oehmichen}, {Gaulme}, {Gelfand}, {Gil-Mar{\'\i}n}, {Gillespie}, {Goddard}, {Gonz{\'a}lez Hern{\'a}ndez}, {Gonzalez-Perez}, {Grabowski}, {Green}, {Grier}, {Gueguen}, {Guo}, {Guy}, {Hagen}, {Hall}, {Harding}, {Hasselquist}, {Hawley}, {Hayes}, {Hearty}, {Hekker}, {Hernandez}, {Hernandez Toledo}, {Hogg}, {Holley-Bockelmann}, {Holtzman}, {Hou}, {Hsieh}, {Hunt}, {Hutchinson}, {Hwang}, {Jimenez Angel}, {Johnson}, {Jones}, {J{\"o}nsson}, {Jullo}, {Khan},
  {Kinemuchi}, {Kirkby}, {Kirkpatrick}, {Kitaura}, {Knapp}, {Kneib}, {Kollmeier}, {Lacerna}, {Lane}, {Lang}, {Law}, {Le Goff}, {Lee}, {Li}, {Li}, {Lian}, {Liang}, {Lima}, {Lin}, {Long}, {Lucatello}, {Lundgren}, {Mackereth}, {MacLeod}, {Mahadevan}, {Maia}, {Majewski}, {Manchado}, {Maraston}, {Mariappan}, {Marques-Chaves}, {Masseron}, {Masters}, {McDermid}, {McGreer}, {Melendez}, {Meneses-Goytia}, {Merloni}, {Merrifield}, {Meszaros}, {Meza}, {Minchev}, {Minniti}, {Mueller}, {Muller-Sanchez}, {Muna}, {Mu{\~n}oz}, {Myers}, {Nair}, {Nandra}, {Ness}, {Newman}, {Nichol}, {Nidever}, {Nitschelm}, {Noterdaeme}, {O'Connell}, {Oelkers}, {Oravetz}, {Oravetz}, {Ort{\'\i}z}, {Osorio}, {Pace}, {Padilla}, {Palanque-Delabrouille}, {Palicio}, {Pan}, {Pan}, {Parikh}, {P{\^a}ris}, {Park}, {Peirani}, {Pellejero-Ibanez}, {Penny}, {Percival}, {Perez-Fournon}, {Petitjean}, {Pieri}, {Pinsonneault}, {Pisani}, {Prada}, {Prakash}, {Queiroz}, {Raddick}, {Raichoor}, {Barboza Rembold}, {Richstein}, {Riffel}, {Riffel}, {Rix}, {Robin},
  {Rodr{\'\i}guez Torres}, {Rom{\'a}n-Z{\'u}{\~n}iga}, {Ross}, {Rossi}, {Ruan}, {Ruggeri}, {Ruiz}, {Salvato}, {S{\'a}nchez}, {S{\'a}nchez}, {Sanchez Almeida}, {S{\'a}nchez-Gallego}, {Santana Rojas}, {Santiago}, {Schiavon}, {Schimoia}, {Schlafly}, {Schlegel}, {Schneider}, {Schuster}, {Schwope}, {Seo}, {Serenelli}, {Shen}, {Shen}, {Shetrone}, {Shull}, {Silva Aguirre}, {Simon}, {Skrutskie}, {Slosar}, {Smethurst}, {Smith}, {Sobeck}, {Somers}, {Souter}, {Souto}, {Spindler}, {Stark}, {Stassun}, {Steinmetz}, {Stello}, {Storchi-Bergmann}, {Streblyanska}, {Stringfellow}, {Su{\'a}rez}, {Sun}, {Szigeti}, {Taghizadeh-Popp}, {Talbot}, {Tang}, {Tao}, {Tayar}, {Tembe}, {Teske}, {Thakar}, {Thomas}, {Tissera}, {Tojeiro}, {Tremonti}, {Troup}, {Urry}, {Valenzuela}, {van den Bosch}, {Vargas-Gonz{\'a}lez}, {Vargas-Maga{\~n}a}, {Vazquez}, {Villanova}, {Vogt}, {Wake}, {Wang}, {Weaver}, {Weijmans}, {Weinberg}, {Westfall}, {Whelan}, {Wilcots}, {Wild}, {Williams}, {Wilson}, {Wood-Vasey}, {Wylezalek}, {Xiao}, {Yan}, {Yang}, {Ybarra},
  {Y{\`e}che}, {Zakamska}, {Zamora}, {Zarrouk}, {Zasowski}, {Zhang}, {Zhao}, {Zhao}, {Zheng}, {Zheng}, {Zhou}, {Zhu}, {Zinn}, \& {Zou}}]{Abolfathi2018}
{Abolfathi}, B., {Aguado}, D.~S., {Aguilar}, G., {et~al.} 2018, \apjs, 235, 42

\bibitem[{{Adachi} {et~al.}(2020){Adachi}, {Aguilar Fa{\'u}ndez}, {Arnold}, {Baccigalupi}, {Barron}, {Beck}, {Bianchini}, {Chapman}, {Cheung}, {Chinone}, {Crowley}, {Dobbs}, {El Bouhargani}, {Elleflot}, {Errard}, {Fabbian}, {Feng}, {Fujino}, {Galitzki}, {Goeckner-Wald}, {Groh}, {Hall}, {Hasegawa}, {Hazumi}, {Hirose}, {Jaffe}, {Jeong}, {Kaneko}, {Katayama}, {Keating}, {Kikuchi}, {Kisner}, {Kusaka}, {Lee}, {Leon}, {Linder}, {Lowry}, {Matsuda}, {Matsumura}, {Minami}, {Navaroli}, {Nishino}, {Pham}, {Poletti}, {Reichardt}, {Segawa}, {Siritanasak}, {Tajima}, {Takakura}, {Takatori}, {Tanabe}, {Teply}, {Tsai}, {Verg{\`e}s}, {Westbrook}, {Zhou}, \& {Polarbear Collaboration}}]{Polarbear2020}
{Adachi}, S., {Aguilar Fa{\'u}ndez}, M.~A.~O., {Arnold}, K., {et~al.} 2020, \apj, 904, 65

\bibitem[{{Ade} {et~al.}(2021){Ade}, {Ahmed}, {Amiri}, {Barkats}, {Thakur}, {Bischoff}, {Beck}, {Bock}, {Boenish}, {Bullock}, {Buza}, {Cheshire}, {Connors}, {Cornelison}, {Crumrine}, {Cukierman}, {Denison}, {Dierickx}, {Duband}, {Eiben}, {Fatigoni}, {Filippini}, {Fliescher}, {Goeckner-Wald}, {Goldfinger}, {Grayson}, {Grimes}, {Hall}, {Halal}, {Halpern}, {Hand}, {Harrison}, {Henderson}, {Hildebrandt}, {Hilton}, {Hubmayr}, {Hui}, {Irwin}, {Kang}, {Karkare}, {Karpel}, {Kefeli}, {Kernasovskiy}, {Kovac}, {Kuo}, {Lau}, {Leitch}, {Lennox}, {Megerian}, {Minutolo}, {Moncelsi}, {Nakato}, {Namikawa}, {Nguyen}, {O'Brient}, {Ogburn}, {Palladino}, {Prouve}, {Pryke}, {Racine}, {Reintsema}, {Richter}, {Schillaci}, {Schwarz}, {Schmitt}, {Sheehy}, {Soliman}, {Germaine}, {Steinbach}, {Sudiwala}, {Teply}, {Thompson}, {Tolan}, {Tucker}, {Turner}, {Umilt{\`a}}, {Verg{\`e}s}, {Vieregg}, {Wandui}, {Weber}, {Wiebe}, {Willmert}, {Wong}, {Wu}, {Yang}, {Yoon}, {Young}, {Yu}, {Zeng}, {Zhang}, {Zhang}, \& {Bicep/Keck
  Collaboration}}]{BICEP2021}
{Ade}, P.~A.~R., {Ahmed}, Z., {Amiri}, M., {et~al.} 2021, \prl, 127, 151301

\bibitem[{{Adil} {et~al.}(2024){Adil}, {Akarsu}, {Malekjani}, {{\'O} Colg{\'a}in}, {Pourojaghi}, {Sen}, \& {Sheikh-Jabbari}}]{Adil2024}
{Adil}, S.~A., {Akarsu}, {\"O}., {Malekjani}, M., {et~al.} 2024, \mnras, 528, L20

\bibitem[{{Ahumada} {et~al.}(2020){Ahumada}, {Allende Prieto}, {Almeida}, {Anders}, {Anderson}, {Andrews}, {Anguiano}, {Arcodia}, {Armengaud}, {Aubert}, {Avila}, {Avila-Reese}, {Badenes}, {Balland}, {Barger}, {Barrera-Ballesteros}, {Basu}, {Bautista}, {Beaton}, {Beers}, {Benavides}, {Bender}, {Bernardi}, {Bershady}, {Beutler}, {Bidin}, {Bird}, {Bizyaev}, {Blanc}, {Blanton}, {Boquien}, {Borissova}, {Bovy}, {Brandt}, {Brinkmann}, {Brownstein}, {Bundy}, {Bureau}, {Burgasser}, {Burtin}, {Cano-D{\'\i}az}, {Capasso}, {Cappellari}, {Carrera}, {Chabanier}, {Chaplin}, {Chapman}, {Cherinka}, {Chiappini}, {Doohyun Choi}, {Chojnowski}, {Chung}, {Clerc}, {Coffey}, {Comerford}, {Comparat}, {da Costa}, {Cousinou}, {Covey}, {Crane}, {Cunha}, {Ilha}, {Dai}, {Damsted}, {Darling}, {Davidson}, {Davies}, {Dawson}, {De}, {de la Macorra}, {De Lee}, {Queiroz}, {Deconto Machado}, {de la Torre}, {Dell'Agli}, {du Mas des Bourboux}, {Diamond-Stanic}, {Dillon}, {Donor}, {Drory}, {Duckworth}, {Dwelly}, {Ebelke}, {Eftekharzadeh}, {Davis
  Eigenbrot}, {Elsworth}, {Eracleous}, {Erfanianfar}, {Escoffier}, {Fan}, {Farr}, {Fern{\'a}ndez-Trincado}, {Feuillet}, {Finoguenov}, {Fofie}, {Fraser-McKelvie}, {Frinchaboy}, {Fromenteau}, {Fu}, {Galbany}, {Garcia}, {Garc{\'\i}a-Hern{\'a}ndez}, {Garma Oehmichen}, {Ge}, {Geimba Maia}, {Geisler}, {Gelfand}, {Goddy}, {Gonzalez-Perez}, {Grabowski}, {Green}, {Grier}, {Guo}, {Guy}, {Harding}, {Hasselquist}, {Hawken}, {Hayes}, {Hearty}, {Hekker}, {Hogg}, {Holtzman}, {Horta}, {Hou}, {Hsieh}, {Huber}, {Hunt}, {Ider Chitham}, {Imig}, {Jaber}, {Jimenez Angel}, {Johnson}, {Jones}, {J{\"o}nsson}, {Jullo}, {Kim}, {Kinemuchi}, {Kirkpatrick}, {Kite}, {Klaene}, {Kneib}, {Kollmeier}, {Kong}, {Kounkel}, {Krishnarao}, {Lacerna}, {Lan}, {Lane}, {Law}, {Le Goff}, {Leung}, {Lewis}, {Li}, {Lian}, {Lin}, {Long}, {Longa-Pe{\~n}a}, {Lundgren}, {Lyke}, {Mackereth}, {MacLeod}, {Majewski}, {Manchado}, {Maraston}, {Martini}, {Masseron}, {Masters}, {Mathur}, {McDermid}, {Merloni}, {Merrifield}, {M{\'e}sz{\'a}ros}, {Miglio}, {Minniti},
  {Minsley}, {Miyaji}, {Mohammad}, {Mosser}, {Mueller}, {Muna}, {Mu{\~n}oz-Guti{\'e}rrez}, {Myers}, {Nadathur}, {Nair}, {Nandra}, {Correa do Nascimento}, {Nevin}, {Newman}, {Nidever}, {Nitschelm}, {Noterdaeme}, {O'Connell}, {Olmstead}, {Oravetz}, {Oravetz}, {Osorio}, {Pace}, {Padilla}, {Palanque-Delabrouille}, {Palicio}, {Pan}, {Pan}, {Parker}, {Paviot}, {Peirani}, {Ram{\'r}ez}, {Penny}, {Percival}, {Perez-Fournon}, {P{\'e}rez-R{\`a}fols}, {Petitjean}, {Pieri}, {Pinsonneault}, {Poovelil}, {Povick}, {Prakash}, {Price-Whelan}, {Raddick}, {Raichoor}, {Ray}, {Rembold}, {Rezaie}, {Riffel}, {Riffel}, {Rix}, {Robin}, {Roman-Lopes}, {Rom{\'a}n-Z{\'u}{\~n}iga}, {Rose}, {Ross}, {Rossi}, {Rowlands}, {Rubin}, {Salvato}, {S{\'a}nchez}, {S{\'a}nchez-Menguiano}, {S{\'a}nchez-Gallego}, {Sayres}, {Schaefer}, {Schiavon}, {Schimoia}, {Schlafly}, {Schlegel}, {Schneider}, {Schultheis}, {Schwope}, {Seo}, {Serenelli}, {Shafieloo}, {Shamsi}, {Shao}, {Shen}, {Shetrone}, {Shirley}, {Silva Aguirre}, {Simon}, {Skrutskie}, {Slosar},
  {Smethurst}, {Sobeck}, {Sodi}, {Souto}, {Stark}, {Stassun}, {Steinmetz}, {Stello}, {Stermer}, {Storchi-Bergmann}, {Streblyanska}, {Stringfellow}, {Stutz}, {Su{\'a}rez}, {Sun}, {Taghizadeh-Popp}, {Talbot}, {Tayar}, {Thakar}, {Theriault}, {Thomas}, {Thomas}, {Tinker}, {Tojeiro}, {Toledo}, {Tremonti}, {Troup}, {Tuttle}, {Unda-Sanzana}, {Valentini}, {Vargas-Gonz{\'a}lez}, {Vargas-Maga{\~n}a}, {V{\'a}zquez-Mata}, {Vivek}, {Wake}, {Wang}, {Weaver}, {Weijmans}, {Wild}, {Wilson}, {Wilson}, {Wolthuis}, {Wood-Vasey}, {Yan}, {Yang}, {Y{\`e}che}, {Zamora}, {Zarrouk}, {Zasowski}, {Zhang}, {Zhao}, {Zhao}, {Zheng}, {Zheng}, {Zhu}, \& {Zou}}]{eBOSSDR16_2020}
{Ahumada}, R., {Allende Prieto}, C., {Almeida}, A., {et~al.} 2020, \apjs, 249, 3

\bibitem[{{Aiola} {et~al.}(2020){Aiola}, {Calabrese}, {Maurin}, {Naess}, {Schmitt}, {Abitbol}, {Addison}, {Ade}, {Alonso}, {Amiri}, {Amodeo}, {Angile}, {Austermann}, {Baildon}, {Battaglia}, {Beall}, {Bean}, {Becker}, {Bond}, {Bruno}, {Calafut}, {Campusano}, {Carrero}, {Chesmore}, {Cho}, {Choi}, {Clark}, {Cothard}, {Crichton}, {Crowley}, {Darwish}, {Datta}, {Denison}, {Devlin}, {Duell}, {Duff}, {Duivenvoorden}, {Dunkley}, {D{\"u}nner}, {Essinger-Hileman}, {Fankhanel}, {Ferraro}, {Fox}, {Fuzia}, {Gallardo}, {Gluscevic}, {Golec}, {Grace}, {Gralla}, {Guan}, {Hall}, {Halpern}, {Han}, {Hargrave}, {Hasselfield}, {Helton}, {Henderson}, {Hensley}, {Hill}, {Hilton}, {Hilton}, {Hincks}, {Hlo{\v{z}}ek}, {Ho}, {Hubmayr}, {Huffenberger}, {Hughes}, {Infante}, {Irwin}, {Jackson}, {Klein}, {Knowles}, {Koopman}, {Kosowsky}, {Lakey}, {Li}, {Li}, {Li}, {Lokken}, {Louis}, {Lungu}, {MacInnis}, {Madhavacheril}, {Maldonado}, {Mallaby-Kay}, {Marsden}, {McMahon}, {Menanteau}, {Moodley}, {Morton}, {Namikawa}, {Nati}, {Newburgh},
  {Nibarger}, {Nicola}, {Niemack}, {Nolta}, {Orlowski-Sherer}, {Page}, {Pappas}, {Partridge}, {Phakathi}, {Pisano}, {Prince}, {Puddu}, {Qu}, {Rivera}, {Robertson}, {Rojas}, {Salatino}, {Schaan}, {Schillaci}, {Sehgal}, {Sherwin}, {Sierra}, {Sievers}, {Sifon}, {Sikhosana}, {Simon}, {Spergel}, {Staggs}, {Stevens}, {Storer}, {Sunder}, {Switzer}, {Thorne}, {Thornton}, {Trac}, {Treu}, {Tucker}, {Vale}, {Van Engelen}, {Van Lanen}, {Vavagiakis}, {Wagoner}, {Wang}, {Ward}, {Wollack}, {Xu}, {Zago}, \& {Zhu}}]{ACTDR42020}
{Aiola}, S., {Calabrese}, E., {Maurin}, L., {et~al.} 2020, \jcap, 2020, 047

\bibitem[{{Alam} {et~al.}(2015){Alam}, {Albareti}, {Allende Prieto}, {Anders}, {Anderson}, {Anderton}, {Andrews}, {Armengaud}, {Aubourg}, {Bailey}, {Basu}, {Bautista}, {Beaton}, {Beers}, {Bender}, {Berlind}, {Beutler}, {Bhardwaj}, {Bird}, {Bizyaev}, {Blake}, {Blanton}, {Blomqvist}, {Bochanski}, {Bolton}, {Bovy}, {Shelden Bradley}, {Brandt}, {Brauer}, {Brinkmann}, {Brown}, {Brownstein}, {Burden}, {Burtin}, {Busca}, {Cai}, {Capozzi}, {Carnero Rosell}, {Carr}, {Carrera}, {Chambers}, {Chaplin}, {Chen}, {Chiappini}, {Chojnowski}, {Chuang}, {Clerc}, {Comparat}, {Covey}, {Croft}, {Cuesta}, {Cunha}, {da Costa}, {Da Rio}, {Davenport}, {Dawson}, {De Lee}, {Delubac}, {Deshpande}, {Dhital}, {Dutra-Ferreira}, {Dwelly}, {Ealet}, {Ebelke}, {Edmondson}, {Eisenstein}, {Ellsworth}, {Elsworth}, {Epstein}, {Eracleous}, {Escoffier}, {Esposito}, {Evans}, {Fan}, {Fern{\'a}ndez-Alvar}, {Feuillet}, {Filiz Ak}, {Finley}, {Finoguenov}, {Flaherty}, {Fleming}, {Font-Ribera}, {Foster}, {Frinchaboy}, {Galbraith-Frew}, {Garc{\'\i}a},
  {Garc{\'\i}a-Hern{\'a}ndez}, {Garc{\'\i}a P{\'e}rez}, {Gaulme}, {Ge}, {G{\'e}nova-Santos}, {Georgakakis}, {Ghezzi}, {Gillespie}, {Girardi}, {Goddard}, {Gontcho}, {Gonz{\'a}lez Hern{\'a}ndez}, {Grebel}, {Green}, {Grieb}, {Grieves}, {Gunn}, {Guo}, {Harding}, {Hasselquist}, {Hawley}, {Hayden}, {Hearty}, {Hekker}, {Ho}, {Hogg}, {Holley-Bockelmann}, {Holtzman}, {Honscheid}, {Huber}, {Huehnerhoff}, {Ivans}, {Jiang}, {Johnson}, {Kinemuchi}, {Kirkby}, {Kitaura}, {Klaene}, {Knapp}, {Kneib}, {Koenig}, {Lam}, {Lan}, {Lang}, {Laurent}, {Le Goff}, {Leauthaud}, {Lee}, {Lee}, {Licquia}, {Liu}, {Long}, {L{\'o}pez-Corredoira}, {Lorenzo-Oliveira}, {Lucatello}, {Lundgren}, {Lupton}, {Mack}, {Mahadevan}, {Maia}, {Majewski}, {Malanushenko}, {Malanushenko}, {Manchado}, {Manera}, {Mao}, {Maraston}, {Marchwinski}, {Margala}, {Martell}, {Martig}, {Masters}, {Mathur}, {McBride}, {McGehee}, {McGreer}, {McMahon}, {M{\'e}nard}, {Menzel}, {Merloni}, {M{\'e}sz{\'a}ros}, {Miller}, {Miralda-Escud{\'e}}, {Miyatake}, {Montero-Dorta}, {More},
  {Morganson}, {Morice-Atkinson}, {Morrison}, {Mosser}, {Muna}, {Myers}, {Nandra}, {Newman}, {Neyrinck}, {Nguyen}, {Nichol}, {Nidever}, {Noterdaeme}, {Nuza}, {O'Connell}, {O'Connell}, {O'Connell}, {Ogando}, {Olmstead}, {Oravetz}, {Oravetz}, {Osumi}, {Owen}, {Padgett}, {Padmanabhan}, {Paegert}, {Palanque-Delabrouille}, {Pan}, {Parejko}, {P{\^a}ris}, {Park}, {Pattarakijwanich}, {Pellejero-Ibanez}, {Pepper}, {Percival}, {P{\'e}rez-Fournon}, {P{\'e}rez-R{\`a}fols}, {Petitjean}, {Pieri}, {Pinsonneault}, {Porto de Mello}, {Prada}, {Prakash}, {Price-Whelan}, {Protopapas}, {Raddick}, {Rahman}, {Reid}, {Rich}, {Rix}, {Robin}, {Rockosi}, {Rodrigues}, {Rodr{\'\i}guez-Torres}, {Roe}, {Ross}, {Ross}, {Rossi}, {Ruan}, {Rubi{\~n}o-Mart{\'\i}n}, {Rykoff}, {Salazar-Albornoz}, {Salvato}, {Samushia}, {S{\'a}nchez}, {Santiago}, {Sayres}, {Schiavon}, {Schlegel}, {Schmidt}, {Schneider}, {Schultheis}, {Schwope}, {Sc{\'o}ccola}, {Scott}, {Sellgren}, {Seo}, {Serenelli}, {Shane}, {Shen}, {Shetrone}, {Shu}, {Silva Aguirre}, {Sivarani},
  {Skrutskie}, {Slosar}, {Smith}, {Sobreira}, {Souto}, {Stassun}, {Steinmetz}, {Stello}, {Strauss}, {Streblyanska}, {Suzuki}, {Swanson}, {Tan}, {Tayar}, {Terrien}, {Thakar}, {Thomas}, {Thomas}, {Thompson}, {Tinker}, {Tojeiro}, {Troup}, {Vargas-Maga{\~n}a}, {Vazquez}, {Verde}, {Viel}, {Vogt}, {Wake}, {Wang}, {Weaver}, {Weinberg}, {Weiner}, {White}, {Wilson}, {Wisniewski}, {Wood-Vasey}, {Ye`che}, {York}, {Zakamska}, {Zamora}, {Zasowski}, {Zehavi}, {Zhao}, {Zheng}, {Zhou}, {Zhou}, {Zou}, \& {Zhu}}]{BOSSDR12_2015}
{Alam}, S., {Albareti}, F.~D., {Allende Prieto}, C., {et~al.} 2015, \apjs, 219, 12

\bibitem[{{Alonso} {et~al.}(2023){Alonso}, {Fabbian}, {Storey-Fisher}, {Eilers}, {Garc{\'\i}a-Garc{\'\i}a}, {Hogg}, \& {Rix}}]{Alonso2023}
{Alonso}, D., {Fabbian}, G., {Storey-Fisher}, K., {et~al.} 2023, \jcap, 2023, 043

\bibitem[{{Amon} {et~al.}(2018){Amon}, {Blake}, {Heymans}, {Leonard}, {Asgari}, {Bilicki}, {Choi}, {Erben}, {Glazebrook}, {Harnois-D{\'e}raps}, {Hildebrandt}, {Hoekstra}, {Joachimi}, {Joudaki}, {Kuijken}, {Lidman}, {Loveday}, {Parkinson}, {Valentijn}, \& {Wolf}}]{Amon2018}
{Amon}, A., {Blake}, C., {Heymans}, C., {et~al.} 2018, \mnras, 479, 3422

\bibitem[{{Amon} {et~al.}(2022){Amon}, {Gruen}, {Troxel}, {MacCrann}, {Dodelson}, {Choi}, {Doux}, {Secco}, {Samuroff}, {Krause}, {Cordero}, {Myles}, {DeRose}, {Wechsler}, {Gatti}, {Navarro-Alsina}, {Bernstein}, {Jain}, {Blazek}, {Alarcon}, {Fert{\'e}}, {Lemos}, {Raveri}, {Campos}, {Prat}, {S{\'a}nchez}, {Jarvis}, {Alves}, {Andrade-Oliveira}, {Baxter}, {Bechtol}, {Becker}, {Bridle}, {Camacho}, {Carnero Rosell}, {Carrasco Kind}, {Cawthon}, {Chang}, {Chen}, {Chintalapati}, {Crocce}, {Davis}, {Diehl}, {Drlica-Wagner}, {Eckert}, {Eifler}, {Elvin-Poole}, {Everett}, {Fang}, {Fosalba}, {Friedrich}, {Gaztanaga}, {Giannini}, {Gruendl}, {Harrison}, {Hartley}, {Herner}, {Huang}, {Huff}, {Huterer}, {Kuropatkin}, {Leget}, {Liddle}, {McCullough}, {Muir}, {Pandey}, {Park}, {Porredon}, {Refregier}, {Rollins}, {Roodman}, {Rosenfeld}, {Ross}, {Rykoff}, {Sanchez}, {Sevilla-Noarbe}, {Sheldon}, {Shin}, {Troja}, {Tutusaus}, {Tutusaus}, {Varga}, {Weaverdyck}, {Yanny}, {Yin}, {Zhang}, {Zuntz}, {Aguena}, {Allam}, {Annis}, {Bacon},
  {Bertin}, {Bhargava}, {Brooks}, {Buckley-Geer}, {Burke}, {Carretero}, {Costanzi}, {da Costa}, {Pereira}, {De Vicente}, {Desai}, {Dietrich}, {Doel}, {Ferrero}, {Flaugher}, {Frieman}, {Garc{\'\i}a-Bellido}, {Gaztanaga}, {Gerdes}, {Giannantonio}, {Gschwend}, {Gutierrez}, {Hinton}, {Hollowood}, {Honscheid}, {Hoyle}, {James}, {Kron}, {Kuehn}, {Lahav}, {Lima}, {Lin}, {Maia}, {Marshall}, {Martini}, {Melchior}, {Menanteau}, {Miquel}, {Mohr}, {Morgan}, {Ogando}, {Palmese}, {Paz-Chinch{\'o}n}, {Petravick}, {Pieres}, {Romer}, {Sanchez}, {Scarpine}, {Schubnell}, {Serrano}, {Smith}, {Soares-Santos}, {Tarle}, {Thomas}, {To}, {Weller}, \& {DES Collaboration}}]{Amon2022}
{Amon}, A., {Gruen}, D., {Troxel}, M.~A., {et~al.} 2022, \prd, 105, 023514

\bibitem[{{Asgari} {et~al.}(2021){Asgari}, {Lin}, {Joachimi}, {Giblin}, {Heymans}, {Hildebrandt}, {Kannawadi}, {St{\"o}lzner}, {Tr{\"o}ster}, {van den Busch}, {Wright}, {Bilicki}, {Blake}, {de Jong}, {Dvornik}, {Erben}, {Getman}, {Hoekstra}, {K{\"o}hlinger}, {Kuijken}, {Miller}, {Radovich}, {Schneider}, {Shan}, \& {Valentijn}}]{KiDS2021}
{Asgari}, M., {Lin}, C.-A., {Joachimi}, B., {et~al.} 2021, \aap, 645, A104

\bibitem[{{Balaguera-Antol{\'\i}nez} {et~al.}(2018){Balaguera-Antol{\'\i}nez}, {Bilicki}, {Branchini}, \& {Postiglione}}]{Balaguera2018}
{Balaguera-Antol{\'\i}nez}, A., {Bilicki}, M., {Branchini}, E., \& {Postiglione}, A. 2018, \mnras, 476, 1050

\bibitem[{{Bianchini} {et~al.}(2015){Bianchini}, {Bielewicz}, {Lapi}, {Gonzalez-Nuevo}, {Baccigalupi}, {de Zotti}, {Danese}, {Bourne}, {Cooray}, {Dunne}, {Dye}, {Eales}, {Ivison}, {Maddox}, {Negrello}, {Scott}, {Smith}, \& {Valiante}}]{Bianchini2015}
{Bianchini}, F., {Bielewicz}, P., {Lapi}, A., {et~al.} 2015, \apj, 802, 64

\bibitem[{{Bianchini} {et~al.}(2016){Bianchini}, {Lapi}, {Calabrese}, {Bielewicz}, {Gonzalez-Nuevo}, {Baccigalupi}, {Danese}, {de Zotti}, {Bourne}, {Cooray}, {Dunne}, {Eales}, \& {Valiante}}]{Bianchini2016}
{Bianchini}, F., {Lapi}, A., {Calabrese}, M., {et~al.} 2016, \apj, 825, 24

\bibitem[{{Bianchini} \& {Reichardt}(2018)}]{Bianchini2018}
{Bianchini}, F. \& {Reichardt}, C.~L. 2018, \apj, 862, 81

\bibitem[{{Blake} {et~al.}(2016){Blake}, {Joudaki}, {Heymans}, {Choi}, {Erben}, {Harnois-Deraps}, {Hildebrandt}, {Joachimi}, {Nakajima}, {van Waerbeke}, \& {Viola}}]{Blake2016}
{Blake}, C., {Joudaki}, S., {Heymans}, C., {et~al.} 2016, \mnras, 456, 2806

\bibitem[{{Blum} {et~al.}(2016){Blum}, {Burleigh}, {Dey}, {Schlegel}, {Meisner}, {Levi}, {Myers}, {Lang}, {Moustakas}, {Patej}, {Valdes}, {Kneib}, {Huanyuan}, {Nord}, {Olsen}, {Delubac}, {Saha}, {James}, {Walker}, \& {DECaLS Team}}]{DECaLS2016}
{Blum}, R.~D., {Burleigh}, K., {Dey}, A., {et~al.} 2016, in American Astronomical Society Meeting Abstracts, Vol. 228, American Astronomical Society Meeting Abstracts \#228, 317.01

\bibitem[{{Cawthon} {et~al.}(2018){Cawthon}, {Davis}, {Gatti}, {Vielzeuf}, {Elvin-Poole}, {Rozo}, {Frieman}, {Rykoff}, {Alarcon}, {Bernstein}, {Bonnett}, {Carnero Rosell}, {Castander}, {Chang}, {da Costa}, {De Vicente}, {DeRose}, {Drlica-Wagner}, {Gaztanaga}, {Giannantonio}, {Gruen}, {Gschwend}, {Hartley}, {Hoyle}, {Lin}, {Maia}, {Miquel}, {Ogando}, {Rau}, {Roodman}, {Ross}, {Sevilla-Noarbe}, {Troxel}, {Wechsler}, {Abbott}, {Abdalla}, {Allam}, {Annis}, {Avila}, {Banerji}, {Bechtol}, {Bernstein}, {Bertin}, {Brooks}, {Burke}, {Carrasco Kind}, {Carretero}, {Cunha}, {D'Andrea}, {DePoy}, {Desai}, {Diehl}, {Doel}, {Eifler}, {Evrard}, {Flaugher}, {Fosalba}, {Garc{\'\i}a-Bellido}, {Gerdes}, {Gruendl}, {Gutierrez}, {Hollowood}, {Honscheid}, {James}, {Jeltema}, {Krause}, {Kuehn}, {Kuhlmann}, {Kuropatkin}, {Lahav}, {Lima}, {Marshall}, {Martini}, {Menanteau}, {Miller}, {Plazas}, {Sanchez}, {Scarpine}, {Schindler}, {Schubnell}, {Sheldon}, {Smith}, {Smith}, {Soares-Santos}, {Sobreira}, {Suchyta}, {Swanson}, {Tarle},
  {Thomas}, {Tucker}, {Walker}, \& {DES Collaboration}}]{REDMAGIC_2018}
{Cawthon}, R., {Davis}, C., {Gatti}, M., {et~al.} 2018, \mnras, 481, 2427

\bibitem[{{Chang} {et~al.}(2023){Chang}, {Omori}, {Baxter}, {Doux}, {Choi}, {Pandey}, {Alarcon}, {Alves}, {Amon}, {Andrade-Oliveira}, {Bechtol}, {Becker}, {Bernstein}, {Bianchini}, {Blazek}, {Bleem}, {Camacho}, {Campos}, {Carnero Rosell}, {Carrasco Kind}, {Cawthon}, {Chen}, {Cordero}, {Crawford}, {Crocce}, {Davis}, {DeRose}, {Dodelson}, {Drlica-Wagner}, {Eckert}, {Eifler}, {Elsner}, {Elvin-Poole}, {Everett}, {Fang}, {Fert{\'e}}, {Fosalba}, {Friedrich}, {Gatti}, {Giannini}, {Gruen}, {Gruendl}, {Harrison}, {Herner}, {Huang}, {Huff}, {Huterer}, {Jarvis}, {Kovacs}, {Krause}, {Kuropatkin}, {Leget}, {Lemos}, {Liddle}, {MacCrann}, {McCullough}, {Muir}, {Myles}, {Navarro-Alsina}, {Park}, {Porredon}, {Prat}, {Raveri}, {Rollins}, {Roodman}, {Rosenfeld}, {Ross}, {Rykoff}, {S{\'a}nchez}, {Sanchez}, {Secco}, {Sevilla-Noarbe}, {Sheldon}, {Shin}, {Troxel}, {Tutusaus}, {Varga}, {Weaverdyck}, {Wechsler}, {Wu}, {Yanny}, {Yin}, {Zhang}, {Zuntz}, {Abbott}, {Aguena}, {Allam}, {Annis}, {Bacon}, {Benson}, {Bertin}, {Bocquet},
  {Brooks}, {Burke}, {Carlstrom}, {Carretero}, {Chang}, {Chown}, {Costanzi}, {da Costa}, {Crites}, {Pereira}, {de Haan}, {De Vicente}, {Desai}, {Diehl}, {Dobbs}, {Doel}, {Everett}, {Ferrero}, {Flaugher}, {Friedel}, {Frieman}, {Garc{\'\i}a-Bellido}, {Gaztanaga}, {George}, {Giannantonio}, {Halverson}, {Hinton}, {Holder}, {Hollowood}, {Holzapfel}, {Honscheid}, {Hrubes}, {James}, {Knox}, {Kuehn}, {Lahav}, {Lee}, {Lima}, {Luong-Van}, {March}, {McMahon}, {Melchior}, {Menanteau}, {Meyer}, {Miquel}, {Mocanu}, {Mohr}, {Morgan}, {Natoli}, {Padin}, {Palmese}, {Paz-Chinch{\'o}n}, {Pieres}, {Plazas Malag{\'o}n}, {Pryke}, {Reichardt}, {Rodr{\'\i}guez-Monroy}, {Romer}, {Ruhl}, {Sanchez}, {Schaffer}, {Schubnell}, {Serrano}, {Shirokoff}, {Smith}, {Staniszewski}, {Stark}, {Suchyta}, {Tarle}, {Thomas}, {To}, {Vieira}, {Weller}, {Williamson}, {DES}, \& {SPT Collaborations}}]{Chang2023}
{Chang}, C., {Omori}, Y., {Baxter}, E.~J., {et~al.} 2023, \prd, 107, 023530

\bibitem[{{Darwish} {et~al.}(2021){Darwish}, {Madhavacheril}, {Sherwin}, {Aiola}, {Battaglia}, {Beall}, {Becker}, {Bond}, {Calabrese}, {Choi}, {Devlin}, {Dunkley}, {D{\"u}nner}, {Ferraro}, {Fox}, {Gallardo}, {Guan}, {Halpern}, {Han}, {Hasselfield}, {Hill}, {Hilton}, {Hilton}, {Hincks}, {Patty Ho}, {Hubmayr}, {Hughes}, {Koopman}, {Kosowsky}, {Van Lanen}, {Louis}, {Lungu}, {MacInnis}, {Maurin}, {McMahon}, {Moodley}, {Naess}, {Namikawa}, {Nati}, {Newburgh}, {Nibarger}, {Niemack}, {Page}, {Partridge}, {Qu}, {Robertson}, {Schillaci}, {Schmitt}, {Sehgal}, {Sif{\'o}n}, {Spergel}, {Staggs}, {Storer}, {van Engelen}, \& {Wollack}}]{Darwish2021}
{Darwish}, O., {Madhavacheril}, M.~S., {Sherwin}, B.~D., {et~al.} 2021, \mnras, 500, 2250

\bibitem[{{Delabrouille} {et~al.}(2003){Delabrouille}, {Cardoso}, \& {Patanchon}}]{Delabrouille2003}
{Delabrouille}, J., {Cardoso}, J.~F., \& {Patanchon}, G. 2003, \mnras, 346, 1089

\bibitem[{{DESI Collaboration} {et~al.}(2016){DESI Collaboration}, {Aghamousa}, {Aguilar}, {Ahlen}, {Alam}, {Allen}, {Allende Prieto}, {Annis}, {Bailey}, {Balland}, {Ballester}, {Baltay}, {Beaufore}, {Bebek}, {Beers}, {Bell}, {Bernal}, {Besuner}, {Beutler}, {Blake}, {Bleuler}, {Blomqvist}, {Blum}, {Bolton}, {Briceno}, {Brooks}, {Brownstein}, {Buckley-Geer}, {Burden}, {Burtin}, {Busca}, {Cahn}, {Cai}, {Cardiel-Sas}, {Carlberg}, {Carton}, {Casas}, {Castander}, {Cervantes-Cota}, {Claybaugh}, {Close}, {Coker}, {Cole}, {Comparat}, {Cooper}, {Cousinou}, {Crocce}, {Cuby}, {Cunningham}, {Davis}, {Dawson}, {de la Macorra}, {De Vicente}, {Delubac}, {Derwent}, {Dey}, {Dhungana}, {Ding}, {Doel}, {Duan}, {Ealet}, {Edelstein}, {Eftekharzadeh}, {Eisenstein}, {Elliott}, {Escoffier}, {Evatt}, {Fagrelius}, {Fan}, {Fanning}, {Farahi}, {Farihi}, {Favole}, {Feng}, {Fernandez}, {Findlay}, {Finkbeiner}, {Fitzpatrick}, {Flaugher}, {Flender}, {Font-Ribera}, {Forero-Romero}, {Fosalba}, {Frenk}, {Fumagalli}, {Gaensicke}, {Gallo},
  {Garcia-Bellido}, {Gaztanaga}, {Pietro Gentile Fusillo}, {Gerard}, {Gershkovich}, {Giannantonio}, {Gillet}, {Gonzalez-de-Rivera}, {Gonzalez-Perez}, {Gott}, {Graur}, {Gutierrez}, {Guy}, {Habib}, {Heetderks}, {Heetderks}, {Heitmann}, {Hellwing}, {Herrera}, {Ho}, {Holland}, {Honscheid}, {Huff}, {Hutchinson}, {Huterer}, {Hwang}, {Illa Laguna}, {Ishikawa}, {Jacobs}, {Jeffrey}, {Jelinsky}, {Jennings}, {Jiang}, {Jimenez}, {Johnson}, {Joyce}, {Jullo}, {Juneau}, {Kama}, {Karcher}, {Karkar}, {Kehoe}, {Kennamer}, {Kent}, {Kilbinger}, {Kim}, {Kirkby}, {Kisner}, {Kitanidis}, {Kneib}, {Koposov}, {Kovacs}, {Koyama}, {Kremin}, {Kron}, {Kronig}, {Kueter-Young}, {Lacey}, {Lafever}, {Lahav}, {Lambert}, {Lampton}, {Landriau}, {Lang}, {Lauer}, {Le Goff}, {Le Guillou}, {Le Van Suu}, {Lee}, {Lee}, {Leitner}, {Lesser}, {Levi}, {L'Huillier}, {Li}, {Liang}, {Lin}, {Linder}, {Loebman}, {Luki{\'c}}, {Ma}, {MacCrann}, {Magneville}, {Makarem}, {Manera}, {Manser}, {Marshall}, {Martini}, {Massey}, {Matheson}, {McCauley}, {McDonald},
  {McGreer}, {Meisner}, {Metcalfe}, {Miller}, {Miquel}, {Moustakas}, {Myers}, {Naik}, {Newman}, {Nichol}, {Nicola}, {Nicolati da Costa}, {Nie}, {Niz}, {Norberg}, {Nord}, {Norman}, {Nugent}, {O'Brien}, {Oh}, {Olsen}, {Padilla}, {Padmanabhan}, {Padmanabhan}, {Palanque-Delabrouille}, {Palmese}, {Pappalardo}, {P{\^a}ris}, {Park}, {Patej}, {Peacock}, {Peiris}, {Peng}, {Percival}, {Perruchot}, {Pieri}, {Pogge}, {Pollack}, {Poppett}, {Prada}, {Prakash}, {Probst}, {Rabinowitz}, {Raichoor}, {Ree}, {Refregier}, {Regal}, {Reid}, {Reil}, {Rezaie}, {Rockosi}, {Roe}, {Ronayette}, {Roodman}, {Ross}, {Ross}, {Rossi}, {Rozo}, {Ruhlmann-Kleider}, {Rykoff}, {Sabiu}, {Samushia}, {Sanchez}, {Sanchez}, {Schlegel}, {Schneider}, {Schubnell}, {Secroun}, {Seljak}, {Seo}, {Serrano}, {Shafieloo}, {Shan}, {Sharples}, {Sholl}, {Shourt}, {Silber}, {Silva}, {Sirk}, {Slosar}, {Smith}, {Smoot}, {Som}, {Song}, {Sprayberry}, {Staten}, {Stefanik}, {Tarle}, {Sien Tie}, {Tinker}, {Tojeiro}, {Valdes}, {Valenzuela}, {Valluri}, {Vargas-Magana},
  {Verde}, {Walker}, {Wang}, {Wang}, {Weaver}, {Weaverdyck}, {Wechsler}, {Weinberg}, {White}, {Yang}, {Yeche}, {Zhang}, {Zhao}, {Zheng}, {Zhou}, {Zhou}, {Zhu}, {Zou}, \& {Zu}}]{DESI_Collab2016}
{DESI Collaboration}, {Aghamousa}, A., {Aguilar}, J., {et~al.} 2016, arXiv e-prints, arXiv:1611.00037

\bibitem[{Dey {et~al.}(2016)Dey, Rabinowitz, Karcher, Bebek, Baltay, Sprayberry, Valdes, Stupak, Donaldson, Emmet, Hurteau, Abareshi, Marshall, Lang, Fitzpatrick, Daly, Joyce, Schlegel, Schweiker, Allen, Blum, \& Levi}]{Mosaic3}
Dey, A., Rabinowitz, D., Karcher, A., {et~al.} 2016, in Ground-based and Airborne Instrumentation for Astronomy VI, ed. C.~J. Evans, L.~Simard, \& H.~Takami, Vol. 9908, International Society for Optics and Photonics (SPIE), 99082C

\bibitem[{{Dey} {et~al.}(2019){Dey}, {Schlegel}, {Lang}, {Blum}, {Burleigh}, {Fan}, {Findlay}, {Finkbeiner}, {Herrera}, {Juneau}, {Landriau}, {Levi}, {McGreer}, {Meisner}, {Myers}, {Moustakas}, {Nugent}, {Patej}, {Schlafly}, {Walker}, {Valdes}, {Weaver}, {Y{\`e}che}, {Zou}, {Zhou}, {Abareshi}, {Abbott}, {Abolfathi}, {Aguilera}, {Alam}, {Allen}, {Alvarez}, {Annis}, {Ansarinejad}, {Aubert}, {Beechert}, {Bell}, {BenZvi}, {Beutler}, {Bielby}, {Bolton}, {Brice{\~n}o}, {Buckley-Geer}, {Butler}, {Calamida}, {Carlberg}, {Carter}, {Casas}, {Castander}, {Choi}, {Comparat}, {Cukanovaite}, {Delubac}, {DeVries}, {Dey}, {Dhungana}, {Dickinson}, {Ding}, {Donaldson}, {Duan}, {Duckworth}, {Eftekharzadeh}, {Eisenstein}, {Etourneau}, {Fagrelius}, {Farihi}, {Fitzpatrick}, {Font-Ribera}, {Fulmer}, {G{\"a}nsicke}, {Gaztanaga}, {George}, {Gerdes}, {Gontcho}, {Gorgoni}, {Green}, {Guy}, {Harmer}, {Hernandez}, {Honscheid}, {Huang}, {James}, {Jannuzi}, {Jiang}, {Joyce}, {Karcher}, {Karkar}, {Kehoe}, {Kneib}, {Kueter-Young}, {Lan},
  {Lauer}, {Le Guillou}, {Le Van Suu}, {Lee}, {Lesser}, {Perreault Levasseur}, {Li}, {Mann}, {Marshall}, {Mart{\'\i}nez-V{\'a}zquez}, {Martini}, {du Mas des Bourboux}, {McManus}, {Meier}, {M{\'e}nard}, {Metcalfe}, {Mu{\~n}oz-Guti{\'e}rrez}, {Najita}, {Napier}, {Narayan}, {Newman}, {Nie}, {Nord}, {Norman}, {Olsen}, {Paat}, {Palanque-Delabrouille}, {Peng}, {Poppett}, {Poremba}, {Prakash}, {Rabinowitz}, {Raichoor}, {Rezaie}, {Robertson}, {Roe}, {Ross}, {Ross}, {Rudnick}, {Safonova}, {Saha}, {S{\'a}nchez}, {Savary}, {Schweiker}, {Scott}, {Seo}, {Shan}, {Silva}, {Slepian}, {Soto}, {Sprayberry}, {Staten}, {Stillman}, {Stupak}, {Summers}, {Sien Tie}, {Tirado}, {Vargas-Maga{\~n}a}, {Vivas}, {Wechsler}, {Williams}, {Yang}, {Yang}, {Yapici}, {Zaritsky}, {Zenteno}, {Zhang}, {Zhang}, {Zhou}, \& {Zhou}}]{Dey2019}
{Dey}, A., {Schlegel}, D.~J., {Lang}, D., {et~al.} 2019, \aj, 157, 168

\bibitem[{{Dor{\'e}} {et~al.}(2014){Dor{\'e}}, {Bock}, {Ashby}, {Capak}, {Cooray}, {de Putter}, {Eifler}, {Flagey}, {Gong}, {Habib}, {Heitmann}, {Hirata}, {Jeong}, {Katti}, {Korngut}, {Krause}, {Lee}, {Masters}, {Mauskopf}, {Melnick}, {Mennesson}, {Nguyen}, {{\"O}berg}, {Pullen}, {Raccanelli}, {Smith}, {Song}, {Tolls}, {Unwin}, {Venumadhav}, {Viero}, {Werner}, \& {Zemcov}}]{SPHEREX2014}
{Dor{\'e}}, O., {Bock}, J., {Ashby}, M., {et~al.} 2014, arXiv e-prints, arXiv:1412.4872

\bibitem[{{Dutcher} {et~al.}(2021){Dutcher}, {Balkenhol}, {Ade}, {Ahmed}, {Anderes}, {Anderson}, {Archipley}, {Avva}, {Aylor}, {Barry}, {Basu Thakur}, {Benabed}, {Bender}, {Benson}, {Bianchini}, {Bleem}, {Bouchet}, {Bryant}, {Byrum}, {Carlstrom}, {Carter}, {Cecil}, {Chang}, {Chaubal}, {Chen}, {Cho}, {Chou}, {Cliche}, {Crawford}, {Cukierman}, {Daley}, {de Haan}, {Denison}, {Dibert}, {Ding}, {Dobbs}, {Everett}, {Feng}, {Ferguson}, {Foster}, {Fu}, {Galli}, {Gambrel}, {Gardner}, {Goeckner-Wald}, {Gualtieri}, {Guns}, {Gupta}, {Guyser}, {Halverson}, {Harke-Hosemann}, {Harrington}, {Henning}, {Hilton}, {Hivon}, {Holder}, {Holzapfel}, {Hood}, {Howe}, {Huang}, {Irwin}, {Jeong}, {Jonas}, {Jones}, {Khaire}, {Knox}, {Kofman}, {Korman}, {Kubik}, {Kuhlmann}, {Kuo}, {Lee}, {Leitch}, {Lowitz}, {Lu}, {Meyer}, {Michalik}, {Millea}, {Montgomery}, {Nadolski}, {Natoli}, {Nguyen}, {Noble}, {Novosad}, {Omori}, {Padin}, {Pan}, {Paschos}, {Pearson}, {Posada}, {Prabhu}, {Quan}, {Raghunathan}, {Rahlin}, {Reichardt}, {Riebel}, {Riedel},
  {Rouble}, {Ruhl}, {Sayre}, {Schiappucci}, {Shirokoff}, {Smecher}, {Sobrin}, {Stark}, {Stephen}, {Story}, {Suzuki}, {Thompson}, {Thorne}, {Tucker}, {Umilta}, {Vale}, {Vanderlinde}, {Vieira}, {Wang}, {Whitehorn}, {Wu}, {Yefremenko}, {Yoon}, {Young}, \& {SPT-3G Collaboration}}]{SPT-3G2021}
{Dutcher}, D., {Balkenhol}, L., {Ade}, P.~A.~R., {et~al.} 2021, \prd, 104, 022003

\bibitem[{{Foreman-Mackey} {et~al.}(2013){Foreman-Mackey}, {Hogg}, {Lang}, \& {Goodman}}]{EMCEE2013}
{Foreman-Mackey}, D., {Hogg}, D.~W., {Lang}, D., \& {Goodman}, J. 2013, \pasp, 125, 306

\bibitem[{{Fry}(1996)}]{Fry1996}
{Fry}, J.~N. 1996, \apjl, 461, L65

\bibitem[{{Giannantonio} {et~al.}(2016){Giannantonio}, {Fosalba}, {Cawthon}, {Omori}, {Crocce}, {Elsner}, {Leistedt}, {Dodelson}, {Benoit-L{\'e}vy}, {Gazta{\~n}aga}, {Holder}, {Peiris}, {Percival}, {Kirk}, {Bauer}, {Benson}, {Bernstein}, {Carretero}, {Crawford}, {Crittenden}, {Huterer}, {Jain}, {Krause}, {Reichardt}, {Ross}, {Simard}, {Soergel}, {Stark}, {Story}, {Vieira}, {Weller}, {Abbott}, {Abdalla}, {Allam}, {Armstrong}, {Banerji}, {Bernstein}, {Bertin}, {Brooks}, {Buckley-Geer}, {Burke}, {Capozzi}, {Carlstrom}, {Carnero Rosell}, {Carrasco Kind}, {Castander}, {Chang}, {Cunha}, {da Costa}, {D'Andrea}, {DePoy}, {Desai}, {Diehl}, {Dietrich}, {Doel}, {Eifler}, {Evrard}, {Neto}, {Fernandez}, {Finley}, {Flaugher}, {Frieman}, {Gerdes}, {Gruen}, {Gruendl}, {Gutierrez}, {Holzapfel}, {Honscheid}, {James}, {Kuehn}, {Kuropatkin}, {Lahav}, {Li}, {Lima}, {March}, {Marshall}, {Martini}, {Melchior}, {Miquel}, {Mohr}, {Nichol}, {Nord}, {Ogando}, {Plazas}, {Romer}, {Roodman}, {Rykoff}, {Sako}, {Saliwanchik}, {Sanchez},
  {Schubnell}, {Sevilla-Noarbe}, {Smith}, {Soares-Santos}, {Sobreira}, {Suchyta}, {Swanson}, {Tarle}, {Thaler}, {Thomas}, {Vikram}, {Walker}, {Wechsler}, \& {Zuntz}}]{Giannantonio2016}
{Giannantonio}, T., {Fosalba}, P., {Cawthon}, R., {et~al.} 2016, \mnras, 456, 3213

\bibitem[{{G{\'o}rski} {et~al.}(2005){G{\'o}rski}, {Hivon}, {Banday}, {Wandelt}, {Hansen}, {Reinecke}, \& {Bartelmann}}]{Gorski2005}
{G{\'o}rski}, K.~M., {Hivon}, E., {Banday}, A.~J., {et~al.} 2005, \apj, 622, 759

\bibitem[{Guth(1981)}]{Guth1980}
Guth, A.~H. 1981, Phys. Rev. D, 23, 347

\bibitem[{{Hang} {et~al.}(2021){Hang}, {Alam}, {Peacock}, \& {Cai}}]{Hang2021}
{Hang}, Q., {Alam}, S., {Peacock}, J.~A., \& {Cai}, Y.-C. 2021, \mnras, 501, 1481

\bibitem[{{Hu}(2000)}]{Hu2000}
{Hu}, W. 2000, \prd, 62, 043007

\bibitem[{{Ilbert} {et~al.}(2009){Ilbert}, {Capak}, {Salvato}, {Aussel}, {McCracken}, {Sanders}, {Scoville}, {Kartaltepe}, {Arnouts}, {Le Floc'h}, {Mobasher}, {Taniguchi}, {Lamareille}, {Leauthaud}, {Sasaki}, {Thompson}, {Zamojski}, {Zamorani}, {Bardelli}, {Bolzonella}, {Bongiorno}, {Brusa}, {Caputi}, {Carollo}, {Contini}, {Cook}, {Coppa}, {Cucciati}, {de la Torre}, {de Ravel}, {Franzetti}, {Garilli}, {Hasinger}, {Iovino}, {Kampczyk}, {Kneib}, {Knobel}, {Kovac}, {Le Borgne}, {Le Brun}, {Le F{\`e}vre}, {Lilly}, {Looper}, {Maier}, {Mainieri}, {Mellier}, {Mignoli}, {Murayama}, {Pell{\`o}}, {Peng}, {P{\'e}rez-Montero}, {Renzini}, {Ricciardelli}, {Schiminovich}, {Scodeggio}, {Shioya}, {Silverman}, {Surace}, {Tanaka}, {Tasca}, {Tresse}, {Vergani}, \& {Zucca}}]{COSMOS_2009}
{Ilbert}, O., {Capak}, P., {Salvato}, M., {et~al.} 2009, \apj, 690, 1236

\bibitem[{Ivezi{\'{c}} {et~al.}(2019{\natexlab{a}})Ivezi{\'{c}}, Kahn, Tyson, Abel, Acosta, Allsman, Alonso, AlSayyad, Anderson, Andrew, Angel, Angeli, Ansari, Antilogus, Araujo, Armstrong, Arndt, Astier, Aubourg, Auza, Axelrod, Bard, Barr, Barrau, Bartlett, Bauer, Bauman, Baumont, Bechtol, Bechtol, Becker, Becla, Beldica, Bellavia, Bianco, Biswas, Blanc, Blazek, Blandford, Bloom, Bogart, Bond, Booth, Borgland, Borne, Bosch, Boutigny, Brackett, Bradshaw, Brandt, Brown, Bullock, Burchat, Burke, Cagnoli, Calabrese, Callahan, Callen, Carlin, Carlson, Chandrasekharan, Charles-Emerson, Chesley, Cheu, Chiang, Chiang, Chirino, Chow, Ciardi, Claver, Cohen-Tanugi, Cockrum, Coles, Connolly, Cook, Cooray, Covey, Cribbs, Cui, Cutri, Daly, Daniel, Daruich, Daubard, Daues, Dawson, Delgado, Dellapenna, de~Peyster, de~Val-Borro, Digel, Doherty, Dubois, Dubois-Felsmann, Durech, Economou, Eifler, Eracleous, Emmons, Neto, Ferguson, Figueroa, Fisher-Levine, Focke, Foss, Frank, Freemon, Gangler, Gawiser, Geary, Gee, Geha,
  Gessner, Gibson, Gilmore, Glanzman, Glick, Goldina, Goldstein, Goodenow, Graham, Gressler, Gris, Guy, Guyonnet, Haller, Harris, Hascall, Haupt, Hernandez, Herrmann, Hileman, Hoblitt, Hodgson, Hogan, Howard, Huang, Huffer, Ingraham, Innes, Jacoby, Jain, Jammes, Jee, Jenness, Jernigan, Jevremovi{\'{c}}, Johns, Johnson, Johnson, Jones, Juramy-Gilles, Juri{\'{c}}, Kalirai, Kallivayalil, Kalmbach, Kantor, Karst, Kasliwal, Kelly, Kessler, Kinnison, Kirkby, Knox, Kotov, Krabbendam, Krughoff, Kub{\'{a}}nek, Kuczewski, Kulkarni, Ku, Kurita, Lage, Lambert, Lange, Langton, Guillou, Levine, Liang, Lim, Lintott, Long, Lopez, Lotz, Lupton, Lust, MacArthur, Mahabal, Mandelbaum, Markiewicz, Marsh, Marshall, Marshall, May, McKercher, McQueen, Meyers, Migliore, Miller, Mills, Miraval, Moeyens, Moolekamp, Monet, Moniez, Monkewitz, Montgomery, Morrison, Mueller, Muller, Arancibia, Neill, Newbry, Nief, Nomerotski, Nordby, O'Connor, Oliver, Olivier, Olsen, O'Mullane, Ortiz, Osier, Owen, Pain, Palecek, Parejko, Parsons, Pease,
  Peterson, Peterson, Petravick, Petrick, Petry, Pierfederici, Pietrowicz, Pike, Pinto, Plante, Plate, Plutchak, Price, Prouza, Radeka, Rajagopal, Rasmussen, Regnault, Reil, Reiss, Reuter, Ridgway, Riot, Ritz, Robinson, Roby, Roodman, Rosing, Roucelle, Rumore, Russo, Saha, Sassolas, Schalk, Schellart, Schindler, Schmidt, Schneider, Schneider, Schoening, Schumacher, Schwamb, Sebag, Selvy, Sembroski, Seppala, Serio, Serrano, Shaw, Shipsey, Sick, Silvestri, Slater, Smith, Smith, Sobhani, Soldahl, Storrie-Lombardi, Stover, Strauss, Street, Stubbs, Sullivan, Sweeney, Swinbank, Szalay, Takacs, Tether, Thaler, Thayer, Thomas, Thornton, Thukral, Tice, Trilling, Turri, Berg, Berk, Vetter, Virieux, Vucina, Wahl, Walkowicz, Walsh, Walter, Wang, Wang, Warner, Wiecha, Willman, Winters, Wittman, Wolff, Wood-Vasey, Wu, Xin, Yoachim, \& Zhan}]{Ivezic2019}
Ivezi{\'{c}}, {\v{Z}}., Kahn, S.~M., Tyson, J.~A., {et~al.} 2019{\natexlab{a}}, ApJ, 873, 111

\bibitem[{Ivezi{\'{c}} {et~al.}(2019{\natexlab{b}})Ivezi{\'{c}}, Kahn, Tyson, Abel, Acosta, Allsman, Alonso, AlSayyad, Anderson, Andrew, Angel, Angeli, Ansari, Antilogus, Araujo, Armstrong, Arndt, Astier, Aubourg, Auza, Axelrod, Bard, Barr, Barrau, Bartlett, Bauer, Bauman, Baumont, Bechtol, Bechtol, Becker, Becla, Beldica, Bellavia, Bianco, Biswas, Blanc, Blazek, Blandford, Bloom, Bogart, Bond, Booth, Borgland, Borne, Bosch, Boutigny, Brackett, Bradshaw, Brandt, Brown, Bullock, Burchat, Burke, Cagnoli, Calabrese, Callahan, Callen, Carlin, Carlson, Chandrasekharan, Charles-Emerson, Chesley, Cheu, Chiang, Chiang, Chirino, Chow, Ciardi, Claver, Cohen-Tanugi, Cockrum, Coles, Connolly, Cook, Cooray, Covey, Cribbs, Cui, Cutri, Daly, Daniel, Daruich, Daubard, Daues, Dawson, Delgado, Dellapenna, de~Peyster, de~Val-Borro, Digel, Doherty, Dubois, Dubois-Felsmann, Durech, Economou, Eifler, Eracleous, Emmons, Neto, Ferguson, Figueroa, Fisher-Levine, Focke, Foss, Frank, Freemon, Gangler, Gawiser, Geary, Gee, Geha,
  Gessner, Gibson, Gilmore, Glanzman, Glick, Goldina, Goldstein, Goodenow, Graham, Gressler, Gris, Guy, Guyonnet, Haller, Harris, Hascall, Haupt, Hernandez, Herrmann, Hileman, Hoblitt, Hodgson, Hogan, Howard, Huang, Huffer, Ingraham, Innes, Jacoby, Jain, Jammes, Jee, Jenness, Jernigan, Jevremovi{\'{c}}, Johns, Johnson, Johnson, Jones, Juramy-Gilles, Juri{\'{c}}, Kalirai, Kallivayalil, Kalmbach, Kantor, Karst, Kasliwal, Kelly, Kessler, Kinnison, Kirkby, Knox, Kotov, Krabbendam, Krughoff, Kub{\'{a}}nek, Kuczewski, Kulkarni, Ku, Kurita, Lage, Lambert, Lange, Langton, Guillou, Levine, Liang, Lim, Lintott, Long, Lopez, Lotz, Lupton, Lust, MacArthur, Mahabal, Mandelbaum, Markiewicz, Marsh, Marshall, Marshall, May, McKercher, McQueen, Meyers, Migliore, Miller, Mills, Miraval, Moeyens, Moolekamp, Monet, Moniez, Monkewitz, Montgomery, Morrison, Mueller, Muller, Arancibia, Neill, Newbry, Nief, Nomerotski, Nordby, O'Connor, Oliver, Olivier, Olsen, O'Mullane, Ortiz, Osier, Owen, Pain, Palecek, Parejko, Parsons, Pease,
  Peterson, Peterson, Petravick, Petrick, Petry, Pierfederici, Pietrowicz, Pike, Pinto, Plante, Plate, Plutchak, Price, Prouza, Radeka, Rajagopal, Rasmussen, Regnault, Reil, Reiss, Reuter, Ridgway, Riot, Ritz, Robinson, Roby, Roodman, Rosing, Roucelle, Rumore, Russo, Saha, Sassolas, Schalk, Schellart, Schindler, Schmidt, Schneider, Schneider, Schoening, Schumacher, Schwamb, Sebag, Selvy, Sembroski, Seppala, Serio, Serrano, Shaw, Shipsey, Sick, Silvestri, Slater, Smith, Smith, Sobhani, Soldahl, Storrie-Lombardi, Stover, Strauss, Street, Stubbs, Sullivan, Sweeney, Swinbank, Szalay, Takacs, Tether, Thaler, Thayer, Thomas, Thornton, Thukral, Tice, Trilling, Turri, Berg, Berk, Vetter, Virieux, Vucina, Wahl, Walkowicz, Walsh, Walter, Wang, Wang, Warner, Wiecha, Willman, Winters, Wittman, Wolff, Wood-Vasey, Wu, Xin, Yoachim, \& Zhan}]{LSST2019}
Ivezi{\'{c}}, {\v{Z}}., Kahn, S.~M., Tyson, J.~A., {et~al.} 2019{\natexlab{b}}, ApJ, 873, 111

\bibitem[{{Joudaki} {et~al.}(2017){Joudaki}, {Blake}, {Heymans}, {Choi}, {Harnois-Deraps}, {Hildebrandt}, {Joachimi}, {Johnson}, {Mead}, {Parkinson}, {Viola}, \& {van Waerbeke}}]{Joudaki2017}
{Joudaki}, S., {Blake}, C., {Heymans}, C., {et~al.} 2017, \mnras, 465, 2033

\bibitem[{{Krolewski} {et~al.}(2021){Krolewski}, {Ferraro}, \& {White}}]{Krolewski2021}
{Krolewski}, A., {Ferraro}, S., \& {White}, M. 2021, \jcap, 2021, 028

\bibitem[{{Laureijs} {et~al.}(2011){Laureijs}, {Amiaux}, {Arduini}, {Augu{\`e}res}, {Brinchmann}, {Cole}, {Cropper}, {Dabin}, {Duvet}, {Ealet}, {Garilli}, {Gondoin}, {Guzzo}, {Hoar}, {Hoekstra}, {Holmes}, {Kitching}, {Maciaszek}, {Mellier}, {Pasian}, {Percival}, {Rhodes}, {Saavedra Criado}, {Sauvage}, {Scaramella}, {Valenziano}, {Warren}, {Bender}, {Castander}, {Cimatti}, {Le F{\`e}vre}, {Kurki-Suonio}, {Levi}, {Lilje}, {Meylan}, {Nichol}, {Pedersen}, {Popa}, {Rebolo Lopez}, {Rix}, {Rottgering}, {Zeilinger}, {Grupp}, {Hudelot}, {Massey}, {Meneghetti}, {Miller}, {Paltani}, {Paulin-Henriksson}, {Pires}, {Saxton}, {Schrabback}, {Seidel}, {Walsh}, {Aghanim}, {Amendola}, {Bartlett}, {Baccigalupi}, {Beaulieu}, {Benabed}, {Cuby}, {Elbaz}, {Fosalba}, {Gavazzi}, {Helmi}, {Hook}, {Irwin}, {Kneib}, {Kunz}, {Mannucci}, {Moscardini}, {Tao}, {Teyssier}, {Weller}, {Zamorani}, {Zapatero Osorio}, {Boulade}, {Foumond}, {Di Giorgio}, {Guttridge}, {James}, {Kemp}, {Martignac}, {Spencer}, {Walton}, {Bl{\"u}mchen}, {Bonoli},
  {Bortoletto}, {Cerna}, {Corcione}, {Fabron}, {Jahnke}, {Ligori}, {Madrid}, {Martin}, {Morgante}, {Pamplona}, {Prieto}, {Riva}, {Toledo}, {Trifoglio}, {Zerbi}, {Abdalla}, {Douspis}, {Grenet}, {Borgani}, {Bouwens}, {Courbin}, {Delouis}, {Dubath}, {Fontana}, {Frailis}, {Grazian}, {Koppenh{\"o}fer}, {Mansutti}, {Melchior}, {Mignoli}, {Mohr}, {Neissner}, {Noddle}, {Poncet}, {Scodeggio}, {Serrano}, {Shane}, {Starck}, {Surace}, {Taylor}, {Verdoes-Kleijn}, {Vuerli}, {Williams}, {Zacchei}, {Altieri}, {Escudero Sanz}, {Kohley}, {Oosterbroek}, {Astier}, {Bacon}, {Bardelli}, {Baugh}, {Bellagamba}, {Benoist}, {Bianchi}, {Biviano}, {Branchini}, {Carbone}, {Cardone}, {Clements}, {Colombi}, {Conselice}, {Cresci}, {Deacon}, {Dunlop}, {Fedeli}, {Fontanot}, {Franzetti}, {Giocoli}, {Garcia-Bellido}, {Gow}, {Heavens}, {Hewett}, {Heymans}, {Holland}, {Huang}, {Ilbert}, {Joachimi}, {Jennins}, {Kerins}, {Kiessling}, {Kirk}, {Kotak}, {Krause}, {Lahav}, {van Leeuwen}, {Lesgourgues}, {Lombardi}, {Magliocchetti}, {Maguire},
  {Majerotto}, {Maoli}, {Marulli}, {Maurogordato}, {McCracken}, {McLure}, {Melchiorri}, {Merson}, {Moresco}, {Nonino}, {Norberg}, {Peacock}, {Pello}, {Penny}, {Pettorino}, {Di Porto}, {Pozzetti}, {Quercellini}, {Radovich}, {Rassat}, {Roche}, {Ronayette}, {Rossetti}, {Sartoris}, {Schneider}, {Semboloni}, {Serjeant}, {Simpson}, {Skordis}, {Smadja}, {Smartt}, {Spano}, {Spiro}, {Sullivan}, {Tilquin}, {Trotta}, {Verde}, {Wang}, {Williger}, {Zhao}, {Zoubian}, \& {Zucca}}]{Euclid2011}
{Laureijs}, R., {Amiaux}, J., {Arduini}, S., {et~al.} 2011, arXiv e-prints, arXiv:1110.3193

\bibitem[{{Lewis} {et~al.}(2000){Lewis}, {Challinor}, \& {Lasenby}}]{Lewis2000}
{Lewis}, A., {Challinor}, A., \& {Lasenby}, A. 2000, \apj, 538, 473

\bibitem[{{Limber}(1953)}]{Limber1953}
{Limber}, D.~N. 1953, \apj, 117, 134

\bibitem[{{Linder}(2005)}]{Linder2005}
{Linder}, E.~V. 2005, \prd, 72, 043529

\bibitem[{{Liske} {et~al.}(2015){Liske}, {Baldry}, {Driver}, {Tuffs}, {Alpaslan}, {Andrae}, {Brough}, {Cluver}, {Grootes}, {Gunawardhana}, {Kelvin}, {Loveday}, {Robotham}, {Taylor}, {Bamford}, {Bland-Hawthorn}, {Brown}, {Drinkwater}, {Hopkins}, {Meyer}, {Norberg}, {Peacock}, {Agius}, {Andrews}, {Bauer}, {Ching}, {Colless}, {Conselice}, {Croom}, {Davies}, {De Propris}, {Dunne}, {Eardley}, {Ellis}, {Foster}, {Frenk}, {H{\"a}u{\ss}ler}, {Holwerda}, {Howlett}, {Ibarra}, {Jarvis}, {Jones}, {Kafle}, {Lacey}, {Lange}, {Lara-L{\'o}pez}, {L{\'o}pez-S{\'a}nchez}, {Maddox}, {Madore}, {McNaught-Roberts}, {Moffett}, {Nichol}, {Owers}, {Palamara}, {Penny}, {Phillipps}, {Pimbblet}, {Popescu}, {Prescott}, {Proctor}, {Sadler}, {Sansom}, {Seibert}, {Sharp}, {Sutherland}, {V{\'a}zquez-Mata}, {van Kampen}, {Wilkins}, {Williams}, \& {Wright}}]{GAMADR2_2015}
{Liske}, J., {Baldry}, I.~K., {Driver}, S.~P., {et~al.} 2015, \mnras, 452, 2087

\bibitem[{{LSST Science Collaboration} {et~al.}(2009){LSST Science Collaboration}, {Abell}, {Allison}, {Anderson}, {Andrew}, {Angel}, {Armus}, {Arnett}, {Asztalos}, {Axelrod}, {Bailey}, {Ballantyne}, {Bankert}, {Barkhouse}, {Barr}, {Barrientos}, {Barth}, {Bartlett}, {Becker}, {Becla}, {Beers}, {Bernstein}, {Biswas}, {Blanton}, {Bloom}, {Bochanski}, {Boeshaar}, {Borne}, {Bradac}, {Brandt}, {Bridge}, {Brown}, {Brunner}, {Bullock}, {Burgasser}, {Burge}, {Burke}, {Cargile}, {Chandrasekharan}, {Chartas}, {Chesley}, {Chu}, {Cinabro}, {Claire}, {Claver}, {Clowe}, {Connolly}, {Cook}, {Cooke}, {Cooray}, {Covey}, {Culliton}, {de Jong}, {de Vries}, {Debattista}, {Delgado}, {Dell'Antonio}, {Dhital}, {Di Stefano}, {Dickinson}, {Dilday}, {Djorgovski}, {Dobler}, {Donalek}, {Dubois-Felsmann}, {Durech}, {Eliasdottir}, {Eracleous}, {Eyer}, {Falco}, {Fan}, {Fassnacht}, {Ferguson}, {Fernandez}, {Fields}, {Finkbeiner}, {Figueroa}, {Fox}, {Francke}, {Frank}, {Frieman}, {Fromenteau}, {Furqan}, {Galaz}, {Gal-Yam}, {Garnavich},
  {Gawiser}, {Geary}, {Gee}, {Gibson}, {Gilmore}, {Grace}, {Green}, {Gressler}, {Grillmair}, {Habib}, {Haggerty}, {Hamuy}, {Harris}, {Hawley}, {Heavens}, {Hebb}, {Henry}, {Hileman}, {Hilton}, {Hoadley}, {Holberg}, {Holman}, {Howell}, {Infante}, {Ivezic}, {Jacoby}, {Jain}, {R}, {Jedicke}, {Jee}, {Garrett Jernigan}, {Jha}, {Johnston}, {Jones}, {Juric}, {Kaasalainen}, {Styliani}, {Kafka}, {Kahn}, {Kaib}, {Kalirai}, {Kantor}, {Kasliwal}, {Keeton}, {Kessler}, {Knezevic}, {Kowalski}, {Krabbendam}, {Krughoff}, {Kulkarni}, {Kuhlman}, {Lacy}, {Lepine}, {Liang}, {Lien}, {Lira}, {Long}, {Lorenz}, {Lotz}, {Lupton}, {Lutz}, {Macri}, {Mahabal}, {Mandelbaum}, {Marshall}, {May}, {McGehee}, {Meadows}, {Meert}, {Milani}, {Miller}, {Miller}, {Mills}, {Minniti}, {Monet}, {Mukadam}, {Nakar}, {Neill}, {Newman}, {Nikolaev}, {Nordby}, {O'Connor}, {Oguri}, {Oliver}, {Olivier}, {Olsen}, {Olsen}, {Olszewski}, {Oluseyi}, {Padilla}, {Parker}, {Pepper}, {Peterson}, {Petry}, {Pinto}, {Pizagno}, {Popescu}, {Prsa}, {Radcka}, {Raddick},
  {Rasmussen}, {Rau}, {Rho}, {Rhoads}, {Richards}, {Ridgway}, {Robertson}, {Roskar}, {Saha}, {Sarajedini}, {Scannapieco}, {Schalk}, {Schindler}, {Schmidt}, {Schmidt}, {Schneider}, {Schumacher}, {Scranton}, {Sebag}, {Seppala}, {Shemmer}, {Simon}, {Sivertz}, {Smith}, {Allyn Smith}, {Smith}, {Spitz}, {Stanford}, {Stassun}, {Strader}, {Strauss}, {Stubbs}, {Sweeney}, {Szalay}, {Szkody}, {Takada}, {Thorman}, {Trilling}, {Trimble}, {Tyson}, {Van Berg}, {Vanden Berk}, {VanderPlas}, {Verde}, {Vrsnak}, {Walkowicz}, {Wandelt}, {Wang}, {Wang}, {Warner}, {Wechsler}, {West}, {Wiecha}, {Williams}, {Willman}, {Wittman}, {Wolff}, {Wood-Vasey}, {Wozniak}, {Young}, {Zentner}, \& {Zhan}}]{LSST2009}
{LSST Science Collaboration}, {Abell}, P.~A., {Allison}, J., {et~al.} 2009, arXiv e-prints, arXiv:0912.0201

\bibitem[{{Macaulay} {et~al.}(2013){Macaulay}, {Wehus}, \& {Eriksen}}]{Macaulay2013}
{Macaulay}, E., {Wehus}, I.~K., \& {Eriksen}, H.~K. 2013, \prl, 111, 161301

\bibitem[{{Marques} \& {Bernui}(2020)}]{Marques2020}
{Marques}, G.~A. \& {Bernui}, A. 2020, \jcap, 2020, 052

\bibitem[{{Miyatake} {et~al.}(2022){Miyatake}, {Harikane}, {Ouchi}, {Ono}, {Yamamoto}, {Nishizawa}, {Bahcall}, {Miyazaki}, \& {Malag{\'o}n}}]{Mitayake2022}
{Miyatake}, H., {Harikane}, Y., {Ouchi}, M., {et~al.} 2022, \prl, 129, 061301

\bibitem[{{Moscardini} {et~al.}(1998){Moscardini}, {Coles}, {Lucchin}, \& {Matarrese}}]{Moscardini1998}
{Moscardini}, L., {Coles}, P., {Lucchin}, F., \& {Matarrese}, S. 1998, \mnras, 299, 95

\bibitem[{{Nakoneczny} {et~al.}(2024){Nakoneczny}, {Alonso}, {Bilicki}, {Schwarz}, {Hale}, {Pollo}, {Heneka}, {Tiwari}, {Zheng}, {Br{\"u}ggen}, {Jarvis}, \& {Shimwell}}]{Nakoneczny2024}
{Nakoneczny}, S.~J., {Alonso}, D., {Bilicki}, M., {et~al.} 2024, \aap, 681, A105

\bibitem[{{Newman} {et~al.}(2013){Newman}, {Cooper}, {Davis}, {Faber}, {Coil}, {Guhathakurta}, {Koo}, {Phillips}, {Conroy}, {Dutton}, {Finkbeiner}, {Gerke}, {Rosario}, {Weiner}, {Willmer}, {Yan}, {Harker}, {Kassin}, {Konidaris}, {Lai}, {Madgwick}, {Noeske}, {Wirth}, {Connolly}, {Kaiser}, {Kirby}, {Lemaux}, {Lin}, {Lotz}, {Luppino}, {Marinoni}, {Matthews}, {Metevier}, \& {Schiavon}}]{DEEP2_2013}
{Newman}, J.~A., {Cooper}, M.~C., {Davis}, M., {et~al.} 2013, \apjs, 208, 5

\bibitem[{{Norberg} {et~al.}(2009){Norberg}, {Baugh}, {Gazta{\~n}aga}, \& {Croton}}]{Norberg2009}
{Norberg}, P., {Baugh}, C.~M., {Gazta{\~n}aga}, E., \& {Croton}, D.~J. 2009, \mnras, 396, 19

\bibitem[{{Pandey} {et~al.}(2022){Pandey}, {Krause}, {DeRose}, {MacCrann}, {Jain}, {Crocce}, {Blazek}, {Choi}, {Huang}, {To}, {Fang}, {Elvin-Poole}, {Prat}, {Porredon}, {Secco}, {Rodriguez-Monroy}, {Weaverdyck}, {Park}, {Raveri}, {Rozo}, {Rykoff}, {Bernstein}, {S{\'a}nchez}, {Jarvis}, {Troxel}, {Zacharegkas}, {Chang}, {Alarcon}, {Alves}, {Amon}, {Andrade-Oliveira}, {Baxter}, {Bechtol}, {Becker}, {Camacho}, {Campos}, {Carnero Rosell}, {Carrasco Kind}, {Cawthon}, {Chen}, {Chintalapati}, {Davis}, {Di Valentino}, {Diehl}, {Dodelson}, {Doux}, {Drlica-Wagner}, {Eckert}, {Eifler}, {Elsner}, {Everett}, {Farahi}, {Fert{\'e}}, {Fosalba}, {Friedrich}, {Gatti}, {Giannini}, {Gruen}, {Gruendl}, {Harrison}, {Hartley}, {Huff}, {Huterer}, {Kovacs}, {Leget}, {McCullough}, {Muir}, {Myles}, {Navarro-Alsina}, {Omori}, {Rollins}, {Roodman}, {Rosenfeld}, {Sevilla-Noarbe}, {Sheldon}, {Shin}, {Troja}, {Tutusaus}, {Varga}, {Wechsler}, {Yanny}, {Yin}, {Zhang}, {Zuntz}, {Abbott}, {Aguena}, {Allam}, {Annis}, {Bacon}, {Bertin}, {Brooks},
  {Burke}, {Carretero}, {Conselice}, {Costanzi}, {da Costa}, {Pereira}, {De Vicente}, {Dietrich}, {Doel}, {Evrard}, {Ferrero}, {Flaugher}, {Frieman}, {Garc{\'\i}a-Bellido}, {Gaztanaga}, {Gerdes}, {Giannantonio}, {Gschwend}, {Gutierrez}, {Hinton}, {Hollowood}, {Honscheid}, {James}, {Jeltema}, {Kuehn}, {Kuropatkin}, {Lahav}, {Lima}, {Lin}, {Maia}, {Marshall}, {Melchior}, {Menanteau}, {Miller}, {Miquel}, {Mohr}, {Morgan}, {Palmese}, {Paz-Chinch{\'o}n}, {Petravick}, {Pieres}, {Plazas Malag{\'o}n}, {Sanchez}, {Scarpine}, {Serrano}, {Smith}, {Soares-Santos}, {Suchyta}, {Tarle}, {Thomas}, {Weller}, \& {DES Collaboration}}]{Pandey2022}
{Pandey}, S., {Krause}, E., {DeRose}, J., {et~al.} 2022, \prd, 106, 043520

\bibitem[{{Peacock} \& {Bilicki}(2018)}]{Peacock&Bilicki2018}
{Peacock}, J.~A. \& {Bilicki}, M. 2018, \mnras, 481, 1133

\bibitem[{{Perlmutter} {et~al.}(1999){Perlmutter}, {Aldering}, {Goldhaber}, {Knop}, {Nugent}, {Castro}, {Deustua}, {Fabbro}, {Goobar}, {Groom}, {Hook}, {Kim}, {Kim}, {Lee}, {Nunes}, {Pain}, {Pennypacker}, {Quimby}, {Lidman}, {Ellis}, {Irwin}, {McMahon}, {Ruiz-Lapuente}, {Walton}, {Schaefer}, {Boyle}, {Filippenko}, {Matheson}, {Fruchter}, {Panagia}, {Newberg}, {Couch}, \& {Project}}]{Perlmutter1999}
{Perlmutter}, S., {Aldering}, G., {Goldhaber}, G., {et~al.} 1999, \apj, 517, 565

\bibitem[{{Philcox} \& {Ivanov}(2022)}]{Philcox2022}
{Philcox}, O. H.~E. \& {Ivanov}, M.~M. 2022, \prd, 105, 043517

\bibitem[{{Planck Collaboration} {et~al.}(2014){Planck Collaboration}, {Ade}, {Aghanim}, {Armitage-Caplan}, {Arnaud}, {Ashdown}, {Atrio-Barandela}, {Aumont}, {Baccigalupi}, {Banday}, {Barreiro}, {Bartlett}, {Battaner}, {Benabed}, {Beno{\^\i}t}, {Benoit-L{\'e}vy}, {Bernard}, {Bersanelli}, {Bielewicz}, {Bobin}, {Bock}, {Bonaldi}, {Bonavera}, {Bond}, {Borrill}, {Bouchet}, {Boulanger}, {Bridges}, {Bucher}, {Burigana}, {Butler}, {Cardoso}, {Castex}, {Catalano}, {Challinor}, {Chamballu}, {Chary}, {Chen}, {Chiang}, {Chiang}, {Christensen}, {Church}, {Clements}, {Colombi}, {Colombo}, {Couchot}, {Coulais}, {Crill}, {Cruz}, {Curto}, {Cuttaia}, {Danese}, {Davies}, {Davis}, {de Bernardis}, {de Rosa}, {de Zotti}, {Delabrouille}, {Delouis}, {D{\'e}sert}, {Dickinson}, {Diego}, {Dobler}, {Dole}, {Donzelli}, {Dor{\'e}}, {Douspis}, {Dunkley}, {Dupac}, {Efstathiou}, {En{\ss}lin}, {Eriksen}, {Falgarone}, {Finelli}, {Forni}, {Frailis}, {Fraisse}, {Franceschi}, {Galeotta}, {Ganga}, {Giard}, {Giardino}, {Giraud-H{\'e}raud},
  {Gonz{\'a}lez-Nuevo}, {G{\'o}rski}, {Gratton}, {Gregorio}, {Gruppuso}, {Hansen}, {Hanson}, {Harrison}, {Helou}, {Henrot-Versill{\'e}}, {Hern{\'a}ndez-Monteagudo}, {Herranz}, {Hildebrandt}, {Hivon}, {Hobson}, {Holmes}, {Hornstrup}, {Hovest}, {Huey}, {Huffenberger}, {Jaffe}, {Jaffe}, {Jewell}, {Jones}, {Juvela}, {Keih{\"a}nen}, {Keskitalo}, {Kisner}, {Kneissl}, {Knoche}, {Knox}, {Kunz}, {Kurki-Suonio}, {Lagache}, {L{\"a}hteenm{\"a}ki}, {Lamarre}, {Lasenby}, {Laureijs}, {Lawrence}, {Le Jeune}, {Leach}, {Leahy}, {Leonardi}, {Lesgourgues}, {Liguori}, {Lilje}, {Linden-V{\o}rnle}, {L{\'o}pez-Caniego}, {Lubin}, {Mac{\'\i}as-P{\'e}rez}, {Maffei}, {Maino}, {Mandolesi}, {Marcos-Caballero}, {Maris}, {Marshall}, {Martin}, {Mart{\'\i}nez-Gonz{\'a}lez}, {Masi}, {Massardi}, {Matarrese}, {Matthai}, {Mazzotta}, {Meinhold}, {Melchiorri}, {Mendes}, {Mennella}, {Migliaccio}, {Mikkelsen}, {Mitra}, {Miville-Desch{\^e}nes}, {Molinari}, {Moneti}, {Montier}, {Morgante}, {Mortlock}, {Moss}, {Munshi}, {Murphy}, {Naselsky}, {Nati},
  {Natoli}, {Netterfield}, {N{\o}rgaard-Nielsen}, {Noviello}, {Novikov}, {Novikov}, {O'Dwyer}, {Osborne}, {Oxborrow}, {Paci}, {Pagano}, {Pajot}, {Paladini}, {Paoletti}, {Partridge}, {Pasian}, {Patanchon}, {Pearson}, {Perdereau}, {Perotto}, {Perrotta}, {Pettorino}, {Piacentini}, {Piat}, {Pierpaoli}, {Pietrobon}, {Plaszczynski}, {Platania}, {Pointecouteau}, {Polenta}, {Ponthieu}, {Popa}, {Poutanen}, {Pratt}, {Pr{\'e}zeau}, {Prunet}, {Puget}, {Rachen}, {Reach}, {Rebolo}, {Reinecke}, {Remazeilles}, {Renault}, {Renzi}, {Ricciardi}, {Riller}, {Ristorcelli}, {Rocha}, {Roman}, {Rosset}, {Roudier}, {Rowan-Robinson}, {Rubi{\~n}o-Mart{\'\i}n}, {Rusholme}, {Salerno}, {Sandri}, {Santos}, {Savini}, {Schiavon}, {Scott}, {Seiffert}, {Shellard}, {Spencer}, {Starck}, {Stompor}, {Sudiwala}, {Sunyaev}, {Sureau}, {Sutton}, {Suur-Uski}, {Sygnet}, {Tauber}, {Tavagnacco}, {Terenzi}, {Toffolatti}, {Tomasi}, {Tristram}, {Tucci}, {Tuovinen}, {T{\"u}rler}, {Umana}, {Valenziano}, {Valiviita}, {Van Tent}, {Varis}, {Viel}, {Vielva},
  {Villa}, {Vittorio}, {Wade}, {Wandelt}, {Wehus}, {Wilkinson}, {Xia}, {Yvon}, {Zacchei}, \& {Zonca}}]{Planck2014XII}
{Planck Collaboration}, {Ade}, P.~A.~R., {Aghanim}, N., {et~al.} 2014, \aap, 571, A12

\bibitem[{{Planck Collaboration} {et~al.}(2020{\natexlab{a}}){Planck Collaboration}, {Aghanim}, {Akrami}, {Ashdown}, {Aumont}, {Baccigalupi}, {Ballardini}, {Banday}, {Barreiro}, {Bartolo}, {Basak}, {Battye}, {Benabed}, {Bernard}, {Bersanelli}, {Bielewicz}, {Bock}, {Bond}, {Borrill}, {Bouchet}, {Boulanger}, {Bucher}, {Burigana}, {Butler}, {Calabrese}, {Cardoso}, {Carron}, {Challinor}, {Chiang}, {Chluba}, {Colombo}, {Combet}, {Contreras}, {Crill}, {Cuttaia}, {de Bernardis}, {de Zotti}, {Delabrouille}, {Delouis}, {Di Valentino}, {Diego}, {Dor{\'e}}, {Douspis}, {Ducout}, {Dupac}, {Dusini}, {Efstathiou}, {Elsner}, {En{\ss}lin}, {Eriksen}, {Fantaye}, {Farhang}, {Fergusson}, {Fernandez-Cobos}, {Finelli}, {Forastieri}, {Frailis}, {Fraisse}, {Franceschi}, {Frolov}, {Galeotta}, {Galli}, {Ganga}, {G{\'e}nova-Santos}, {Gerbino}, {Ghosh}, {Gonz{\'a}lez-Nuevo}, {G{\'o}rski}, {Gratton}, {Gruppuso}, {Gudmundsson}, {Hamann}, {Handley}, {Hansen}, {Herranz}, {Hildebrandt}, {Hivon}, {Huang}, {Jaffe}, {Jones}, {Karakci},
  {Keih{\"a}nen}, {Keskitalo}, {Kiiveri}, {Kim}, {Kisner}, {Knox}, {Krachmalnicoff}, {Kunz}, {Kurki-Suonio}, {Lagache}, {Lamarre}, {Lasenby}, {Lattanzi}, {Lawrence}, {Le Jeune}, {Lemos}, {Lesgourgues}, {Levrier}, {Lewis}, {Liguori}, {Lilje}, {Lilley}, {Lindholm}, {L{\'o}pez-Caniego}, {Lubin}, {Ma}, {Mac{\'\i}as-P{\'e}rez}, {Maggio}, {Maino}, {Mandolesi}, {Mangilli}, {Marcos-Caballero}, {Maris}, {Martin}, {Martinelli}, {Mart{\'\i}nez-Gonz{\'a}lez}, {Matarrese}, {Mauri}, {McEwen}, {Meinhold}, {Melchiorri}, {Mennella}, {Migliaccio}, {Millea}, {Mitra}, {Miville-Desch{\^e}nes}, {Molinari}, {Montier}, {Morgante}, {Moss}, {Natoli}, {N{\o}rgaard-Nielsen}, {Pagano}, {Paoletti}, {Partridge}, {Patanchon}, {Peiris}, {Perrotta}, {Pettorino}, {Piacentini}, {Polastri}, {Polenta}, {Puget}, {Rachen}, {Reinecke}, {Remazeilles}, {Renzi}, {Rocha}, {Rosset}, {Roudier}, {Rubi{\~n}o-Mart{\'\i}n}, {Ruiz-Granados}, {Salvati}, {Sandri}, {Savelainen}, {Scott}, {Shellard}, {Sirignano}, {Sirri}, {Spencer}, {Sunyaev}, {Suur-Uski},
  {Tauber}, {Tavagnacco}, {Tenti}, {Toffolatti}, {Tomasi}, {Trombetti}, {Valenziano}, {Valiviita}, {Van Tent}, {Vibert}, {Vielva}, {Villa}, {Vittorio}, {Wandelt}, {Wehus}, {White}, {White}, {Zacchei}, \& {Zonca}}]{Planck2020VI}
{Planck Collaboration}, {Aghanim}, N., {Akrami}, Y., {et~al.} 2020{\natexlab{a}}, \aap, 641, A6

\bibitem[{{Planck Collaboration} {et~al.}(2020{\natexlab{b}}){Planck Collaboration}, {Aghanim}, {Akrami}, {Ashdown}, {Aumont}, {Baccigalupi}, {Ballardini}, {Banday}, {Barreiro}, {Bartolo}, {Basak}, {Benabed}, {Bernard}, {Bersanelli}, {Bielewicz}, {Bock}, {Bond}, {Borrill}, {Bouchet}, {Boulanger}, {Bucher}, {Burigana}, {Calabrese}, {Cardoso}, {Carron}, {Challinor}, {Chiang}, {Colombo}, {Combet}, {Crill}, {Cuttaia}, {de Bernardis}, {de Zotti}, {Delabrouille}, {Di Valentino}, {Diego}, {Dor{\'e}}, {Douspis}, {Ducout}, {Dupac}, {Efstathiou}, {Elsner}, {En{\ss}lin}, {Eriksen}, {Fantaye}, {Fernandez-Cobos}, {Finelli}, {Forastieri}, {Frailis}, {Fraisse}, {Franceschi}, {Frolov}, {Galeotta}, {Galli}, {Ganga}, {G{\'e}nova-Santos}, {Gerbino}, {Ghosh}, {Gonz{\'a}lez-Nuevo}, {G{\'o}rski}, {Gratton}, {Gruppuso}, {Gudmundsson}, {Hamann}, {Handley}, {Hansen}, {Herranz}, {Hivon}, {Huang}, {Jaffe}, {Jones}, {Karakci}, {Keih{\"a}nen}, {Keskitalo}, {Kiiveri}, {Kim}, {Knox}, {Krachmalnicoff}, {Kunz}, {Kurki-Suonio}, {Lagache},
  {Lamarre}, {Lasenby}, {Lattanzi}, {Lawrence}, {Le Jeune}, {Levrier}, {Lewis}, {Liguori}, {Lilje}, {Lindholm}, {L{\'o}pez-Caniego}, {Lubin}, {Ma}, {Mac{\'\i}as-P{\'e}rez}, {Maggio}, {Maino}, {Mandolesi}, {Mangilli}, {Marcos-Caballero}, {Maris}, {Martin}, {Mart{\'\i}nez-Gonz{\'a}lez}, {Matarrese}, {Mauri}, {McEwen}, {Melchiorri}, {Mennella}, {Migliaccio}, {Miville-Desch{\^e}nes}, {Molinari}, {Moneti}, {Montier}, {Morgante}, {Moss}, {Natoli}, {Pagano}, {Paoletti}, {Partridge}, {Patanchon}, {Perrotta}, {Pettorino}, {Piacentini}, {Polastri}, {Polenta}, {Puget}, {Rachen}, {Reinecke}, {Remazeilles}, {Renzi}, {Rocha}, {Rosset}, {Roudier}, {Rubi{\~n}o-Mart{\'\i}n}, {Ruiz-Granados}, {Salvati}, {Sandri}, {Savelainen}, {Scott}, {Sirignano}, {Sunyaev}, {Suur-Uski}, {Tauber}, {Tavagnacco}, {Tenti}, {Toffolatti}, {Tomasi}, {Trombetti}, {Valiviita}, {Van Tent}, {Vielva}, {Villa}, {Vittorio}, {Wandelt}, {Wehus}, {White}, {White}, {Zacchei}, \& {Zonca}}]{Planck2020VIII}
{Planck Collaboration}, {Aghanim}, N., {Akrami}, Y., {et~al.} 2020{\natexlab{b}}, \aap, 641, A8

\bibitem[{{Pullen} {et~al.}(2016){Pullen}, {Alam}, {He}, \& {Ho}}]{Pullen2016}
{Pullen}, A.~R., {Alam}, S., {He}, S., \& {Ho}, S. 2016, MNRAS, 460, 4098

\bibitem[{{Riess} {et~al.}(1998){Riess}, {Filippenko}, {Challis}, {Clocchiatti}, {Diercks}, {Garnavich}, {Gilliland}, {Hogan}, {Jha}, {Kirshner}, {Leibundgut}, {Phillips}, {Reiss}, {Schmidt}, {Schommer}, {Smith}, {Spyromilio}, {Stubbs}, {Suntzeff}, \& {Tonry}}]{Riess1998}
{Riess}, A.~G., {Filippenko}, A.~V., {Challis}, P., {et~al.} 1998, \aj, 116, 1009

\bibitem[{{Robertson} {et~al.}(2021){Robertson}, {Alonso}, {Harnois-D{\'e}raps}, {Darwish}, {Kannawadi}, {Amon}, {Asgari}, {Bilicki}, {Calabrese}, {Choi}, {Devlin}, {Dunkley}, {Dvornik}, {Erben}, {Ferraro}, {Fortuna}, {Giblin}, {Han}, {Heymans}, {Hildebrandt}, {Hill}, {Hilton}, {Ho}, {Hoekstra}, {Hubmayr}, {Hughes}, {Joachimi}, {Joudaki}, {Knowles}, {Kuijken}, {Madhavacheril}, {Moodley}, {Miller}, {Namikawa}, {Nati}, {Niemack}, {Page}, {Partridge}, {Schaan}, {Schillaci}, {Schneider}, {Sehgal}, {Sherwin}, {Sif{\'o}n}, {Staggs}, {Tr{\"o}ster}, {van Engelen}, {Valentijn}, {Wollack}, {Wright}, \& {Xu}}]{Robertson2021}
{Robertson}, N.~C., {Alonso}, D., {Harnois-D{\'e}raps}, J., {et~al.} 2021, \aap, 649, A146

\bibitem[{{Saraf} {et~al.}(2022){Saraf}, {Bielewicz}, \& {Chodorowski}}]{Saraf2022}
{Saraf}, C.~S., {Bielewicz}, P., \& {Chodorowski}, M. 2022, \mnras, 515, 1993

\bibitem[{{Scodeggio} {et~al.}(2018){Scodeggio}, {Guzzo}, {Garilli}, {Granett}, {Bolzonella}, {de la Torre}, {Abbas}, {Adami}, {Arnouts}, {Bottini}, {Cappi}, {Coupon}, {Cucciati}, {Davidzon}, {Franzetti}, {Fritz}, {Iovino}, {Krywult}, {Le Brun}, {Le F{\`e}vre}, {Maccagni}, {Ma{\l}ek}, {Marchetti}, {Marulli}, {Polletta}, {Pollo}, {Tasca}, {Tojeiro}, {Vergani}, {Zanichelli}, {Bel}, {Branchini}, {De Lucia}, {Ilbert}, {McCracken}, {Moutard}, {Peacock}, {Zamorani}, {Burden}, {Fumana}, {Jullo}, {Marinoni}, {Mellier}, {Moscardini}, \& {Percival}}]{VIPERSDR2_2018}
{Scodeggio}, M., {Guzzo}, L., {Garilli}, B., {et~al.} 2018, \aap, 609, A84

\bibitem[{{Secco} {et~al.}(2022){Secco}, {Samuroff}, {Krause}, {Jain}, {Blazek}, {Raveri}, {Campos}, {Amon}, {Chen}, {Doux}, {Choi}, {Gruen}, {Bernstein}, {Chang}, {DeRose}, {Myles}, {Fert{\'e}}, {Lemos}, {Huterer}, {Prat}, {Troxel}, {MacCrann}, {Liddle}, {Kacprzak}, {Fang}, {S{\'a}nchez}, {Pandey}, {Dodelson}, {Chintalapati}, {Hoffmann}, {Alarcon}, {Alves}, {Andrade-Oliveira}, {Baxter}, {Bechtol}, {Becker}, {Brandao-Souza}, {Camacho}, {Carnero Rosell}, {Carrasco Kind}, {Cawthon}, {Cordero}, {Crocce}, {Davis}, {Di Valentino}, {Drlica-Wagner}, {Eckert}, {Eifler}, {Elidaiana}, {Elsner}, {Elvin-Poole}, {Everett}, {Fosalba}, {Friedrich}, {Gatti}, {Giannini}, {Gruendl}, {Harrison}, {Hartley}, {Herner}, {Huang}, {Huff}, {Jarvis}, {Jeffrey}, {Kuropatkin}, {Leget}, {Muir}, {Mccullough}, {Navarro Alsina}, {Omori}, {Park}, {Porredon}, {Rollins}, {Roodman}, {Rosenfeld}, {Ross}, {Rykoff}, {Sanchez}, {Sevilla-Noarbe}, {Sheldon}, {Shin}, {Troja}, {Tutusaus}, {Varga}, {Weaverdyck}, {Wechsler}, {Yanny}, {Yin}, {Zhang},
  {Zuntz}, {Abbott}, {Aguena}, {Allam}, {Annis}, {Bacon}, {Bertin}, {Bhargava}, {Bridle}, {Brooks}, {Buckley-Geer}, {Burke}, {Carretero}, {Costanzi}, {da Costa}, {De Vicente}, {Diehl}, {Dietrich}, {Doel}, {Ferrero}, {Flaugher}, {Frieman}, {Garc{\'\i}a-Bellido}, {Gaztanaga}, {Gerdes}, {Giannantonio}, {Gschwend}, {Gutierrez}, {Hinton}, {Hollowood}, {Honscheid}, {Hoyle}, {James}, {Jeltema}, {Kuehn}, {Lahav}, {Lima}, {Lin}, {Maia}, {Marshall}, {Martini}, {Melchior}, {Menanteau}, {Miquel}, {Mohr}, {Morgan}, {Ogando}, {Palmese}, {Paz-Chinch{\'o}n}, {Petravick}, {Pieres}, {Plazas Malag{\'o}n}, {Rodriguez-Monroy}, {Romer}, {Sanchez}, {Scarpine}, {Schubnell}, {Scolnic}, {Serrano}, {Smith}, {Soares-Santos}, {Suchyta}, {Swanson}, {Tarle}, {Thomas}, {To}, \& {DES Collaboration}}]{Secco2022}
{Secco}, L.~F., {Samuroff}, S., {Krause}, E., {et~al.} 2022, \prd, 105, 023515

\bibitem[{{Shekhar Saraf} \& {Bielewicz}(2024)}]{Saraf2024}
{Shekhar Saraf}, C. \& {Bielewicz}, P. 2024, \aap, 687, A150

\bibitem[{{Singh} {et~al.}(2017){Singh}, {Mandelbaum}, \& {Brownstein}}]{Singh2017}
{Singh}, S., {Mandelbaum}, R., \& {Brownstein}, J.~R. 2017, \mnras, 464, 2120

\bibitem[{{Skara} \& {Perivolaropoulos}(2020)}]{Skara2020}
{Skara}, F. \& {Perivolaropoulos}, L. 2020, \prd, 101, 063521

\bibitem[{{Solarz} {et~al.}(2015){Solarz}, {Pollo}, {Takeuchi}, {Ma{\l}ek}, {Matsuhara}, {White}, {P{\c{e}}piak}, {Goto}, {Wada}, {Oyabu}, {Takagi}, {Ohyama}, {Pearson}, {Hanami}, {Ishigaki}, \& {Malkan}}]{Solarz2015}
{Solarz}, A., {Pollo}, A., {Takeuchi}, T.~T., {et~al.} 2015, \aap, 582, A58

\bibitem[{{Spergel} {et~al.}(2013){Spergel}, {Gehrels}, {Breckinridge}, {Donahue}, {Dressler}, {Gaudi}, {Greene}, {Guyon}, {Hirata}, {Kalirai}, {Kasdin}, {Moos}, {Perlmutter}, {Postman}, {Rauscher}, {Rhodes}, {Wang}, {Weinberg}, {Centrella}, {Traub}, {Baltay}, {Colbert}, {Bennett}, {Kiessling}, {Macintosh}, {Merten}, {Mortonson}, {Penny}, {Rozo}, {Savransky}, {Stapelfeldt}, {Zu}, {Baker}, {Cheng}, {Content}, {Dooley}, {Foote}, {Goullioud}, {Grady}, {Jackson}, {Kruk}, {Levine}, {Melton}, {Peddie}, {Ruffa}, \& {Shaklan}}]{WFIRST2013}
{Spergel}, D., {Gehrels}, N., {Breckinridge}, J., {et~al.} 2013, arXiv e-prints, arXiv:1305.5422

\bibitem[{{St{\"o}lzner} {et~al.}(2021){St{\"o}lzner}, {Joachimi}, {Korn}, {Hildebrandt}, \& {Wright}}]{Stolzner2021}
{St{\"o}lzner}, B., {Joachimi}, B., {Korn}, A., {Hildebrandt}, H., \& {Wright}, A.~H. 2021, \aap, 650, A148

\bibitem[{{Sun} {et~al.}(2022){Sun}, {Yao}, {Dong}, {Yang}, {Zhang}, \& {Zhang}}]{Sun2022}
{Sun}, Z., {Yao}, J., {Dong}, F., {et~al.} 2022, \mnras, 511, 3548

\bibitem[{{The Dark Energy Survey Collaboration}(2005)}]{DES2005}
{The Dark Energy Survey Collaboration}. 2005, arXiv e-prints, arXiv:astro ph/0510346

\bibitem[{{Trimble}(1987)}]{Trimble1987}
{Trimble}, V. 1987, \araa, 25, 425

\bibitem[{{Wang} {et~al.}(2023){Wang}, {Yao}, {Liu}, {Liu}, {Fan}, \& {Hu}}]{Zhengyi2023}
{Wang}, Z., {Yao}, J., {Liu}, X., {et~al.} 2023, \mnras, 523, 3001

\bibitem[{{White} {et~al.}(2022){White}, {Zhou}, {DeRose}, {Ferraro}, {Chen}, {Kokron}, {Bailey}, {Brooks}, {Garc{\'\i}a-Bellido}, {Guy}, {Honscheid}, {Kehoe}, {Kremin}, {Levi}, {Palanque-Delabrouille}, {Poppett}, {Schlegel}, \& {Tarle}}]{White2022}
{White}, M., {Zhou}, R., {DeRose}, J., {et~al.} 2022, \jcap, 2022, 007

\bibitem[{{Wright} {et~al.}(2010){Wright}, {Eisenhardt}, {Mainzer}, {Ressler}, {Cutri}, {Jarrett}, {Kirkpatrick}, {Padgett}, {McMillan}, {Skrutskie}, {Stanford}, {Cohen}, {Walker}, {Mather}, {Leisawitz}, {Gautier}, {McLean}, {Benford}, {Lonsdale}, {Blain}, {Mendez}, {Irace}, {Duval}, {Liu}, {Royer}, {Heinrichsen}, {Howard}, {Shannon}, {Kendall}, {Walsh}, {Larsen}, {Cardon}, {Schick}, {Schwalm}, {Abid}, {Fabinsky}, {Naes}, \& {Tsai}}]{Wright2010}
{Wright}, E.~L., {Eisenhardt}, P. R.~M., {Mainzer}, A.~K., {et~al.} 2010, \aj, 140, 1868

\bibitem[{{Xavier} {et~al.}(2016){Xavier}, {Abdalla}, \& {Joachimi}}]{Flask2016}
{Xavier}, H.~S., {Abdalla}, F.~B., \& {Joachimi}, B. 2016, \mnras, 459, 3693

\bibitem[{{Yu} {et~al.}(2022){Yu}, {Ferraro}, {Knight}, {Knox}, \& {Sherwin}}]{Yu2022}
{Yu}, B., {Ferraro}, S., {Knight}, Z.~R., {Knox}, L., \& {Sherwin}, B.~D. 2022, \mnras, 513, 1887

\bibitem[{{Zhang} {et~al.}(2017){Zhang}, {Yu}, \& {Zhang}}]{Zhang2017}
{Zhang}, L., {Yu}, Y., \& {Zhang}, P. 2017, \apj, 848, 44

\bibitem[{{Zhang} {et~al.}(2010){Zhang}, {Pen}, \& {Bernstein}}]{Zhang2010}
{Zhang}, P., {Pen}, U.-L., \& {Bernstein}, G. 2010, \mnras, 405, 359

\bibitem[{{Zou} {et~al.}(2017){Zou}, {Zhou}, {Fan}, {Zhang}, {Zhou}, {Nie}, {Peng}, {McGreer}, {Jiang}, {Dey}, {Fan}, {He}, {Jiang}, {Lang}, {Lesser}, {Ma}, {Mao}, {Schlegel}, \& {Wang}}]{BASS2017}
{Zou}, H., {Zhou}, X., {Fan}, X., {et~al.} 2017, \pasp, 129, 064101

\end{thebibliography}

\FloatBarrier 
\begin{appendix}
\onecolumn
\section{Power spectra from simulations}\label{sec_appndx:appndx_a}
\begin{figure}[hbt!]
    \begin{subfigure}[b]{0.88\linewidth}
        \centering
        \includegraphics[width=14.8cm]{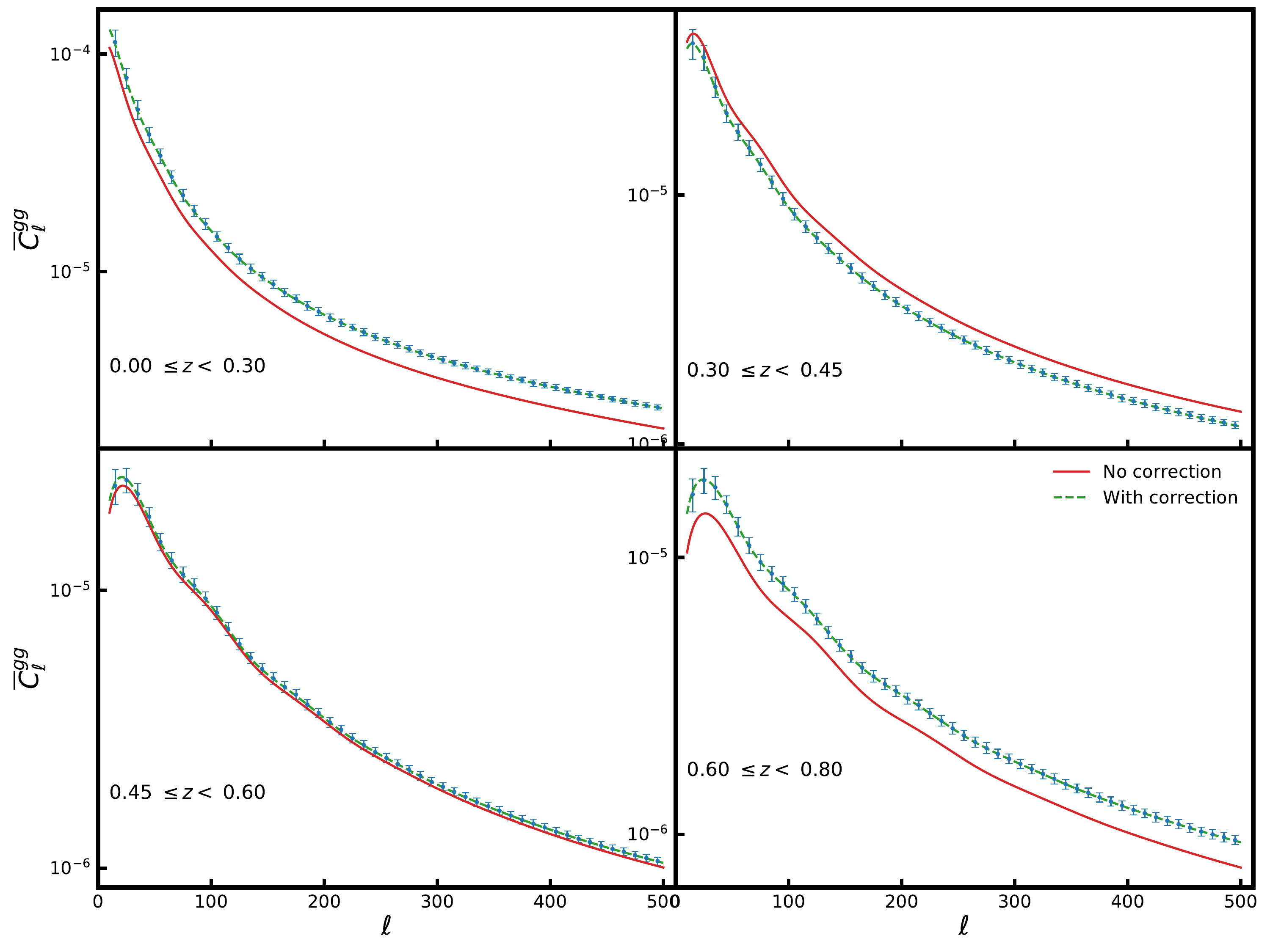}
        \captionsetup{labelformat=empty}
    \end{subfigure}\\
    \begin{subfigure}[b]{0.88\linewidth}
        \centering
        \includegraphics[width=14.8cm]{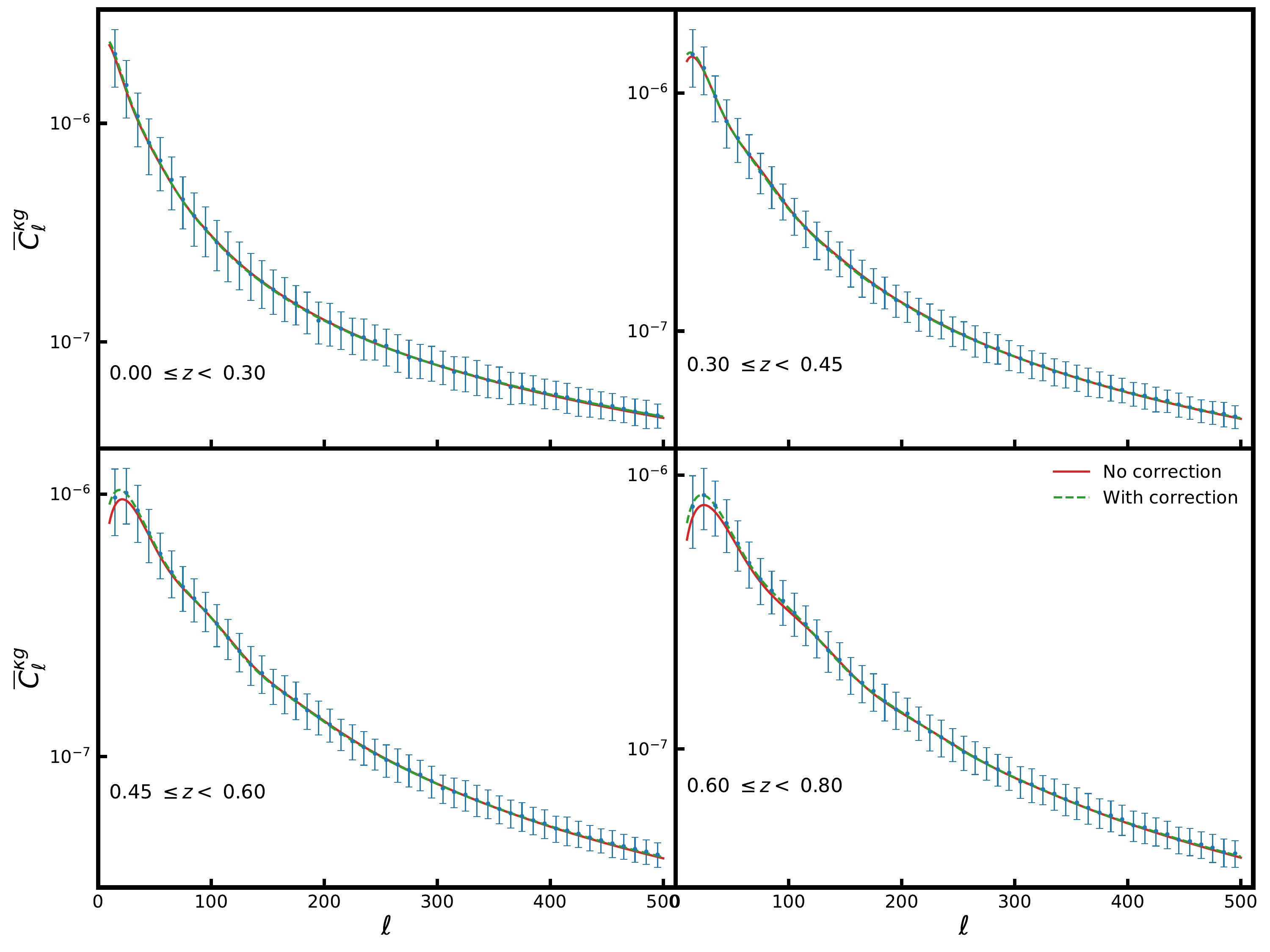}
        \captionsetup{labelformat=empty}
    \end{subfigure}%
    \caption{{(Top:)} Galaxy auto-power spectra and {(bottom:)} cross-power spectra averaged from $300$ simulations. The red solid lines are the theoretical power spectra estimated from Eq. (\ref{eq:true_dist_conv_paper3}) and the green dashed lines are the leakage corrected theoretical power spectra. {The standard error bars on the data points are estimated from Eq.\,(\ref{eq:err_data_points_group_hang}).}}
    \label{fig:power_spectra_simualtions_desi_avg_gg_and_kg}
\end{figure}

\begin{figure}[hbt!]
    \begin{subfigure}[b]{0.95\linewidth}
        \centering
        \includegraphics[width=15cm]{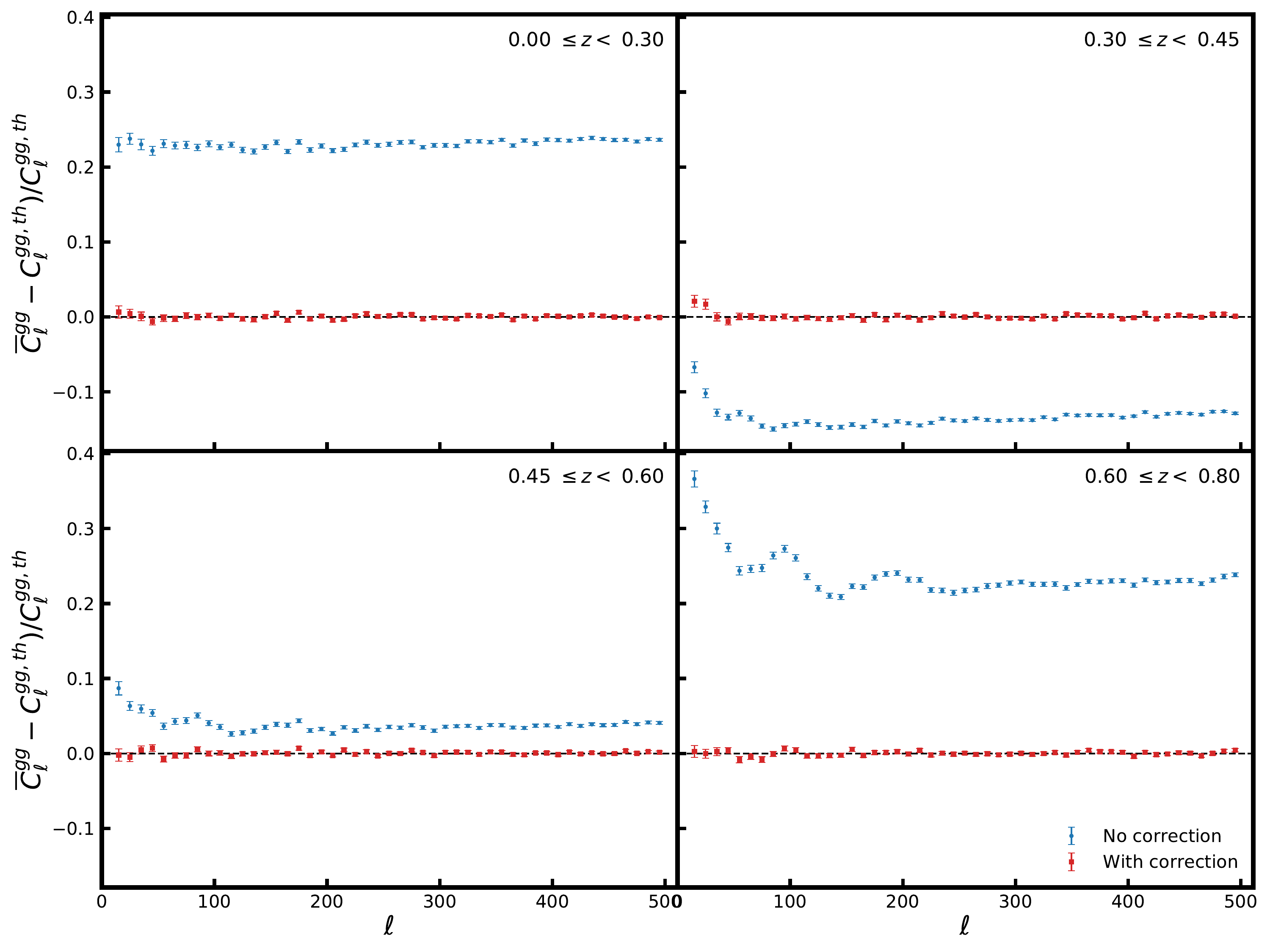}
        \captionsetup{labelformat=empty}
    \end{subfigure}\\
    \begin{subfigure}[b]{0.95\linewidth}
        \centering
        \includegraphics[width=15cm]{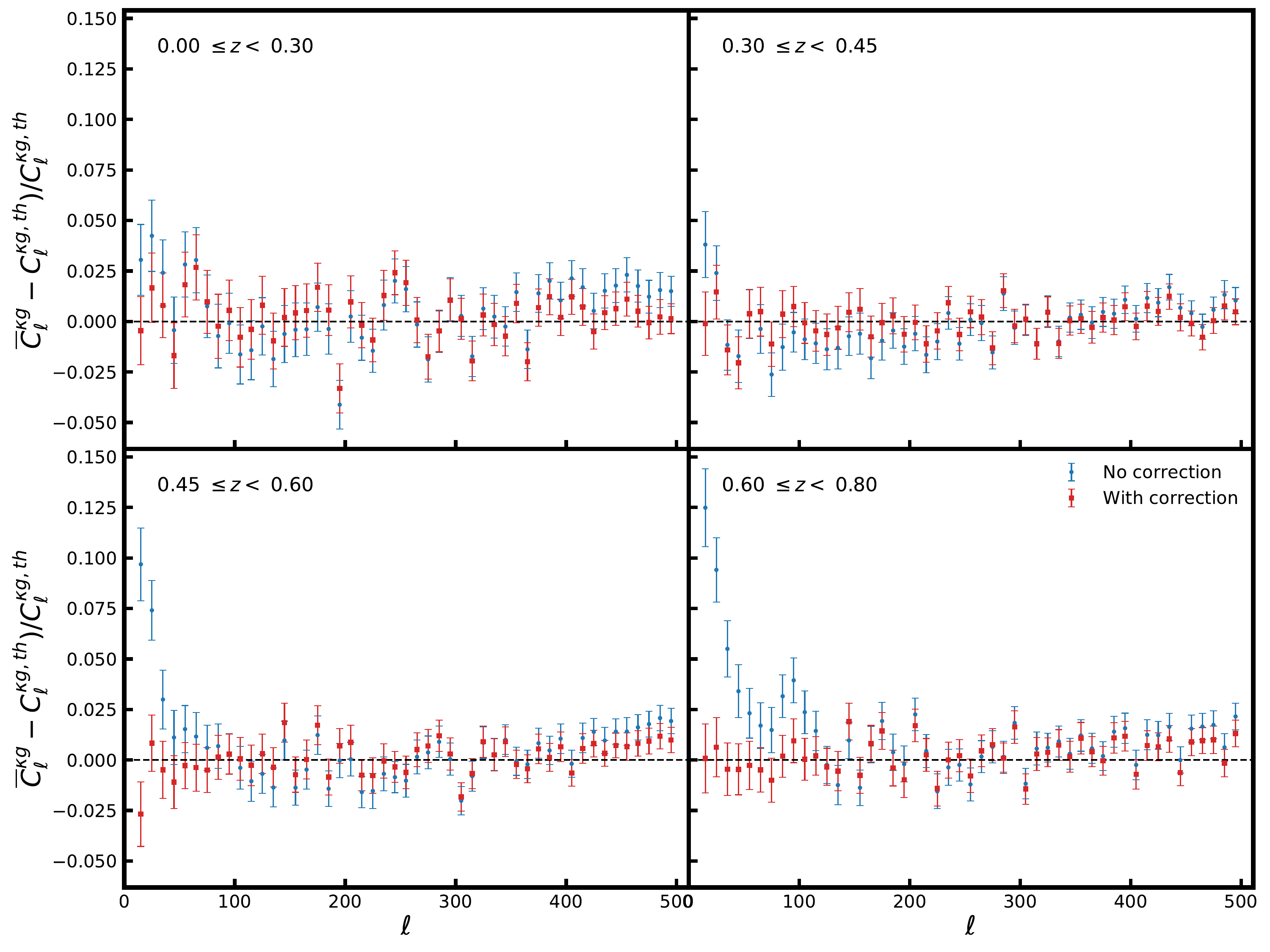}
        \captionsetup{labelformat=empty}
    \end{subfigure}
    \caption{Relative errors on the average ({top}) galaxy auto-power spectra and ({bottom}) cross-power spectra without (blue circles) and after (red squares) correcting for the redshift bin mismatch. {The standard error bars are estimated from Eq.\,(\ref{eq:err_data_points_group_hang}).}}
    \label{fig:power_spectra_simualtions_desi_avg_gg_and_kg_rel_err}
\end{figure}
\end{appendix}
\end{document}